\documentclass[a4paper,fleqn]{./elsevier_class/cas-dc}

\usepackage[authoryear]{natbib}
\usepackage{subfigure}
\usepackage{hyperref}
\usepackage{xspace}
\usepackage{makecell}
\usepackage{caption}
\newcommand{\repeatcaption}[2]{%
  \renewcommand{\thetable}{\ref{tab:table_parameters}}%
  \captionsetup{list=no}%
  \caption{(continued)}%
  \addtocounter{table}{-1}% So that next table after the repeat gets the right number.
}
\def\tsc#1{\csdef{#1}{\textsc{\lowercase{#1}}\xspace}}
\tsc{WGM}
\tsc{QE}
\begin{document}
\let\WriteBookmarks\relax
\def\floatpagepagefraction{1}
\def\textpagefraction{.001}

\shorttitle{Star-planet interaction and its impact on the stellar rotation}    

\shortauthors{T. M. Santiago et al.}  

\title[mode = title]{Star-planet interaction and its impact on the stellar rotation}  

%\tnotemark[<tnote number>] 

%\tnotetext[<tnote number>]{<tnote text>} 


\author[1,2]{Thiago M. Santiago}[]

%\cormark[<corr mark no>]

%\fnmark[<footnote mark no>]

%\ead{thiago@fisica.ufc.br}

%\ead[url]{<URL>}

\credit{<Credit authorship details>}

\author[1]{Sarah G. A. Barbosa}[]

%\cormark[<corr mark no>]

%\fnmark[<footnote mark no>]

%\ead{thiago@fisica.ufc.br}

%\ead[url]{<URL>}

%\credit{<Credit authorship details>}

\affiliation[1]{organization={Departamento de Física, Universidade Federal do Ceará},
            addressline={Caixa Postal 6030, Campus do Pici}, 
            city={Fortaleza},
%          citysep={}, % Uncomment if no comma needed between city and postcode
            postcode={60455-900}, 
            state={Ceará},
            country={Brazil}}

\affiliation[2]{organization={Centro de Ciências e Tecnologia, Universidade Federal do Cariri},
            addressline={Campus Juazeiro do Norte}, 
            city={Juazeiro do Norte},
%          citysep={}, % Uncomment if no comma needed between city and postcode
            postcode={63048-080}, 
            state={Ceará},
            country={Brazil}}

\author[3]{Francisco J. Cavalcante}[]

% Footnote of the second author

% Email id of the second author
%\ead{}


\credit{}

\affiliation[3]{organization={Departamento de Física, Instituto Federal de Educação, Ciência e Tecnologia do Ceará},
            addressline={Campus Tianguá}, 
            city={Tianguá},
%          citysep={}, % Uncomment if no comma needed between city and postcode
            postcode={62324-075}, 
            state={Ceará},
            country={Brazil}}

\author[1]{Daniel B. de Freitas}[]

%\cormark[<corr mark no>]

%\fnmark[<footnote mark no>]

\ead{danielbrito@fisica.ufc.br}

%\ead[url]{<URL>}

%\credit{<Credit authorship details>}

\cortext[1]{Corresponding author}

\fntext[1]{}


\begin{abstract}
The evolution of the star-planet systems involves an intricate interplay of tidal interaction and magnetic braking. The stellar rotation has an essential role in modifying the structure of the star and, therefore, the way these different interplays arise. On the other hand, changes in orbits impact the star's rotation and its evolution. The evolution of the star's rotation accounts for the angular momentum exchange with the planet and follows the effects of the internal transport of angular momentum and metallicity. Several models in the literature have aimed to find a theoretical way to study these interactions between the planet's orbital evolution and the star's rotation. Our work is a promising attempt to investigate these interactions from a model based on a new statistical approach. To this end, we propose a ``tidal interaction index'' that carries all the parameters of the star-planet system that can affect the transport of angular momentum and, consequently, the evolution of stellar rotation. This index is similar to the ``magnetic braking index'' whose most successful value equals 3, which expresses the seminal Skumunich law.
Our model is computed for masses of the host star less than the Kraft limit for three orbital-rotation period regimes and the semi-major axis less than 1 AU. We consider planets with masses between 0.4M$_{\oplus}$ and 20M$_{\rm J}$ with orbital periods between 0.3 and 225 days.
We show that the tidal index $q$ segregated by stellar mass without wind magnetic braking during the main-sequence phase is strongly anti-correlated with planetary mass. Finally, we conclude that in cases where planets retain less than 84\% of the total angular momentum within the system, the magnetic braking mechanism proves to be more effective than tidal interactions, irrespective of whether the planets' angular momentum surpasses that of the host star. Conversely, our analysis demonstrates that in planetary systems such as WASP-89, tidal interactions notably influence the evolution of stellar rotation.
\end{abstract}


%\begin{HIGHLIGHTS}
%\item
%\item 
%\item 
%\end{HIGHLIGHTS}

% Keywords
\begin{keywords}
 stellar rotation\sep star-planet interaction \sep \sep
\end{keywords}

\maketitle

% Main text
%\section{}\label{}

% Use the style of numbering in square brackets.
% If nothing is used, default style will be taken.
%\begin{enumerate}[a)]
%\item 
%\item 
%\item 
%\end{enumerate}  

%\begin{itemize}
%\item 
%\item 
%\item 
%\end{itemize}  

%\begin{description}
%\item[]
%\item[] 
%\item[] 
%\end{description}  

% % Figure
% \begin{figure}[<options>]
% 	\centering
% 		\includegraphics[<options>]{}
% 	  \caption{}\label{fig1}
% \end{figure}

% \begin{table}[<options>]
% \begin{tabular*}{\tblwidth}{@{}LL@{}}
% \toprule
% \midrule
%  & \\
%  & \\
%  & \\
%  & \\
% \bottomrule
% \end{tabular*}
% \end{table}

% Uncomment and use as the case may be
%\begin{theorem} 
%\end{theorem}

% Uncomment and use as the case may be
%\begin{lemma} 
%\end{lemma}


\section{Introduction}\label{sec:level1}

The evolution of stellar rotation in star-planet systems involves the intricate interplay of tidal interaction and magnetic braking \citep{privitera2016}. As a star with close-in exoplanet experiences tidal forces, the planet's gravitational pull can cause the star to deform, leading to tidal interactions that influence both the star and the planet's orbits \citep{privitera2016,busetti2018,benba}. Additionally, magnetic braking, driven by the star's magnetic field interacting with its stellar wind, can slow down its rotation. This dynamic interaction between tidal forces and magnetic braking affects the rotational evolution of the star and its orbital dynamics with the planet. Other research focuses on understanding these processes and their impact on the long-term evolution of star-planet systems, as seen in \cite{nascimento2014} and \cite{benba}.

It is widely accepted that magnetic braking is a fundamental concept for understanding angular momentum losses due to magnetic stellar winds for several classes of stars, such as main-sequence field and cluster stars. This mechanism was initially suggested by \citet{Schatzman}, who pointed out that slow rotators have convective envelopes. As mentioned by \citet{Kraft}, the behavior of the mean rotational velocity of low-mass-main-sequence stars below 1.5$M_{\odot}$ (spectral type F0) is preferentially due to magnetic wind. A few years later, \citet{sku}'s pioneering work argued that stellar rotation, activity, and lithium abundances for solar-like stars obey a simple relationship given by $\left(\mathrm dJ/\mathrm dt\right)_{\rm wind} \propto\Omega_{*}^{3}$, where $t$ is time, $J$ is the angular momentum and $\Omega_{*}$ denotes the stellar angular velocity \citep{deFreitas2013,deFreitas2021,deFreitas2022,Silvaetal13}. The authors mentioned above established early on that stellar rotation and age should be related to cool main-sequence stars since magnetic braking is modeled with a Skumanich-type law and, therefore, a torque of magnitude $\Omega^{3}$ \citep{defreitas2014,defreitasetal2015}.

Within this scenario, something is exciting. Understanding how short-period giant planets, often known as hot Jupiters, came to be in their present orbit is one of the long-standing issues in the study of exoplanets \citep{barker2009}. The dominant idea states that massive planets cannot form inside a protoplanetary disk until they reach the snow line, usually a few astronomical units away from the star \citep{dobbs2004}. According to \cite{damiani2015}, because of their 
0.1 AU semi-major axis, hot Jupiters have to have migrated somehow. In the literature, numerous migration ideas are examined, including various braking mechanisms that may be verified by contrasting the theoretical predictions based on the observable orbital characteristics of exoplanets \citep[more recently papers, as][analyze this issue using M-dwarf systems observed by TESS mission]{alvarado2022}.

\subsection{\label{sec:level11}Types of star-planet interactions}
Star-planet interactions can lead to observable signatures on the star and the planet. These interactions can be categorized into different types. As mentioned by \cite{vidotto2020}, four different interactions between a planet and its host star can characterize an exoplanetary system.

Different outcomes from these interactions could affect the planet or the star in other ways, such as orbital migration or the creation of chromospheric hotspots. In turn, these effects can produce a variety of observables across the electromagnetic spectrum, ranging from radio to ultraviolet/X-rays. They are dependent upon the evolutionary phase as well as the physical properties of the system, including the host stars' activity, orbital distance, and age \citep[see][]{vidotto2014}. These interactions can be subdivided into two groups depending on where the observable signatures occur.
        
In the first group are the interactions where the observable signatures are on the star. The first is tidal interaction, which arises when the gravitational forces between the star and the planet cause tidal forces, leading to changes in the star's shape and observable signatures \citep{Strugarek}. Secondly, magnetic interaction, on the other hand, is observed when the star and the planet are sufficiently close-in \citep{Strugarek}. An essential consideration in this interaction is the orbital distance relative to the Alfvén surface of the star. The Alfvén surface serves as a boundary that separates the region of the stellar wind where magnetic forces dominate (within the surface) from the region where inertial forces prevail (beyond the surface). Within the Alfvén surface, the stellar wind is characterized as sub-Alfvenic, and Alfvén ``wing'' currents form, establishing a direct connection between the star and the planet. However, in the super-Alfvenic regime, this direct connection ceases to exist \citep[cf.][]{Strug2015}.
        
In the second group are the interactions where the observable signatures are on the planet. Stellar wind-planet interaction is another type that can produce observable signatures on the planet. This interaction can lead to phenomena such as planetary radio emission, the formation of ``sideways'' bow shocks, and planetary migration \citep{Shkolnik,Ahuir}. In particular, the solar wind surrounding Earth and the stellar wind surrounding a nearby exoplanet (hot Jupiter) differ significantly. Since close-in planets are ``parked'' inside the acceleration zone of stellar winds, their close-in locations suggest that they interact with higher density winds, higher ambient magnetic fields, and generally lower wind velocities than Earth \citep{vidotto2014,vidotto2020}. Lastly, radiative star-planet interaction can also produce observable signatures on the planet. This type of interaction can lead to phenomena such as the photo-evaporation of the planetary atmosphere and the inflation of planets \citep{he}.

\subsection{\label{sec:level12}Spin-down of stars and the tidal interaction}
As mentioned by \cite{lanza2015,damiani2015}, tidal contact between the planet and star might trigger subsequent secular changes in exoplanet orbits even when the original migration process is ineffective. Since the tidal torque scales as the inverse of the semi-major axis's sixth power, it is particularly significant for hot Jupiters. Estimating the effectiveness of tidal dissipation and its consequences across the star's evolutionary history is essential to evaluate the migration possibilities \citep{barker2009}. However, because of the gaps in our understanding of the precise process causing tidal dissipation, its effectiveness, and the consequences of the host's angular momentum loss owing to magnetic braking, it is challenging to draw firm conclusions.
        
Even without a detailed knowledge of the angular momentum loss law or tidal dissipation mechanisms, there is a way to assess the general outcome of tidal evolution, considering that the total angular momentum of the star-planet system is not conserved because magnetic braking exerts a torque on the star \citep{lanza2015,benba}. In this way, a relationship similar to the torque due to the magnetized wind (shown just above) can be used to represent the torque law for tidal interaction effects given by $\left(\mathrm dJ/\mathrm dt\right)_{\rm tidal} \propto\Omega_{*}^{q}$. It is important to emphasize that the exponent $q$ of this expression differs from that proposed by \cite{deFreitas2013}. The angular momentum loss rate in the two cases is expected to vary and involve different parameters. While $\left(\mathrm dJ/\mathrm dt\right)_{\rm wind}$ for systems without planets involves only stellar parameters, $\left(\mathrm dJ/\mathrm dt\right)_{\rm tidal}$ for systems with planets includes components of the star's spin and planetary orbital parameters. 
        
Several authors \cite[e.g.,][]{lanza2015,ahuir2021} consider that the stellar spin controls the evolution of stellar rotation when the wind torque has a substantially greater amplitude than the tidal torque. This behavior is the case for typical stellar rotation rates of planets and young stars not too close to the 2:1 mean motion resonance, as reported by \cite{damiani2015}. In addition to this condition, there are possibilities that the tidal torque and wind torque are equal in magnitude or that the tidal torque exceeds the wind torque. In the first case, the planet's migration can be slowed down by the system entering a stationary state where tidal evolution proceeds at a nearly constant stellar spin frequency \citep{lanza2022}. As long as there is sufficient orbital angular momentum relative to the stellar rotational angular momentum to maintain the torque balance, the tidal torque must be opposite in sign and comparable in magnitude to the wind torque for the stationary state to be established \citep{spada2011}. In the latter case, tidal dissipation sets the characteristic time associated with the final stages of the evolution when the planet spirals into the star. This condition happens at almost constant angular momentum. As a consequence, there is thus a very low probability of massive planets being observed in this phase of evolution. On the other hand, low-mass planets can have a spiral time longer than the principal sequence lifetime of their host star \citep{barker2009}.

More details about these three conditions can be found in \cite{lanza2015}, where the authors revisit Darwin's model \citep{darwin1879,darwin}. A more detailed study of the behavior between tidal interaction and magnetized winds is crucial, especially when it comes to close-in systems, such as M dwarfs, where the habitable zone (HZ) is close to the star, and these winds can lead to atmospheric erosion \citep{chebly2022}.
        
More recently, \cite{deFreitas2022} showed that the deviation from Skumanich's law is driven primarily by stellar density and rotation period for systems without planets. The authors also revealed that the exponent $q$ that controls the loss of angular momentum due to wind is a powerful proxy for distinguishing stars in evolutionary phases. The authors also concluded that said exponent is limited to a narrow range between 1 and 3. This behavior can be quite adverse when the problem is star-planet interaction. In the present study, we perform several correlations between stellar and planetary parameters, and we investigate their behavior with the index $q$, for which we aim to measure the effects of star-planet interaction on stellar rotation. In summary, we characterize the evolution of stellar rotation under tidal dissipation due to the presence of exoplanets. 

The paper is organized as follows: Section \ref{sec:level2} shows our working sample, considering systems constituted by stars with masses less than 1.5$M_{\odot}$ star with one or more planets. In Section 3, we model the effects of tidal interaction and its physical mechanisms that could affect stellar rotation. In Section 4, we test our set of stellar targets within the framework of our model and investigate different scenarios that mainly involve the behavior of the spin-orbit state with parameters such as synchronization time and corotation radius. Finally, in the last Section, we discuss our main conclusions.
            
\section{\label{sec:level2}Working Sample}
Data related to planetary systems were obtained from the repository of the NASA Exoplanet Archive\footnote{\url{https://exoplanetarchive.ipac.caltech.edu/index.html}}. The research selected datasets with information regarding crucial parameters such as stellar rotation period, eccentricity, stellar and planetary radius, orbital period, stellar and planetary mass, and the semi-major axis. The stellar rotation period is scarce data in the planetary systems catalog; therefore, filtering systems containing this parameter dramatically reduces the size of our sample. In Table \ref{tab:table_flags} is shown the criteria or flags used for the filtering.
%aqui acima, adicionei planetary radius

The planetary and stellar parameter sets were determined following the references mostly provided by \citet{Bonono2017} along with \citet{Barros2022}. Employing this criteria-based filtering process and allowing the use of non-standard parameter sets, an original dataset comprising 210 different exoplanets distributed in 162 planetary systems was compiled. These records were extracted up until November 2023. 

\begin{table}
    \caption{Observables and its flags for datasets filtering, where the abbreviations dset\#01, dset\#02 and dset\#03 correspond to \texttt{dataset\#01}, \texttt{dataset\#02} and \texttt{dataset\#03}, respectively. When the database search did not make any restrictions, the flag field was left blank, and here we represent it by the word \emph{any}. The value 1 for the \emph{number of stars} filter option indicates the number of stars in a given system, while for the \emph{default parameter set} filtering option, it indicates that the search should be restricted to default values only. The filtering option represented by the name of the observable followed by the word \emph{limit} refers to the type of value assigned to that quantity. In these cases, when a flag equal to 0 is assigned to this type of filter, the average value followed by the lower and upper limits of uncertainty is accepted for that quantity. If no flag is assigned (indicated by the word \emph{any}), the data set can only contain estimates for the minimum or maximum values of this quantity. The \emph{not null} flag forces the database search to return only those records containing non-missing data for a given observable, while the flag 0 for the \emph{controversial flag} option indicates that only values without dispute in the literature were selected for the quantities.}\label{tab:table_flags}
    \begin{tabular}{llll}
        \toprule
        Filter option & dset\#01 & dset\#02 & dset\#03\\
        \midrule
        default parameter set           & any       & 1         & 1         \\
        number of stars                 & 1         & 1         & 1         \\
        controversial flag              & any       & 0         & 0         \\
        orbital period ($P_{\rm orb}$)      & not null  & not null  & not null  \\
        orbital period limit            & any       & any       & 0         \\
        semi-major axis ($a$)           & not null  & not null  & not null  \\
        semi-major axis limit           & any       & any       & 0         \\
        planet radius ($R_{p})$         & not null  & not null  & not null  \\
        planet radius  limit            & any       & any       & 0         \\
        planet mass ($M_{p}$)           & not null  & not null  & not null  \\
        planet mass limit               & any       & any       & 0         \\
        eccentricity ($e$)              & not null  & not null  & not null  \\
        eccentricity limit              & any       & any       & 0         \\
        stellar radius ($R_{*}$)        & not null  & not null  & not null  \\
        stellar radius limit            & any       & any       & 0         \\
        stellar mass ($M_{*}$)          & not null  & not null  & not null  \\
        stellar mass limit              & any       & any       & 0         \\
        stellar age                     & any       & not null  & not null  \\
        stellar age limit               & any       & any       & 0         \\
        stellar rot. period ($P_{\rm rot}$) & not null  & not null  & not null  \\
        stellar rot. per. limit         & any       & any       & 0        \\ 
        \bottomrule
    \end{tabular}
\end{table}

\begin{table*}[width=0.95\textwidth,cols=13,pos=h]
    \caption{Planetary systems parameters as follows: Orbital period ($P_{\rm orb}$), Semi-major axis ($a$), Planetary radius ($R_{p}$), Planetary mass ($M_{p}$), eccentricity ($e$), Effective temperature ($T_{\rm eff}$), Stellar radius ($R_{*}$), Stellar mass ($M_{*}$), Rotation period ($P_{\rm rot}$), Stellar age (Age), Tidal interaction index ($q$), and ratio of the orbital angular and stellar spin momenta ($L_{*}/L_{\rm orb}$). The superscript indexes indicate the datasets in which the exoplanet is present.}\label{tab:table_parameters}
    \begin{tabular*}{\tblwidth}{L@{\hspace{-0.15in}} LLLLLLLLLLLLL@{} }
        \toprule
       Planet & \makecell{$P_{\rm orb}$\\(days)} & \makecell{$a$\\(AU)} & \makecell{$R_{p}$\\($R_{\oplus}$)} & \makecell{$M_{p}$\\ ($M_{\oplus}$)} & $e$\hspace{0.3in} & \makecell{$T_{\rm eff}$\\(K)} & \makecell{$R_{*}$\\ ($R_{\odot}$)} & \makecell{$M_{*}$\\ ($M_{\odot}$)} & \makecell{$P_{\rm rot}$\\ (days)} & \makecell{Age\\ (Gyr)} & $q$ &$L_{*}/L_{\rm orb}$\\
        \midrule
        AU Mic b$^{1}$             & 8.46 & 0.06 & 4.07 & 17.00 & 0.00 & 3700 & 0.75 & 0.50 & 4.86 & 0.02 & 242.06 & 19.98\\
        AU Mic c$^{1}$             & 18.86 & 0.11 & 3.24 & 13.60 & 0.00 & 3700 & 0.75 & 0.50 & 4.86 & 0.02 & 230.88 & 19.13\\
        CoRoT-18 b$^{1}$           & 1.90 & 0.03 & 14.68 & 1099.69 & 0.03 & 5440 & 1.00 & 0.95 & 5.40 & 0.60 & 11.55 & 1.01\\
        CoRoT-4 b$^{1}$            & 9.20 & 0.09 & 13.34 & 223.43 & 0.14 & 6190 & 1.17 & 1.16 & 8.90 & 1.00 & 56.84 & 2.63\\
        CoRoT-6 b$^{1}$            & 8.89 & 0.09 & 13.07 & 937.60 & 0.18 & 6090 & 1.02 & 1.05 & 6.40 & 2.50 & 11.32 & 0.65\\
        EPIC 249893012 b$^{1,2,3}$ & 3.60 & 0.05 & 1.95 & 8.75 & 0.06 & 5430 & 1.71 & 1.05 & 41.00 & 9.00 & 494.46 & 41.10\\
        EPIC 249893012 c$^{1,2,3}$ & 15.62 & 0.13 & 3.67 & 14.67 & 0.07 & 5430 & 1.71 & 1.05 & 41.00 & 9.00 & 187.58 & 15.61\\
        EPIC 249893012 d$^{1,2,3}$ & 35.75 & 0.22 & 3.94 & 10.18 & 0.15 & 5430 & 1.71 & 1.05 & 41.00 & 9.00 & 207.35 & 16.79\\
        G 9-40 b$^{1,2}$           & 5.75 & 0.04 & 2.02 & 11.70 & 0.00 & 3348 & 0.31 & 0.29 & 29.85 & 9.90 & 8.28 & 0.71\\
        GJ 1132 b$^{1}$            & 1.63 & 0.02 & 1.13 & 1.66 & 0.22 & 3270 & 0.21 & 0.18 & 122.30 & - - - & 10.02 & 0.81\\
        GJ 1214 b$^{1}$            & 1.58 & 0.01 & 2.74 & 8.17 & 0.06 & 3250 & 0.21 & 0.18 & 124.70 &  - - - & 1.88 & 0.16\\
        GJ 1252 b$^{1}$            & 0.52 & 0.01 & 1.19 & 2.09 & 0.00 & 3458 & 0.39 & 0.38 & 64.00 &  - - - & 92.84 & 7.74\\
        GJ 367 b$^{1}$             & 0.32 & 0.01 & 0.72 & 0.55 & 0.00 & 3522 & 0.46 & 0.45 & 58.00 &  - - - & 680.72 & 56.73\\
        GJ 486 b$^{1}$             & 1.47 & 0.02 & 1.30 & 2.82 & 0.05 & 3340 & 0.33 & 0.32 & 130.10 &  - - - & 16.25 & 1.35\\
        GJ 9827 b$^{1,2}$          & 1.21 & 0.02 & 1.58 & 5.14 & 0.06 & 4305 & 0.60 & 0.61 & 28.90 & 10.00 & 174.93 & 14.54\\
        GJ 9827 c$^{1,2}$          & 3.65 & 0.04 & 1.24 & 1.30 & 0.09 & 4305 & 0.60 & 0.61 & 28.90 & 10.00 & 481.63 & 39.87\\
        GJ 9827 d$^{1,2}$          & 6.20 & 0.06 & 2.02 & 3.53 & 0.13 & 4305 & 0.60 & 0.61 & 28.90 & 10.00 & 150.05 & 12.35\\
        HAT-P-19 b$^{1}$           & 4.01 & 0.05 & 12.69 & 90.26 & 0.02 & 4990 & 0.82 & 0.84 & 35.50 & 8.80 & 11.16 & 0.94\\
        HAT-P-20 b$^{1}$           & 2.88 & 0.04 & 9.72 & 2310.62 & 0.02 & 4595 & 0.69 & 0.76 & 14.48 & - - - & 0.58 & 0.07\\
        HAT-P-21 b$^{1}$           & 4.12 & 0.05 & 11.48 & 1261.79 & 0.22 & 5588 & 1.10 & 0.95 & 15.90 & 10.20 & 3.22 & 0.29\\
        HAT-P-68 b$^{1,2}$         & 2.30 & 0.03 & 12.02 & 230.11 & 0.04 & 4508 & 0.67 & 0.68 & 24.59 & 11.10 & 4.68 & 0.40\\
        HATS-16 b$^{1,2}$          & 2.69 & 0.04 & 14.57 & 1039.30 & 0.00 & 5738 & 1.24 & 0.97 & 12.35 & 9.50 & 7.43 & 0.64\\
        HATS-18 b$^{1}$            & 0.84 & 0.02 & 14.99 & 629.30 & 0.17 & 5600 & 1.02 & 1.04 & 9.80 & 4.20 & 16.88 & 1.38\\
        HATS-2 b$^{1}$             & 1.35 & 0.02 & 13.09 & 486.28 & 0.29 & 5227 & 0.90 & 0.88 & 24.98 & 9.70 & 5.79 & 0.46\\
        HATS-47 b$^{1,2}$          & 3.92 & 0.04 & 12.52 & 117.28 & 0.09 & 4512 & 0.66 & 0.67 & 6.42 & 8.10 & 27.90 & 2.43\\
        HATS-57 b$^{1,2}$          & 2.35 & 0.03 & 12.77 & 1000.21 & 0.03 & 5587 & 0.96 & 1.03 & 12.71 & 2.50 & 4.75 & 0.41\\
        HATS-71 b$^{1,2,3}$        & 3.80 & 0.04 & 11.48 & 117.60 & 0.00 & 3405 & 0.48 & 0.49 & 41.72 & 3.20 & 2.04 & 0.18\\
        HATS-72 b$^{1,2}$          & 7.33 & 0.07 & 8.10 & 39.86 & 0.01 & 4656 & 0.72 & 0.73 & 48.73 & 12.17 & 11.02 & 0.93\\
        HATS-75 b$^{1,2}$          & 2.79 & 0.03 & 9.91 & 156.05 & 0.06 & 3790 & 0.58 & 0.60 & 35.04 & 14.90 & 3.27 & 0.28\\
        HATS-76 b$^{1,2}$          & 1.94 & 0.03 & 12.09 & 835.57 & 0.06 & 4016 & 0.63 & 0.66 & 15.16 & 4.60 & 1.84 & 0.17\\
        HD 110113 b$^{1,2,3}$      & 2.54 & 0.04 & 2.05 & 4.55 & 0.09 & 5732 & 0.97 & 1.00 & 20.80 & 4.00 & 639.64 & 52.95\\
        HD 114082 b$^{1,2,3}$        & 109.75 & 0.51 & 11.21 & 2542.63 & 0.40 & 6651 & 1.49 & 1.47 & 1.92 & 0.01 & 12.56 & 0.89\\
        HD 136352 b$^{1,2,3}$      & 11.58 & 0.10 & 1.66 & 4.72 & 0.00 & 5664 & 1.06 & 0.87 & 23.80 & 12.30 & 383.79 & 32.06\\
        HD 136352 c$^{1,2,3}$      & 27.59 & 0.17 & 2.92 & 11.24 & 0.00 & 5664 & 1.06 & 0.87 & 23.80 & 12.30 & 128.30 & 10.09\\
        HD 136352 d$^{1,2,3}$      & 107.25 & 0.42 & 2.56 & 8.82 & 0.00 & 5664 & 1.06 & 0.87 & 23.80 & 12.30 & 99.28 & 8.17\\
        HD 15337 b$^{1,2,3}$       & 4.76 & 0.05 & 1.64 & 7.51 & 0.09 & 5125 & 0.86 & 0.90 & 36.53 & 5.10 & 137.99 & 11.44\\
        HD 15337 c$^{1,2,3}$       & 17.18 & 0.13 & 2.39 & 8.11 & 0.05 & 5125 & 0.86 & 0.90 & 36.53 & 5.10 & 84.57 & 7.10\\
        HD 183579 b$^{1}$          & 17.47 & 0.13 & 3.53 & 11.20 & 0.00 & 5706 & 0.97 & 1.03 & 23.20 & 2.60 & 121.42 & 10.37\\
        HD 207496 b$^{1,2,3}$      & 6.44 & 0.06 & 2.25 & 6.10 & 0.23 & 4819 & 0.77 & 0.80 & 12.36 & 0.52 & 385.49 & 30.39\\
        HD 207897 b$^{1,2,3}$      & 16.20 & 0.12 & 2.50 & 14.40 & 0.05 & 5070 & 0.78 & 0.80 & 37.00 & 7.10 & 37.21 & 3.16\\
        HD 209458 b$^{1}$          & 3.52 & 0.05 & 15.23 & 216.76 & 0.01 & 6065 & 1.16 & 1.12 & 10.65 & 3.10 & 35.47 & 3.00\\
        HD 213885 b$^{1,2,3}$      & 1.01 & 0.02 & 1.75 & 8.83 & 0.00 & 5978 & 1.10 & 1.07 & 18.58 & 3.80 & 680.02 & 56.67\\
        HD 23472 b$^{1}$           & 17.67 & 0.12 & 2.00 & 8.32 & 0.07 & 4684 & 0.71 & 0.67 & 40.10 &  - - - & 45.48 & 3.84\\
        HD 23472 c$^{1}$           & 29.80 & 0.16 & 1.87 & 3.41 & 0.06 & 4684 & 0.71 & 0.67 & 40.10 &  - - - & 91.97 & 7.86\\
        HD 23472 d$^{1}$           & 3.98 & 0.04 & 0.75 & 0.55 & 0.07 & 4684 & 0.71 & 0.67 & 40.10 &  - - - & 1148.36 & 95.34\\
        HD 23472 e$^{1}$           & 7.91 & 0.07 & 0.82 & 0.72 & 0.07 & 4684 & 0.71 & 0.67 & 40.10 &  - - - & 697.91 & 57.95\\
        HD 23472 f$^{1}$           & 12.16 & 0.09 & 1.14 & 0.77 & 0.07 & 4684 & 0.71 & 0.67 & 40.10 &  - - - & 565.23 & 46.94\\
        HD 260655 b$^{1,2,3}$      & 2.77 & 0.03 & 1.24 & 2.14 & 0.04 & 3803 & 0.44 & 0.44 & 37.50 & 5.00 & 118.18 & 9.84\\
        HD 260655 c$^{1,2,3}$      & 5.71 & 0.05 & 1.53 & 3.09 & 0.04 & 3803 & 0.44 & 0.44 & 37.50 & 5.00 & 64.18 & 5.36\\
        HD 3167 b$^{1}$            & 0.96 & 0.02 & 1.57 & 5.69 & 0.00 & 5286 & 0.83 & 0.88 & 23.52 & 5.00 & 434.13 & 36.18\\
        HD 3167 c$^{1}$            & 29.85 & 0.18 & 2.74 & 8.33 & 0.05 & 5286 & 0.83 & 0.88 & 23.52 & 5.00 & 102.98 & 8.20\\
        HD 5278 b$^{1,2}$          & 14.34 & 0.12 & 2.45 & 7.80 & 0.08 & 6203 & 1.19 & 1.13 & 16.80 & 3.00 & 420.64 & 34.99\\
        HD 73583 c$^{1,2,3}$       & 18.88 & 0.12 & 2.39 & 9.70 & 0.08 & 4511 & 0.65 & 0.73 & 12.20 & 0.48 & 111.63 & 9.07\\
        HD 80653 b$^{1,2,3}$       & 0.72 & 0.02 & 1.61 & 5.72 & 0.00 & 5959 & 1.22 & 1.18 & 19.55 & 2.67 & 1419.33 & 118.28\\
        K2-111 b$^{1}$             & 5.35 & 0.06 & 1.82 & 5.29 & 0.13 & 5775 & 1.25 & 0.84 & 29.20 & 12.30 & 508.79 & 41.83\\
        K2-131 b$^{1,2,3}$         & 0.37 & 0.01 & 1.69 & 7.90 & 0.00 & 5120 & 0.76 & 0.80 & 9.19 & 5.30 & 931.50 & 77.63\\
        \bottomrule
        continued on the next page
    \end{tabular*}
\end{table*}

\begin{table*}[width=\textwidth,cols=13,pos=h]
    \repeatcaption{tab:table_parameters}{}
    \begin{tabular*}{\tblwidth}{L@{} LLLLLLLLLLLLL@{} }
        \toprule
       Planet & \makecell{$P_{\rm orb}$\\(days)} & \makecell{$a$\\(AU)} & \makecell{$R_{p}$\\($R_{\oplus}$)} & \makecell{$M_{p}$\\ ($M_{\oplus}$)} & $e$\hspace{0.3in} & \makecell{$T_{\rm eff}$\\(K)} & \makecell{$R_{*}$\\ ($R_{\odot}$)} & \makecell{$M_{*}$\\ ($M_{\odot}$)} & \makecell{$P_{\rm rot}$\\ (days)} & \makecell{Age\\ (Gyr)} & $q$ &$L_{*}/L_{\rm orb}$\\
        \midrule
        K2-140 b$^{1,2,3}$         & 6.57 & 0.07 & 13.56 & 295.58 & 0.00 & 5585 & 1.06 & 0.96 & 14.60 & 9.80 & 11.63 & 1.04\\
        K2-141 b$^{1,2,3}$          & 0.28 & 0.01 & 1.51 & 4.97 & 0.00 & 4570 & 0.68 & 0.71 & 15.17 & 6.30 & 755.00 & 62.92\\
        K2-141 c$^{1,2}$            & 7.75 & 0.07 & 7.00 & 8.00 & 0.09 & 4570 & 0.68 & 0.71 & 15.17 & 6.30 & 156.18 & 12.98\\
        K2-18 b$^{1}$               & 32.94 & 0.14 & 2.37 & 8.92 & 0.20 & 3457 & 0.41 & 0.36 & 39.63 &  - - - & 5.69 & 0.81\\
        K2-180 b$^{1}$              & 8.87 & 0.07 & 2.24 & 11.44 & 0.00 & 5110 & 0.69 & 0.71 & 15.70 & 9.50 & 102.27 & 8.63\\
        K2-229 b$^{1}$              & 0.58 & 0.01 & 1.16 & 2.59 & 0.00 & 5185 & 0.79 & 0.84 & 18.10 & 5.40 & 1356.68 & 113.06\\
        K2-229 c$^{1}$              & 8.33 & 0.08 & 2.12 & 21.30 & 0.00 & 5185 & 0.79 & 0.84 & 18.10 & 5.40 & 67.19 & 5.67\\
        K2-237 b$^{1}$              & 2.18 & 0.04 & 18.50 & 508.53 & 0.00 & 6257 & 1.43 & 1.28 & 5.07 & 2.55 & 61.42 & 5.18\\
        K2-25 b$^{1,2,3}$           & 3.48 & 0.03 & 3.44 & 24.50 & 0.43 & 3207 & 0.29 & 0.26 & 1.88 & 0.73 & 81.05 & 6.41\\
        K2-3 b$^{1}$                & 10.05 & 0.08 & 2.25 & 6.47 & 0.09 & 3835 & 0.60 & 0.62 & 40.71 & 1.00 & 49.12 & 4.09\\
        K2-3 c$^{1}$                & 24.65 & 0.14 & 1.69 & 3.30 & 0.10 & 3835 & 0.60 & 0.62 & 40.71 & 1.00 & 70.59 & 5.95\\
        K2-3 d$^{1}$                & 44.56 & 0.21 & 1.62 & 1.60 & 0.10 & 3835 & 0.60 & 0.62 & 40.71 & 1.00 & 129.31 & 10.06\\
        K2-33 b$^{1}$               & 5.43 & 0.04 & 5.76 & 1144.19 & 0.00 & 3410 & 1.10 & 0.31 & 6.30 & 0.01 & -0.35 & 0.49\\
        K2-36 b$^{1}$               & 1.42 & 0.02 & 1.43 & 3.90 & 0.00 & 4916 & 0.72 & 0.79 & 16.90 & 1.40 & 569.49 & 47.47\\
        K2-36 c$^{1}$               & 5.34 & 0.05 & 3.20 & 7.80 & 0.00 & 4916 & 0.72 & 0.79 & 16.90 & 1.40 & 183.23 & 15.31\\
        Kepler-102 b$^{1,2}$        & 5.29 & 0.06 & 0.46 & 1.10 & 0.10 & 4909 & 0.72 & 0.80 & 27.95 & 1.10 & 824.06 & 68.13\\
        Kepler-102 c$^{1,2}$        & 7.07 & 0.07 & 0.57 & 1.70 & 0.09 & 4909 & 0.72 & 0.80 & 27.95 & 1.10 & 483.12 & 39.99\\
        Kepler-102 d$^{1,2}$        & 10.31 & 0.09 & 1.15 & 3.00 & 0.09 & 4909 & 0.72 & 0.80 & 27.95 & 1.10 & 241.02 & 19.98\\
        Kepler-102 e$^{1,2}$        & 16.15 & 0.12 & 2.17 & 4.70 & 0.09 & 4909 & 0.72 & 0.80 & 27.95 & 1.10 & 131.57 & 10.98\\
        Kepler-102 f$^{1,2}$        & 27.45 & 0.17 & 0.86 & 4.30 & 0.10 & 4909 & 0.72 & 0.80 & 27.95 & 1.10 & 86.10 & 10.07\\
        Kepler-107 b$^{1}$          & 3.18 & 0.05 & 1.54 & 3.51 & 0.00 & 5854 & 1.45 & 1.24 & 20.30 & 4.29 & 1947.71 & 162.32\\
        Kepler-107 c$^{1}$          & 4.90 & 0.06 & 1.60 & 9.39 & 0.00 & 5854 & 1.45 & 1.24 & 20.30 & 4.29 & 630.11 & 52.54\\
        Kepler-107 d$^{1}$          & 7.96 & 0.08 & 0.86 & 3.80 & 0.00 & 5854 & 1.45 & 1.24 & 20.30 & 4.29 & 1324.74 & 110.45\\
        Kepler-107 e$^{1}$          & 14.75 & 0.13 & 2.90 & 8.60 & 0.00 & 5854 & 1.45 & 1.24 & 20.30 & 4.29 & 474.12 & 39.73\\
        Kepler-39 b$^{1}$           & 21.09 & 0.16 & 13.90 & 6388.38 & 0.13 & 6350 & 1.40 & 1.29 & 4.50 & 2.10 & 3.75 & 0.21\\
        Kepler-419 b$^{1}$          & 69.75 & 0.37 & 10.76 & 820.00 & 0.80 & 6430 & 1.75 & 1.39 & 4.49 & 2.80 & 61.57 & 2.83\\
        Kepler-423 b$^{1}$          & 2.68 & 0.04 & 13.36 & 192.60 & 0.08 & 5560 & 0.95 & 0.85 & 22.05 & 11.00 & 13.07 & 1.10\\
        Kepler-43 b$^{1}$           & 3.02 & 0.04 & 13.00 & 997.99 & 0.03 & 6050 & 1.38 & 1.27 & 12.95 & 2.30 & 9.69 & 0.83\\
        Kepler-45 b$^{1}$           & 2.46 & 0.03 & 10.76 & 153.83 & 0.23 & 3820 & 0.55 & 0.59 & 15.80 & 0.80 & 7.33 & 0.60\\
        Kepler-539 b$^{1}$          & 125.63 & 0.50 & 8.37 & 308.30 & 0.39 & 5820 & 0.95 & 1.05 & 11.77 &  - - - & 6.44 & 0.41\\
        Kepler-75 b$^{1}$           & 8.88 & 0.08 & 11.77 & 3210.08 & 0.57 & 5200 & 0.89 & 0.91 & 19.18 & 6.20 & 0.12 & 0.06\\
        KOI-4777.01$^{1,2}$         & 0.41 & 0.01 & 0.51 & 99.20 & 0.00 & 3515 & 0.40 & 0.41 & 44.00 & 4.00 & 3.28 & 0.27\\
        L 168-9 b$^{1}$             & 1.40 & 0.02 & 1.39 & 4.60 & 0.21 & 3800 & 0.60 & 0.62 & 29.80 &  - - - & 191.62 & 15.43\\
        L 98-59 b$^{1,2}$           & 2.25 & 0.02 & 0.85 & 0.40 & 0.10 & 3415. & 0.30 & 0.27 & 80.90 & 0.80 & 126.61 & 10.47\\
        L 98-59 c$^{1}$             & 3.69 & 0.03 & 1.39 & 2.22 & 0.10 & 3415 & 0.30 & 0.27 & 80.90 & 0.80 & 19.28 & 1.60\\
        L 98-59 d$^{1}$             & 7.45 & 0.05 & 1.52 & 1.94 & 0.07 & 3415 & 0.30 & 0.27 & 80.90 & 0.80 & 17.30 & 1.44\\
        LHS 1140 b$^{1}$            & 24.74 & 0.09 & 1.43 & 6.65 & 0.29 & 3131 & 0.19 & 0.15 & 131.00 & 5.00 & 0.53 & 0.06\\
        LHS 1140 c$^{1}$            & 3.78 & 0.03 & 1.28 & 1.81 & 0.31 & 3216 & 0.21 & 0.18 & 131.00 & 5.00 & 6.89 & 0.53\\
        LHS 1815 b$^{1}$            & 3.81 & 0.04 & 1.09 & 8.70 & 0.00 & 3643 & 0.50 & 0.50 & 47.80 &  - - - & 29.33 & 2.45\\
        LP 714-47 b$^{1}$           & 4.05 & 0.04 & 4.70 & 30.80 & 0.04 & 3950 & 0.58 & 0.59 & 33.00 &  - - - & 15.63 & 1.31\\
        LTT 9779 b$^{1,2,3}$        & 0.79 & 0.02 & 4.72 & 29.32 & 0.00 & 5443 & 0.95 & 0.77 & 45.00 & 1.90 & 66.97 & 5.58\\
        NGTS-10 b$^{1,2,3}$         & 0.77 & 0.01 & 13.51 & 687.15 & 0.00 & 4600 & 0.70 & 0.70 & 17.29 & 10.40 & 3.51 & 0.30\\
        NGTS-21 b$^{1,2,3}$         & 1.54 & 0.02 & 14.91 & 750.08 & 0.00 & 4660 & 0.86 & 0.72 & 17.89 & 10.02 & 3.80 & 0.32\\
        Qatar-1 b$^{1}$             & 1.42 & 0.02 & 13.23 & 419.85 & 0.01 & 4910 & 0.80 & 0.85 & 23.70 & 11.60 & 4.84 & 0.41\\
        Qatar-2 b$^{1}$             & 1.34 & 0.02 & 12.89 & 826.36 & 0.00 & 4645 & 0.71 & 0.74 & 18.50 & 1.40 & 2.38 & 0.21\\
        Qatar-3 b$^{1,2,3}$         & 2.51 & 0.04 & 12.29 & 1369.85 & 0.00 & 6007 & 1.27 & 1.15 & 6.31 & 0.31 & 12.34 & 1.08\\
        Qatar-4 b$^{1,2,3}$         & 1.81 & 0.03 & 12.72 & 1938.76 & 0.00 & 5215 & 0.85 & 0.90 & 6.05 & 0.17 & 3.99 & 0.37\\
        Qatar-5 b$^{1,2,3}$         & 2.88 & 0.04 & 12.41 & 1373.03 & 0.00 & 5747 & 1.08 & 1.13 & 12.10 & 0.53 & 4.34 & 0.39\\
        Qatar-7 b$^{1,2,3}$         & 2.03 & 0.04 & 19.05 & 597.52 & 0.00 & 6387 & 1.56 & 1.41 & 5.90 & 1.69 & 54.72 & 4.60\\
        TIC 257060897 b$^{1,2,3}$     & 3.66 & 0.05 & 16.70 & 212.95 & 0.03 & - - - & 1.82 & 1.32 & 92.00 & 3.47 & 10.86 & 0.91\\
        TOI-1052 b$^{1,2}$          & 9.14 & 0.09 & 2.87 & 16.90 & 0.18 & 6146 & 1.26 & 1.20 & 12.80 & 2.30 & 357.80 & 28.61\\
        TOI-1062 b$^{1,2,3}$        & 4.11 & 0.05 & 2.27 & 10.15 & 0.18 & 5328 & 0.84 & 0.94 & 21.80 & 2.50 & 194.02 & 15.77\\
        TOI-1107 b$^{1,2,3}$        & 4.08 & 0.06 & 14.57 & 1064.73 & 0.03 & 6311 & 1.81 & 1.35 & 6.20 & 2.60 & 29.65 & 2.63\\
        TOI-1235 b$^{1}$            & 3.44 & 0.04 & 1.74 & 6.91 & 0.15 & 3872 & 0.63 & 0.64 & 44.70 &  - - - & 68.13 & 5.58\\
        TOI-1260 b$^{1,2,3}$        & 3.13 & 0.04 & 2.41 & 8.56 & 0.00 & 4227 & 0.67 & 0.68 & 30.63 & 6.70 & 92.70 & 7.73\\
        TOI-1260 c$^{1,2,3}$        & 7.49 & 0.07 & 2.76 & 13.20 & 0.00 & 4227 & 0.67 & 0.68 & 30.63 & 6.70 & 44.65 & 3.75\\
        \bottomrule
        continued on the next page
    \end{tabular*}
\end{table*}

\begin{table*}[width=\textwidth,cols=13,pos=h]
    \repeatcaption{tab:table_parameters}{}
    \begin{tabular*}{\tblwidth}{L@{} LLLLLLLLLLLLL@{} }
        \toprule
       Planet & \makecell{$P_{\rm orb}$\\(days)} & \makecell{$a$\\(AU)} & \makecell{$R_{p}$\\($R_{\oplus}$)} & \makecell{$M_{p}$\\ ($M_{\oplus}$)} & $e$\hspace{0.3in} & \makecell{$T_{\rm eff}$\\(K)} & \makecell{$R_{*}$\\ ($R_{\odot}$)} & \makecell{$M_{*}$\\ ($M_{\odot}$)} & \makecell{$P_{\rm rot}$\\ (days)} & \makecell{Age\\ (Gyr)} & $q$ &$L_{*}/L_{\rm orb}$\\
        \midrule
        TOI-1260 d$^{1,2,3}$        & 16.61 & 0.11 & 3.12 & 11.84 & 0.00 & 4227 & 0.67 & 0.68 & 30.63 & 6.70 & 37.24 & 3.20\\
        TOI-1268 b$^{1,2,3}$        & 8.16 & 0.07 & 9.10 & 96.40 & 0.10 & 5300 & 0.92 & 0.96 & 10.90 & 0.24 & 30.23 & 2.72\\
        TOI-1272 b$^{1,2,3}$        & 3.32 & 0.04 & 4.14 & 24.60 & 0.34 & 4985 & 0.79 & 0.85 & 28.30 & 3.65 & 60.10 & 4.56\\
        TOI-1416 b$^{1,2,3}$        & 1.07 & 0.02 & 1.62 & 3.48 & 0.00 & 4884 & 0.79 & 0.80 & 17.60 & 6.90 & 836.03 & 69.67\\
        TOI-1470 b$^{1,2}$          & 2.53 & 0.03 & 2.18 & 7.32 & 0.30 & 3709 & 0.47 & 0.47 & 29.00 & 1.30 & 61.11 & 4.74\\
        TOI-1470 c$^{1,2}$          & 18.09 & 0.11 & 2.47 & 7.24 & 0.50 & 3709 & 0.47 & 0.47 & 29.00 & 1.30 & 44.76 & 2.74\\
        TOI-1695 b$^{1}$            & 3.13 & 0.03 & 1.90 & 6.36 & 0.10 & 3690 & 0.52 & 0.51 & 47.70 &  - - - & 44.54 & 3.69\\
        TOI-1710 b$^{1,2,3}$          & 24.28 & 0.16 & 5.34 & 28.30 & 0.16 & 5665 & 0.97 & 0.98 & 22.50 & 4.20 & 55.33 & 3.80\\
        TOI-1759 b$^{1}$            & 18.85 & 0.12 & 3.06 & 6.80 & 0.00 & 3972 & 0.63 & 0.61 & 35.65 & 5.00 & 46.07 & 3.93\\
        TOI-1807 b$^{1}$            & 0.55 & 0.01 & 1.37 & 2.57 & 0.00 & 4730 & 0.69 & 0.76 & 8.80 & 0.30 & 2124.25 & 177.03\\
        TOI-1820 b$^{1,2,3}$        & 4.86 & 0.06 & 12.78 & 731.01 & 0.04 & 5734 & 1.51 & 1.04 & 25.00 & 11.00 & 7.01 & 0.60\\
        TOI-2025 b$^{1}$            & 8.87 & 0.09 & 12.52 & 1398.44 & 0.41 & 5880 & 1.56 & 1.32 & 13.20 & 1.70 & 6.76 & 0.58\\
        TOI-2095 b$^{1,2}$          & 17.66 & 0.10 & 1.25 & 4.10 & 0.00 & 3759 & 0.44 & 0.44 & 40.00 & 1.00 & 30.39 & 2.60\\
        TOI-2095 c$^{1,2}$          & 28.17 & 0.14 & 1.33 & 7.40 & 0.00 & 3759 & 0.44 & 0.44 & 40.00 & 1.00 & 12.31 & 1.22\\
        TOI-2136 b$^{1,2,3}$        & 7.85 & 0.06 & 2.19 & 6.37 & 0.00 & 3342 & 0.34 & 0.34 & 75.00 & 4.60 & 8.00 & 0.68\\
        TOI-2158 b$^{1,2}$          & 8.60 & 0.07 & 10.76 & 260.62 & 0.07 & 5673 & 1.41 & 1.12 & 19.00 & 8.00 & 15.47 & 1.35\\
        TOI-251 b$^{1,2}$           & 4.94 & 0.06 & 2.74 & 317.83 & 0.00 & 5875 & 0.88 & 1.04 & 3.84 & 0.18 & 38.58 & 2.84\\
        TOI-2641 b$^{1,2}$          & 4.88 & 0.06 & 18.10 & 122.68 & 0.18 & 6100 & 1.34 & 1.16 & 35.60 & 10.80 & 24.49 & 2.00\\
        TOI-332 b$^{1,2,3}$         & 0.78 & 0.02 & 3.20 & 57.20 & 0.00 & 5251 & 0.87 & 0.88 & 35.60 & 5.00 & 35.12 & 2.93\\
        TOI-4010 b$^{1,2,3}$        & 1.35 & 0.02 & 3.02 & 11.00 & 0.03 & 4960 & 0.83 & 0.88 & 37.70 & 6.10 & 130.45 & 10.87\\
        TOI-4010 c$^{1,2,3}$        & 5.41 & 0.06 & 5.93 & 20.31 & 0.03 & 4960 & 0.83 & 0.88 & 37.70 & 6.10 & 44.41 & 3.71\\
        TOI-4010 d$^{1,2,3}$        & 14.71 & 0.11 & 6.18 & 38.15 & 0.07 & 4960 & 0.83 & 0.88 & 37.70 & 6.10 & 16.48 & 1.42\\
        TOI-431 b$^{1}$             & 0.49 & 0.01 & 1.28 & 3.07 & 0.00 & 4850 & 0.73 & 0.78 & 30.50 &  - - - & 606.26 & 50.52\\
        TOI-431 d$^{1}$             & 12.46 & 0.10 & 3.29 & 9.90 & 0.00 & 4850 & 0.73 & 0.78 & 30.50 &  - - - & 63.43 & 5.34\\
        TOI-4562 b$^{1,2}$          & 225.12 & 0.77 & 12.53 & 732.00 & 0.76 & 6096 & 1.15 & 1.19 & 3.86 & 0.70 & 19.40 & 0.95\\
        TOI-5398 b$^{1,2}$          & 10.59 & 0.10 & 10.30 & 58.70 & 0.13 & 6000 & 1.05 & 1.15 & 7.34 & 0.65 & 113.20 & 9.19\\
        TOI-5398 c$^{1,2}$          & 4.77 & 0.06 & 3.52 & 11.80 & 0.14 & 6000 & 1.05 & 1.15 & 7.34 & 0.65 & 725.77 & 59.11\\
        TOI-615 b$^{1,2}$           & 4.66 & 0.07 & 18.98 & 138.26 & 0.39 & 6850 & 1.73 & 1.45 & 5.38 & 1.70 & 426.31 & 24.47\\
        TOI-622 b$^{1,2}$           & 6.40 & 0.07 & 9.24 & 96.30 & 0.42 & 6400 & 1.42 & 1.31 & 3.77 & 0.90 & 317.80 & 26.05\\
        TOI-674 b$^{1}$             & 1.98 & 0.03 & 5.25 & 23.60 & 0.00 & 3514 & 0.42 & 0.42 & 52.00 &  - - - & 8.36 & 0.70\\
        TOI-763 b$^{1}$             & 5.61 & 0.06 & 2.28 & 9.79 & 0.04 & 5450 & 0.90 & 0.92 & 27.00 &  - - - & 151.59 & 12.64\\
        TOI-763 c$^{1}$             & 12.27 & 0.10 & 2.63 & 9.32 & 0.04 & 5450 & 0.90 & 0.92 & 27.00 &  - - - & 121.95 & 10.22\\
        TOI-776 b$^{1,2,3}$         & 8.25 & 0.07 & 1.85 & 4.00 & 0.06 & 3709 & 0.54 & 0.54 & 34.40 & 7.80 & 77.46 & 6.46\\
        TOI-776 c$^{1,2,3}$         & 15.67 & 0.10 & 2.02 & 5.30 & 0.04 & 3709 & 0.54 & 0.54 & 34.40 & 7.80 & 46.44 & 3.93\\
        TOI-778 b$^{1,2,3}$         & 4.63 & 0.06 & 15.40 & 878.00 & 0.21 & 6643 & 1.71 & 1.40 & 2.58 & 1.95 & 81.57 & 6.58\\
        TOI-836 b$^{1,2,3}$         & 3.82 & 0.04 & 1.70 & 4.53 & 0.05 & 4552 & 0.67 & 0.68 & 21.99 & 5.40 & 230.82 & 19.21\\
        TOI-836 c$^{1,2,3}$         & 8.60 & 0.07 & 2.59 & 9.60 & 0.08 & 4552 & 0.67 & 0.68 & 21.99 & 5.40 & 85.83 & 7.16\\
        TOI-908 b$^{1,2,3}$         & 3.18 & 0.04 & 3.19 & 16.14 & 0.13 & 5626 & 1.03 & 0.95 & 21.93 & 4.60 & 185.23 & 15.23\\
        WASP-10 b$^{1}$             & 3.09 & 0.04 & 11.96 & 1020.23 & 0.06 & 4680 & 0.70 & 0.75 & 11.91 & 7.00 & 1.94 & 0.19\\
        WASP-113 b$^{1,2}$          & 4.54 & 0.06 & 15.79 & 150.97 & 0.00 & 5890 & 1.61 & 1.32 & 10.56 & 6.20 & 96.57 & 8.11\\
        WASP-118 b$^{1,2,3}$        & 4.05 & 0.05 & 16.14 & 163.36 & 0.00 & 6410 & 1.70 & 1.32 & 6.12 & 1.17 & 178.03 & 15.00\\
        WASP-121 b$^{1}$            & 1.27 & 0.03 & 19.65 & 367.73 & 0.00 & - - -  & 1.46 & 1.36 & 1.13 & - - - & 491.35 & 40.21\\
        WASP-132 c$^{1,2}$          & 1.01 & 0.02 & 1.85 & 37.35 & 0.13 & 4714 & 0.75 & 0.78 & 33.00 & 3.20 & 38.73 & 3.19\\
        %WASP-159 b$^{1}$            & 12.4 & 0.1 & 6.4 & 98.5 & 0.2 & 5050 & 1.3 & 1.2 & 2.9 & 3.4 \\
        WASP-166 b$^{1,2,3}$        & 5.44 & 0.06 & 7.06 & 32.10 & 0.00 & 6050 & 1.22 & 1.19 & 12.10 & 2.10 & 207.64 & 17.37\\
        WASP-182 b$^{1,2,3}$        & 3.38 & 0.05 & 9.53 & 47.04 & 0.00 & 5638 & 1.34 & 1.08 & 30.00 & 5.95 & 78.38 & 6.54\\
        WASP-19 b$^{1}$             & 0.79 & 0.02 & 14.68 & 367.73 & 0.01 & 5500 & 0.94 & 0.96 & 10.50 & 5.50 & 22.02 & 1.84\\
        WASP-22 b$^{1}$             & 3.53 & 0.05 & 12.55 & 177.98 & 0.02 & 6000 & 1.13 & 1.10 & 16.00 & 3.00 & 27.18 & 2.29\\
        WASP-32 b$^{1}$             & 2.72 & 0.04 & 12.33 & 1109.23 & 0.00 & 6140 & 1.09 & 1.07 & 11.60 & 2.22 & 5.82 & 0.51\\
        WASP-4 b$^{1}$              & 1.34 & 0.02 & 14.91 & 371.86 & 0.00 & 5436 & 0.92 & 0.89 & 22.20 & 7.00 & 8.03 & 0.67\\
        WASP-41 b$^{1}$             & 3.05 & 0.04 & 13.45 & 299.08 & 0.12 & 5450 & 0.90 & 0.93 & 18.41 & 9.80 & 8.93 & 0.75\\
        WASP-43 b$^{1}$             & 0.81 & 0.02 & 11.61 & 651.55 & 0.01 & 4400 & 0.67 & 0.72 & 15.60 & 7.00 & 3.78 & 0.32\\
        WASP-46 b$^{1}$             & 1.43 & 0.02 & 13.16 & 603.88 & 0.02 & 5600 & 0.86 & 0.83 & 16.00 & 9.60 & 5.68 & 0.48\\
        WASP-47 b$^{1}$             & 4.16 & 0.05 & 12.90 & 383.00 & 0.00 & 5576 & 1.16 & 1.11 & 31.20 & 6.50 & 6.35 & 0.54\\
        WASP-47 d$^{1}$             & 9.10 & 0.09 & 3.71 & 16.80 & 0.01 & 5576 & 1.16 & 1.11 & 31.20 & 6.50 & 114.35 & 9.56\\
        WASP-47 e$^{1}$             & 0.79 & 0.02 & 1.82 & 9.10 & 0.03 & 5576 & 1.16 & 1.11 & 31.20 & 6.50 & 480.21 & 39.99\\
        WASP-5 b$^{1}$              & 1.63 & 0.03 & 13.17 & 505.35 & 0.01 & 5770 & 1.09 & 1.03 & 16.20 & 5.60 & 11.17 & 0.94\\
        \bottomrule 
        continued on the next page
    \end{tabular*}
\end{table*}

\begin{table*}[width=\textwidth,cols=13,pos=h]
    \repeatcaption{tab:table_parameters}{}
    \begin{tabular*}{\tblwidth}{L@{\hspace{0.62in}} LLLLLLLLLLLLL@{} }
        \toprule
       Planet & \makecell{$P_{\rm orb}$\\(days)} & \makecell{$a$\\(AU)} & \makecell{$R_{p}$\\($R_{\oplus}$)} & \makecell{$M_{p}$\\ ($M_{\oplus}$)} & $e$\hspace{0.3in} & \makecell{$T_{\rm eff}$\\(K)} & \makecell{$R_{*}$\\ ($R_{\odot}$)} & \makecell{$M_{*}$\\ ($M_{\odot}$)} & \makecell{$P_{\rm rot}$\\ (days)} & \makecell{Age\\ (Gyr)} & $q$ &$L_{*}/L_{\rm orb}$\\
        \midrule
        WASP-50 b$^{1}$             & 1.96 & 0.03 & 12.92 & 464.03 & 0.02 & 5400 & 0.84 & 0.89 & 16.30 & 7.00 & 6.36 & 0.54\\
        WASP-6 b$^{1}$              & 3.36 & 0.04 & 13.72 & 153.51 & 0.07 & 5450 & 0.87 & 0.88 & 23.80 & 11.00 & 11.92 & 1.00\\
        WASP-69 b$^{1}$ & 3.87 & 0.05 & 11.85 & 79.46 & 0.11 & 4700 & 0.81 & 0.83 & 23.07 & 7.00 & 19.46 & 1.62\\
        WASP-84 b$^{1}$ & 8.52 & 0.08 & 10.94 & 218.35 & 0.00 & 5280 & 0.77 & 0.85 & 14.36 & 2.10 & 6.48 & 0.66\\
        WASP-89 b$^{1}$ & 3.36 & 0.04 & 11.66 & 1843.41 & 0.19 & 5000 & 0.88 & 0.92 & 20.20 & 12.50 & 1.09 & 0.10\\
        WASP-92 b$^{1,2,3}$ & 2.17 & 0.03 & 16.38 & 255.85 & 0.00 & 6280 & 1.34 & 1.19 & 10.07 & 2.29 & 51.25 & 4.29\\
        WASP-93 b$^{1,2,3}$ & 2.73 & 0.04 & 17.90 & 467.21 & 0.00 & 6700 & 1.52 & 1.33 & 1.45 & 0.70 & 244.93 & 20.23\\
        \bottomrule 
    \end{tabular*}
\end{table*}

Although we have 5,599 confirmed exoplanets, many are in systems whose host star does not have a cataloged age or rotation period. It is also possible to find stars with well-defined stellar parameters but in systems whose parameters are not yet known or have not yet been considered as a single reference value in the literature. For instance, it is possible to find catalogs with different age values for the same host star. When this occurs, the data is flagged to differentiate the values of the observables considered as a default parameter set. In our initial dataset, records for a given star or planetary system often contained values considered default but missing important information about stellar observables investigated in the present work, such as age or rotational period. The system WASP-89b is an example. 

For this system, the set of parameters considered default is attributed to \cite{hellier2015}, and the star's age is 1.3 Gyr, but there is no information about the star's rotational period. In the same catalog consulted, we observed that for WASP-89 b, a set of non-default parameters attributed to the catalog by \cite{Bonono2017} contained data for both age and stellar period. In the latter case, the WASP-89 has a well-defined stellar age expressed by the measurement $t_{age}=12.5\substack{+1.5 \\ -3.1}$\ Gyr. In this case, we chose the latter as it contained all the information necessary for our study. We further restricted the original dataset, selecting only systems with a default parameter set to increase confidence in our analyses and investigate whether this could influence our results. 

\begin{figure}
        \begin{center}
            \includegraphics[scale=0.55]{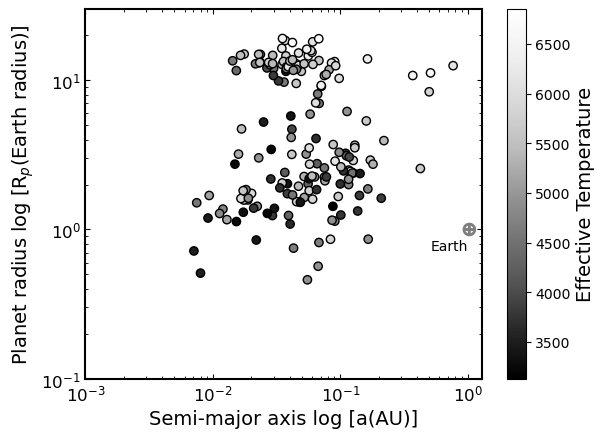}
        \end{center}
        \caption{Our initial exoplanet sample (dataset\#01), plotted as a function of distance to the star (up to $\sim$1 AU) and planetary radii. The figure suggests a continuous distribution of planetary sizes from sub-Earths to super-Jupiters ($\sim$2 Jupiter mass) and stellar effective temperatures (in K) that span three orders of magnitude.}
        \label{fig0a}
    \end{figure}

The primary space missions contributing to identifying these exoplanets were the Transiting Exoplanet Survey Satellite (TESS), responsible for detecting 89 exoplanets, and the Kepler Space Telescope, which detected 57. Both missions utilized the transit method for detection. Upon analyzing these planetary systems, it was observed that a substantial majority consisted of single stars, encompassing approximately 82\% of the total. The remaining 18\%  corresponds to multi-star systems, with three being circumbinary systems (Kepler-1647 b, Kepler-1661 b, and TOI-1338 b). As shown in Fig. \ref{fig0a}, stars with higher effective temperatures tend to host planets with a more significant radius than Jupiter-like planets. Furthermore, it is worth noting that the dataset predominantly comprises single-planet systems, accounting for 108 cases.

We removed planetary systems with more than one star from the original dataset, as this is a very complex subject and will be dealt with in a forthcoming communication. Then, subtracting the 29 multiple systems, our sample was reduced to 175 exoplanets distributed in 133 systems. This data will be referred to as \texttt{dataset\#01} and is presented in Table \ref{tab:table_parameters}.

\begin{figure}[h!]
        	\begin{center}
        		\includegraphics[scale=0.58]{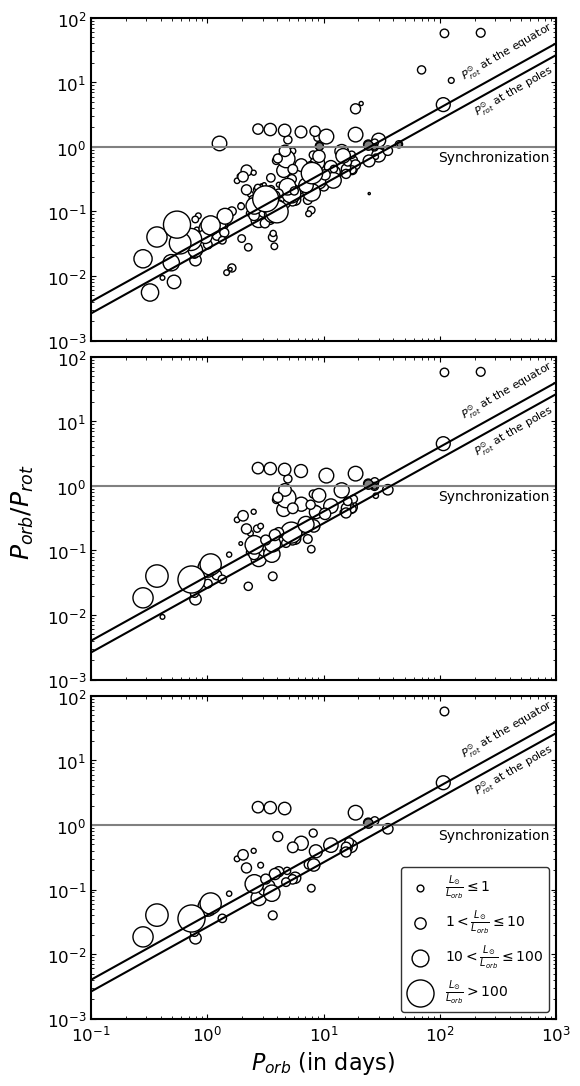}
        	\end{center}
        	\caption{The distribution of the period ratio $P_{\rm orb}/P_{\rm rot}$ vs. the orbital period $P_{\rm orb}$ as a function of ratio $L_{\rm *}/L_{\rm orb}$ (see the legend located in the bottom right corner) considering the three datasets in sequence. Synchronization regime (horizontal line) occurs at $P_{\rm orb}/P_{\rm rot}=1.0\pm0.1$ represented by dark gray circles. In the top panel, the planets CoRoT-4b, K2-3d, Kepler-102f, and TOI-1710b are in synchronous systems as for \texttt{dataset\#02} (middle panel). The synchronous systems are denoted by exoplanets Kepler-102f and TOI-1710b. In the \texttt{dataset\#03} (bottom panel), only TOI-1710b is synchronized. The subsynchronous and supersynchronous regimes are defined by $P_{\rm orb}/P_{\rm rot}<0.9$ and $P_{\rm orb}/P_{\rm rot}>1.1$, respectively. The diagonal lines indicate the values of solar rotation periods measured at the equator and poles. }
        	\label{figOR}
            \end{figure}

As \texttt{dataset\#01} contains some planetary systems whose parameters are not considered standard in the literature and also has 18 planetary systems without a cataloged stellar age, we decided to refine our sample further, creating a subset formed only by planetary systems with standardized sets of parameters from \texttt{dataset\#01}, giving rise to the \texttt{dataset\#02} with 91 exoplanets distributed in 69 systems, all of them with available stellar age data. As also can be seen in Fig. \ref{figOR} and discussed later in Sect. \ref{sec:level41}, \texttt{dataset\#01} (top panel) shows four synchronous systems (dark gray circles). After filtering, two of them, along with another 62 systems that did not meet the filtering criteria, are excluded from the \texttt{dataset\#02} (middle panel) that solely encompasses the two systems mentioned above. In \texttt{dataset\#01}, characterized by the utilization of data flagged both 0 or 1 to the Default Parameter set, the TOI-1710 system is visibly accompanied by the planets CoRoT-4b and K2-3d both flagged as 0 for the default parameter set, and Kepler-102f flagged as 1 for the default parameter set. As Kepler-102f does not have a well-defined mass value, only estimated by upper mass, it is flagged as 1 for the planet mass limit. For this, it is not observed in \texttt{dataset\#03}, where Default Parameter \textit{Flag=1} and Planet Mass Limit \textit{Flag=0} is employed. There is only one synchronous systems present in all datasets denoted by exoplanet TOI-1710b. 

Aiming to use more accurate data to evaluate the model presented here, we further refined the data from \texttt{dataset\#02}, creating a new subset, \texttt{dataset\#03}, consisting only of planetary systems whose parameters had well-defined values. This more restrictive dataset contains 56 exoplanets distributed in 44 systems. Conversely, upon imposing the constraints delineated in \texttt{dataset\#03}, only one of the synchronized systems manifests within the reduced sample.
  
Our sample comprises rotation and orbital period data required for the model described below. In particular, Kepler and TESS rotation period measurements are skewed to values less than 45 days \citep{CantoMartins2023}. In this context, our sample contains a tiny number of 19 stars with rotation periods longer than the stated limit. According to \cite{CantoMartins2023}, main-sequence stars with rotation periods higher than 45 days are becoming scarce, based on several observational investigations and theoretical projections. 

Finally, the age data extracted mostly from the works of \cite{Bonono2017,Bonomo2019,bonomo2023}, \cite{psaridi2023}, \cite{Barros2022}, \cite{alsubai2017} and \cite{hay2016}  contributed 34, 3, 1, 3 and 3 stars with well-defined ages, totaling 33\% of the \texttt{dataset\#01} sample, respectively. This information is crucial for understanding the composition of the datasets and the distribution of age data within them.  These authors obtained the majority of stellar ages in this study using a Bayesian differential evolution Monte Carlo method. This approach simultaneously tunes stellar parameters such as radius, mass, and age while also fitting the stellar spectral energy distribution and employing the MESA Isochrones and Stellar Tracks \citep{Paxton_2015}. Finally, all planets are close to the host star at $a<1$ AU. 

\section{\label{sec:level3}Modeling the effects of tidal evolution due to close-in planets} 

Modeling the effects of tidal evolution due to close-in planets involves understanding the intricate dynamics between planets and their host stars. The tidal interaction index, a new theoretical parameter developed in this Section, plays a crucial role in quantifying the impact of tidal forces on the orbital evolution of planets. Studies \citep[e.g.,][]{gallet2017} have shown that close-in planets can induce significant tidal effects on stellar rotation, leading to planet migration and potential collisions. The evolution of these systems is influenced by factors such as the mass ratio between the planet and the star, the efficiency of inward migration, and the excitation of tidal inertial waves and magnetic fields. By incorporating the tidal interaction index into models, we can better predict the long-term evolution of close-in planets and their effects on stellar rotation, providing valuable insights into the dynamics of exoplanetary systems.

\subsection{\label{sec:level31}Tidal interaction index $q$}
    
First, this paper is the final installment in a series of three papers \citep{deFreitas2021,deFreitas2022} that present a new approach to the evolution of stellar rotation. The present work is crucial for understanding the physical mechanisms that control the spin-down of stars hosting planets since stars and their exoplanets evolve together. By employing the equation proposed by \citet{deFreitas2021}, we can assume that the angular momentum loss rate is a power law-like $\left(\mathrm dJ/\mathrm dt\right)_{\rm tidal} \propto\Omega_{*}^{q}$ and, therefore, the evolution of stellar rotation, considering the inertial moment as a constant, is given by
 
        \begin{equation}
            \label{eq.1}
            \dot{\Omega}_{*}=-\frac{\Omega_{0}}{\tau}\left(\frac{\Omega_{*}}{\Omega_{0}}\right)^{q},\quad q\geq 1
        \end{equation}
where $\tau$ is defined as a characteristic time, $\Omega_{0}$ represents the angular velocity at time $t = 0$ and $\Omega_{*}$ is the stellar angular velocity at time $t=t_{\rm age}$, i.e., now. 
        
The conceptual origin of the parameter $q$ is associated with the degree of non-extensivity that emerges within thermodynamic formalism proposed in \citet{tsallis1988}. As mentioned by \cite{deFreitas2021}, the non-extensive formalism is an influential theory for studying stellar systems, including open clusters and field stars, as seen here. Several studies \citep[e.g.,][]{deFreitas2013, deFreitas2021, deFreitas2022} have demonstrated that the mechanism of stellar rotational braking can be explained by investigating the behavior of the entropic index $q$. Consequently, \citet{deFreitas2013} proposed that increasing index $q$ behavior between the saturated $(q = 1)$ and unsaturated $(1 < q < 3)$ magnetic field regimes, as a function of spectral, may explain the process of magnetic transition that occurs in the stellar dynamo.
        
In the literature cited above, the index $q$ is commonly assumed to be a constant that measures the efficiency of magnetic braking during the lifespan of stars. This braking index can be defined using the second derivative of $\Omega_{*}$ as a function of time. Considering that $q$ does not depend on the time, it can be expressed as
        \begin{equation}\label{eq.2}
              q=\frac{\ddot{\Omega}_{*}\Omega_{*}}{\dot{\Omega}_{*}^{2}}.
                \end{equation}
It is essential to highlight that the expression for $q$ in agreement with $\left(\mathrm dJ/\mathrm dt\right)_{\rm tidal}$ follows this same procedure. 
        
\cite{deFreitas2013} showed that the index $q$ is associated with the dynamo relationship and the measurement of magnetic field geometry by expression 
\begin{equation}\label{eqSku}
q=1+\frac{4aN}{3}. 
\end{equation}
In this scenario, for $a=1$, a value of $N=3/2$ yields the \cite{sku} law ($\Omega_{*} \propto t^{-1/2}$) \citep{kawaler1988}. However, $q$-index extracted from eq. \ref{eq.2} can be quite different when the star-planet interaction is considered.

\subsection{\label{sec:level32}The model \textit{per se}}
    
Based on eq. \ref{eq.2}, the model determines $\Omega_{*}$ and its derivatives thus finds an expression for the \textit{tidal interaction index} $q$. We need a model that describes tidal dissipation in stars and planets. To this end, we choose the \citet{ogilvie2014} model, which considers two bodies (p. ex., one star-planet system) orbiting about their mutual center of mass based on the language of linear tidal theory extracted from \citet{darwin1879} model. 
    
The \citet{ogilvie2014} model introduces the concept of tidal dissipation and its significance in understanding the evolution of stellar and planetary systems. It explores the tidal interaction between a star or planet and its companion, which results in the transfer of angular momentum and energy. This interaction causes tidal deformation and generates internal tidal waves within the body. For the present study, we assume some simplifications are as follows: 
    
According to the mentioned model, assuming all parameters $a$, $e$ and $\Omega_{*}$ having the same time lag $\tau$, Love numbers, that quantify the deformation of the gravity field of a planet in response to the parent star, can be defined by $k_{l,m,n}= k_{l}\tau (n\Omega_{\rm orb}-m\Omega_{*})$ \citep[see also][]{lanza2015}. In particular, the Love number $k_{2,2,2}$ (hereafter $k_{2}$) is of special interest. It measures an object's central condensation level: the more homogeneous the planet in mass distribution, the bigger the Love number $k_{2}$ \citep{refId0}. With these conditions, we obtain the simpler results, as described by the equations below:
            \begin{align}
            \label{eq.12_paper}
                \frac{\dot{a}}{a}&=-6k_{2}\tau(\Omega_{\rm orb}-\Omega_{*})\frac{M_{p}}{M_{*}}\left(\frac{R_{*}}{a}\right)^{5}\Omega_{\rm orb} \\
            \label{eq.14_paper}
                \frac{\dot{e}}{e}&=-\frac{3}{2}k_{2}\tau(18\Omega_{\rm orb}-11\Omega_{*})\frac{M_{p}}{M_{*}}\left(\frac{R_{*}}{a}\right)^{5}\Omega_{\rm orb} \\
            \label{eq.13_paper}
                \frac{\dot\Omega_{*}}{\Omega_{*}}&=3k_{2}\tau(\Omega_{\rm orb}-\Omega_{*})\frac{L_{\rm orb}}{L_{*}}\frac{M_{p}}{M_{*}}\left(\frac{R_{*}}{a}\right)^{5}\Omega_{\rm orb}.
            \end{align}
        From eqs. \ref{eq.12_paper} and \ref{eq.13_paper}, we have
            \begin{equation}
                \label{eq.3}
                \frac{\dot{a}}{a}\frac{\dot\Omega_{*}}{\Omega_{*}}=-2\left(\frac{L_{\rm orb}}{L_{*}}\right).
            \end{equation}
            Now, dividing eq. \ref{eq.14_paper} by eq. \ref{eq.13_paper} we obtain a similar expression to the evolution of $a$:
            \begin{equation}
                \label{eq.4}
                \frac{\dot{e}}{e}\frac{\dot\Omega_{*}}{\Omega_{*}}=-\frac{1}{2}\left(\frac{L_{*}}{L_{\rm orb}}\right)\left(\frac{18\Omega_{\rm orb}-11\Omega_{*}}{\Omega_{\rm orb}-\Omega_{*}}\right).
            \end{equation}     
When the stellar moment of inertia $I_{*}$ is kept constant, the angular momentum $L_{*}=I_{*}\Omega_{*}$ and its first derivative $\dot L_{*}$ can be related by 
             \begin{equation}
                \label{eq.5}
               \frac{\dot L_{*}}{L_{*}}=\frac{\dot \Omega_{*}}{\Omega_{*}}.
            \end{equation}
A similar expression for the orbital momentum and its first derivative can be obtained from eq. \ref{eq.11_paper}:
            \begin{equation}
                \label{eq.6}
               \frac{\dot L_{\rm orb}}{L_{\rm orb}}=-\left(\frac{e^{2}}{1-e^{2}}\right)\frac{\dot e}{e}-\frac{\dot a}{a}. 
            \end{equation}
        
For this work, only the outer convective envelope will provide rotational angular momentum in the star. In other words, we ignore any angular momentum transfer from the stellar inner rigid body-like core. Thus, differentiating the eq. \ref{eq.13_paper} concerning time and dividing this result by $\dot \Omega_{*}$, after some algebra, we can obtain the following expression for the ratio of first and second derivatives of angular velocity:
            \begin{equation}
                \label{eq.7}
               \frac{\ddot \Omega_{*}}{\dot \Omega_{*}}=-5\left(\frac{\dot a}{a}\right)+\frac{\dot L_{\rm orb}}{L_{\rm orb}}-\frac{\dot L_{*}}{L_{*}}+\frac{\dot \Omega_{*}}{\Omega_{*}}\left(\frac{\Omega_{\rm orb}-2\Omega_{*}}{\Omega_{\rm orb}-\Omega_{*}}\right). 
            \end{equation}
        
Replacing eqs. \ref{eq.5} and \ref{eq.6} into eq. \ref{eq.7} and dividing the resulting expression by $\dot{\Omega}_{*}/\Omega_{*}$, we have the following:
            \begin{equation}
                \label{eq.8}
               \frac{\ddot \Omega_{*} \Omega_{*}}{\dot \Omega^{2}_{*}}=-6\left(\frac{\dot a}{a}\right)\frac{\Omega_{*}}{\dot \Omega_{*}}-\left(\frac{e^{2}}{1-e^{2}}\right)\frac{\dot e}{e}\frac{\Omega_{*}}{\dot{\Omega}_{*}}-\frac{\Omega_{*}}{\Omega_{\rm orb}-\Omega_{*}}. 
            \end{equation}
Using eqs. \ref{eq.3} and \ref{eq.4}, the above equation can be rewritten in the form
            \begin{equation}
            \begin{aligned}
                \label{eq.9}
               \frac{\ddot \Omega_{*} \Omega_{*}}{\dot \Omega^{2}_{*}}&=\frac{L_{*}}{L_{\rm orb}}\Biggl[12+\frac{1}{2}\left(\frac{e^{2}}{1-e^{2}}\right)\left(\frac{18\Omega_{\rm orb}-11\Omega_{*}}{\Omega_{\rm orb}-\Omega_{*}}\right)\Biggr]\\
               &\qquad-\frac{\Omega_{*}}{\Omega_{\rm orb}-\Omega_{*}}. 
            \end{aligned}
            \end{equation}
        
Finally, as defined in eq. \ref{eq.2}, $q$-index can be written in the following form
            \begin{equation}
                \label{eq.10}
               q=\frac{1}{1-\frac{\Omega_{\rm orb}}{\Omega_{*}}}+\Biggl[12+\frac{1}{2}\left(\frac{e^{2}}{1-e^{2}}\right)\left(\frac{18\Omega_{\rm orb}-11\Omega_{*}}{\Omega_{\rm orb}-\Omega_{*}}\right)\Biggr]\frac{L_{*}}{L_{\rm orb}}. 
            \end{equation}
It can be rewritten in terms of $P_{\rm rot}/P_{\rm orb}$ ratio by replacing $\Omega_{\rm orb}/\Omega_{*}$ with $\left(P_{\rm orb}/P_{\rm rot}\right)^{-1}$, we have that
            \begin{equation}
                \label{eq.11}
               q=\frac{\frac{P_{\rm orb}}{P_{\rm rot}}}{\frac{P_{\rm orb}}{P_{\rm rot}}-1}+\Biggl[12+\frac{11}{2}\left(\frac{e^{2}}{1-e^{2}}\right)\left(\frac{\frac{P_{\rm orb}}{P_{\rm rot}}-\frac{18}{11}}{\frac{P_{\rm orb}}{P_{\rm rot}}-1}\right)\Biggr]\frac{L_{*}}{L_{\rm orb}}, 
            \end{equation}
this means that the expression is asymptotic for synchronized systems, i.e., the $P_{\rm orb}/P_{\rm rot}$ ratio must differ from 1. 
            
To investigate the angular momentum, it is necessary to define the ratio of the orbital angular momentum $L_{\rm orb}$ to the spin angular momentum of a star $L_{*}$ (with a fixed $I_{*}$), given by 
             \begin{equation}
                \label{eq.11_paper}
                \frac{L_{*}}{L_{\rm orb}}=\frac{I_{*}\Omega_{*}\Omega_{\rm orb}a}{GM_{*}M_{p}\sqrt{1-e^{2}}}.
            \end{equation}

Rewriting the above equation in Earth and Sun units, we have
    \begin{equation}
                \label{eq.12a}
               \frac{L_{*}}{L_{\rm orb}}=\frac{384984}{\sqrt{1-e^2}}\frac{\left(\frac{a}{1\rm{AU}}\right)\left(\frac{R_{*}}{R_{\odot}}\right)^{2}}{\left(\frac{M_{p}}{M_{\oplus}}\right)\left(\frac{P_{\rm orb}}{1\;\rm day}\right)\left(\frac{P_{\rm rot}}{1\;\rm day}\right)},
            \end{equation}
where, we assume that $I_{*}=\frac{2}{5}M_{*}R^{2}_{*}$ \citep[see Ref.][]{Gurumath}, $R_{\odot}=6.96\times 10^{8}$ m, $M_{\oplus}=5.97\times 10^{24}$ kg, and the Universal Gravitational constant $G=6.67\times 10^{-11}$ Nm$^{2}$kg$^{-2}$.
    
As we can see in Fig. \ref{figOR}, the majority of the stars in our sample have orbital mean motions that are much faster than the host star's rotation period (i.e., $P_{\rm orb}/P_{\rm rot}\ll 1$). The planetary rotation periods, $P_{p}$, are frequently unknown, even though these two parameters have been measured for numerous compact systems \citep{montes}. Owing to the proximity of planets to their parent stars, it is assumed that tidal interactions have brought them to a synchronous orbital state (i.e., $P_{\rm orb}\sim P_{p}$) over a period that varies from system to system depending on various physical factors \citep{1977plsa}. Additionally, we disregard any ($P_{p},P_{\rm orb}$) resonances \citep{Winn_2005}.

All parameter symbols emphasized in this subsection are defined in the caption of Table \ref{tab:table_parameters} or throughout the text.
        
\section{\label{sec:leve4}Results and Discussions}
This study's findings are rooted in examining various planetary and stellar characteristics. The ensuing discussion will highlight the most pertinent parameters to our investigation, thereby facilitating a deeper comprehension of the proposed model. The ratio between the orbital and rotational periods is particularly interesting, as the rotational aspect holds a pivotal position within the model under scrutiny. Exploring these interrelations is instrumental in elucidating the physical implications of the tidal index $q$.

\subsection{\label{sec:level41}Spin-orbit states and other planetary parameters}
    
We have utilized $P_{\rm orb}/P_{\rm rot}$ ratio, which is a measure of the degree of synchronization for a particular star-planet system, to examine the orbital-rotation period regimes \citep[cf.][]{CantoMartins2023}. In broad terms, $P_{\rm orb}/P_{\rm rot}$ is the ratio of the planet's orbital period to the star's rotation period measured in days.
    
Figure \ref{figOR} shows the parameter $P_{\rm orb}/P_{\rm rot}$ as a function of the planetary orbital period for a set of planet-hosting stars composing the three datasets as described in Section \ref{sec:level2}. We identify three subsamples, composed of synchronized ($P_{\rm orb}/P_{\rm rot}=1.0\pm0.1$), subsynchronized ($P_{\rm orb}/P_{\rm rot}<0.9$), and supersynchronized ($P_{\rm orb}/P_{\rm rot}>1.1$) star-planet systems which emerge from Fig. \ref{figOR}. This Figure also shows that these three subsamples delineate a continuous transition from subsynchronized to supersynchronized regimes. However, considering the orbital period range, the behavior of these subsamples is different. 
        
Only subsynchronous star-planet systems are observed for orbital periods shorter than one day. The three cited subsamples are presented for orbital periods greater than one day. For orbital periods greater than 40 days, only supersynchronous systems are observed. It is essential to highlight that our sample is hegemonically subsynchronized with only 19 systems within the supersynchronized regime and four synchronous star-planet systems, named CoRoT-4b, K2-3d, Kepler-102f, and TOI-1710b, considering the \texttt{dataset\#01}. On the other hand, two synchronous star-planet systems are in \texttt{dataset\#02}, composed by Kepler-102f and TOI-1710b exoplanets, and only TOI-1710b is synchronized in the \texttt{dataset\#03}. 

Figure \ref{figOR} further demonstrates that systems with a significant concentration of angular momentum within the host stars are primarily located within or in close proximity to the boundaries of the solar rotation period, regardless of the dataset being analyzed. This observation contradicts the distribution observed in the Solar System, where most angular momentum is concentrated within the planets. However, this discrepancy arises due to the limitation of our sample to planets close to their host stars, resulting in a diminished impact. In systems where planets retain the angular momentum, a similar trend is observed, albeit with a wider dispersion, in contrast to the former, primarily confined to subsynchronous systems.

\subsubsection{\label{sec:level42}Synchronization timescale}
Tidal interactions between the star and the planet can lead to the synchronization of their rotation and orbit. The timescales for spin-orbit synchronization, spin-orbit alignment, and orbit decay or expansion can be expressed in the tidal evolution timescales \citep{mardling2010,mardling2011}. The timescales for synchronization can vary depending on the system parameters (see eq. \ref{eq.syn} below). In some cases, the tidal synchronization timescales can be much longer than the ages of the stars \citep{lanza2010}.

        %%%%
    	\begin{figure}
        	\begin{center}	\includegraphics[scale=0.58]{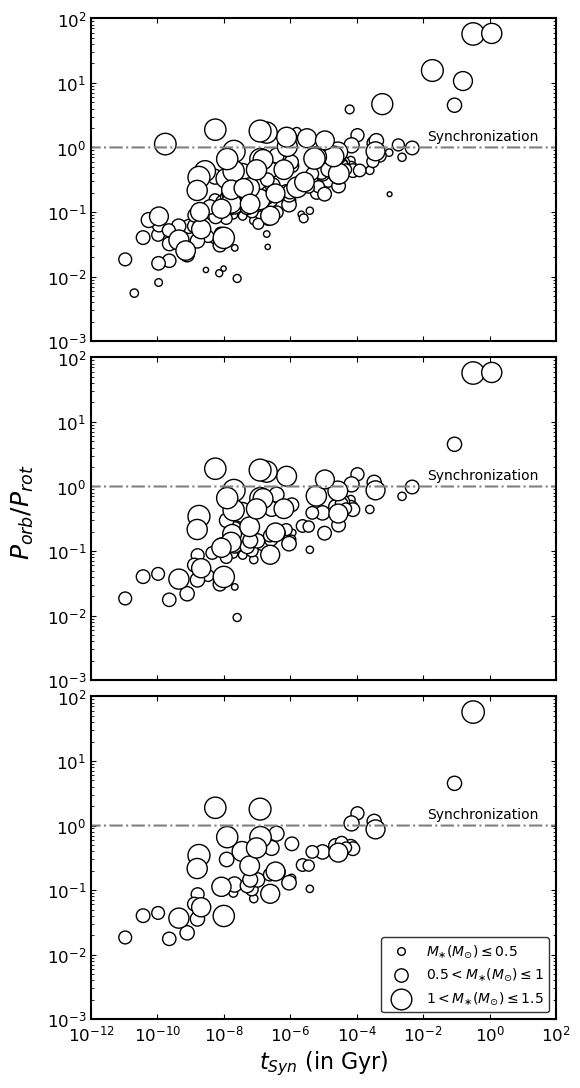}
        	\end{center}
        	\caption{The distribution of the period ratio $P_{\rm orb}/P_{\rm rot}$ vs. synchronization time $t_{\rm Syn}$ as a function of stellar mass shows an almost linear evolution extending from the subsynchronized to the supersynchronized regime, considering the three datasets.}
        	\label{figtSyn}
            \end{figure}
         
         %%%%
         
        \begin{figure}
        	\begin{center}	\includegraphics[scale=0.58]{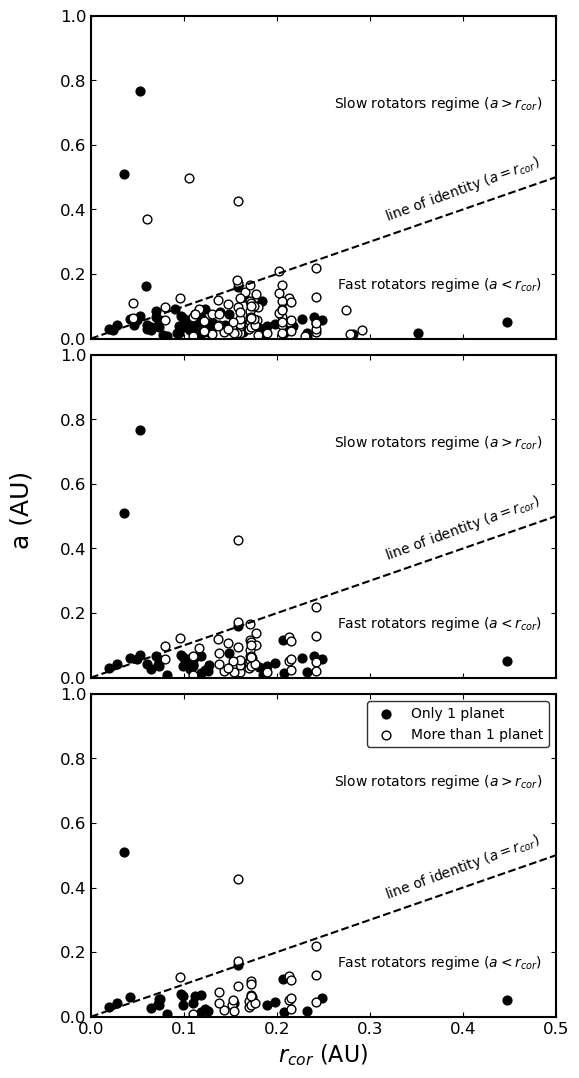}
        	\end{center}
        	\caption{Semi-major axis vs. corotation radius shows two distinct rotation regimes considering the three datasets.}
        	\label{fig0}
        \end{figure}
        
The presence of a close-in planet can also affect the rotation and activity of the host star. Recent theoretical models suggest that stars dissipate tidal energy from interactions with their planets very slowly, with relevant stellar quality factors on the order of $10^{8}-10^{9}$ \citep{poppen2014}. This indicates that the tidal energy dissipation timescales in star-planet systems may be higher than previously thought.
        
The evolution of the planetary spin (synchronization and alignment) takes place first in a star-planet system because the angular momentum contained in the planetary spin is less than that of the orbital and stellar ones. In fact, according to the conventional idea of equilibrium tides, a planet should synchronize on a timescale equal to:
\begin{equation}
                \label{eq.syn}
               t_{\rm syn}=20\left(\frac{a}{1\rm AU}\right)^{6}\left(\frac{M_{*}}{M_{\odot}}\right)^{-2}\left(\frac{M_{\rm p}}{M_{\oplus}}\right)\left(\frac{R_{\rm p}}{R_{\oplus}}\right)^{-3} \rm Gyr,
            \end{equation}
where the value 20 was derived from the expression:
\begin{equation}
\frac{r^{2}_{g}M_{\rm p}(1\rm AU)^6}{3G(k_{2}\Delta t)M^{2}_{\odot}R^{3}_{\oplus}},
\end{equation}
$k_{2}$ is the tidal Love number of degree 2 (as seen previously), and $\Delta t$ time lag, which characterizes the efficiency of the tidal dissipation into the interior of the planet (the higher $t_{\rm Syn}$, the higher the dissipation), and $r_{g}$ is the dimensionless gyration radius, where for a homogeneous interior, $r^{2}_{g}=2/5$. The time lag obtained for the Earth from the study of Lunar Laser Ranging studies, $k_{2}\Delta t=0.305\times 629$ s \citep{neron}, can provide one a general notion of the orders of magnitude involved, even though the precise size of the tidal dissipation in terrestrial planets is still difficult to determine. This results in a synchronization period for the Earth of 20 Gyr, Venus of 3 Gyr, and Mercury of 80 Myr \citep{forget}. 
        
As shown in Fig. \ref{figtSyn}, supersynchronized systems take longer to achieve synchronization than subsynchronized ones. The Figure also reveals that less massive host stars, for the most part, dissipate tidal energy much more slowly than more massive stars for the same synchronization regime. This phenomenon is observable across all three datasets. Nevertheless, with a decrease in sample size (from \texttt{dataset\#01} to \texttt{dataset\#03}), the distribution predominantly exhibits sub-synchronization. This skew is attributed to the sampling methodology outlined in Section \ref{sec:level2}.

\subsubsection{\label{sec:level43}Corotation radius}
            
The corotation radius ($r_{\rm cor}$) corresponds to the orbital radius for which the orbital period is equal to the rotation period of the star, i.e., when system synchronization occurs \citep{gallet2018}. In other words,
        \begin{equation}
                    \label{eq.rcorot}
                   r_{\rm cor}=\left(\frac{GM_{*}P^{2}_{\rm rot}}{4\pi^{2}}\right)^{1/3}=0.02\left(\frac{P_{\rm rot}}{1\;\rm day}\right)^{2/3}\left(\frac{M_{*}}{M_{\odot}}\right)^{1/3} \rm AU,
                \end{equation}
for example, for a solar mass star at 1-day rotation, $r_{\rm cor}$ is at 0.02 AU. Thus, a planet orbiting outside this radius takes away angular momentum from the star, causing the star to spin down and the planet to be pushed away from the star \citep{vidotto2014}. An important application of this equation is when we assume the Sun's rotation period; in this case, the $r_{\rm cor}$ is close to 2 AU. Thus, Earth is within this limit and interacts magnetically with the Sun, whereas Jupiter takes angular momentum away from the Sun.

        \begin{figure}
            \begin{center}
                \includegraphics[scale=0.58]{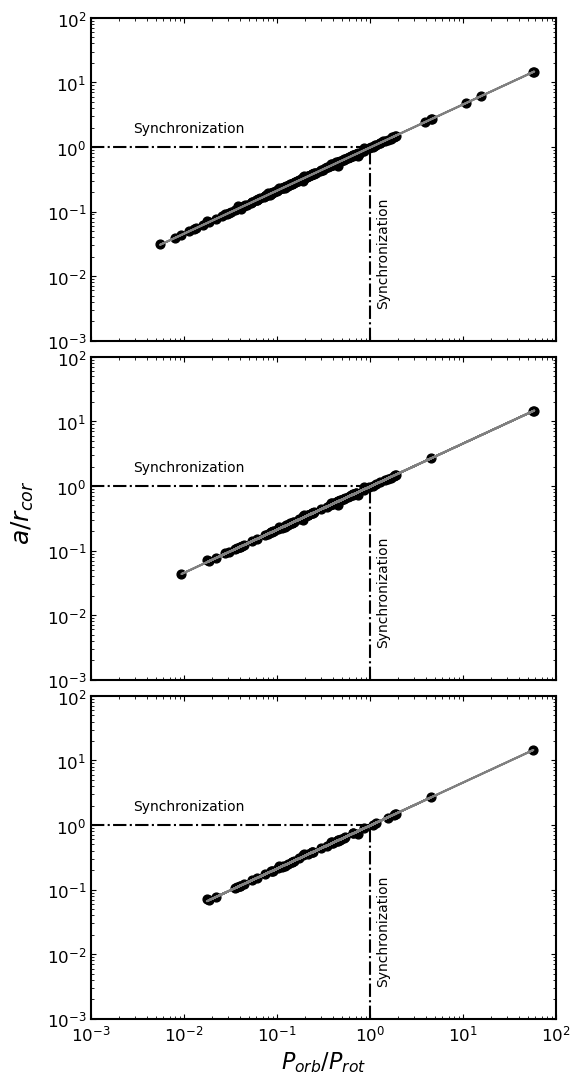}
            \end{center}
            \caption{Correlation between $a/r_{\rm cor}$ and $P_{\rm orb}/P_{\rm rot}$ considering the three datasets. The linear fit was obtained using the eq. \ref{eq.12abc} in log-log axis, where the slope is $\sim 2/3$.}
            \label{figaCor}
        \end{figure}
                        
In this context, when the semi-major axis of the planet-star system is smaller than the corotation radius ($a<r_{\rm cor}$), the tidal forces can transfer angular momentum from the planetary orbit to the star, causing the star to accelerate its rotation and, ultimately, possibly leading to the engulfment of the planet. On the other hand, when the semi-major axis of the system is larger than the corotation radius ($a>r_{\rm cor}$), tidal forces can transfer angular momentum from the star to the planet, doing the star spins more slowly and, consequently, pushing the planet away from the star \citep{privitera2016,vidotto1}. Hence, stars born as fast rotators could have pushed away their inner planets at the beginning of their lives. This scenario can be compiled using eq. \ref{eq.rcorot} and third Kepler law, as follows:
        \begin{equation}
                    \label{eq.12ab}
                    \frac{a^{3}}{P^{2}_{\rm orb}}=\frac{GM_{*}}{4\pi^{2}},
                \end{equation}
        hence
        \begin{equation}
                    \label{eq.12abc}
                    \frac{a}{r_{\rm cor}}=\left(\frac{P_{\rm orb}}{P_{\rm rot}}\right)^{2/3}.
                \end{equation}

This strong correlation can be seen clearly in Fig. \ref{figaCor}, where a power-like law $y=ax^{b}$ (here, the values calculated of $a$ and $b$ are, respectively, $0.978\pm 0.001$ and $0.6671\pm 0.0002$) show that for synchronous systems, the exoplanet is located precisely in the corotation radius.

Thus, the Sun could have been born as a fast rotator, but its rotation decreased as Jupiter moved away from it. This could explain why the Sun's rotation is anomalous compared to stars with the same spectral type but do not host planets. Likewise, stars in the red giant phase with high rotation would result from the increase in the corotation radius due to the expansion of the convective envelope, allowing planets to have their orbit reduced to the point of being engulfed by the star. In line with Fig. \ref{fig0}, the outcomes for the three datasets are mutually consistent, albeit with a decrease in the occurrence of supersynchronous systems, leading to a corresponding reduction in the number of planetary systems situated at greater distances from the host star.

\subsubsection{\label{sec:level44}Roche radius}
In our sample, all planetary orbits are larger than their critical Roche radius, $a_{\rm Roche}$, which is the distance at which the tidal pressures of a host star begin to overwhelm a planet's self-gravity \citep{montes}. In this work the $a_{\rm Roche}$ is defined, in Sun and Earth units, as
            \begin{equation}
            \label{eq.roche}
            a_{\rm Roche}=0.0008\left(\frac{M_{*}}{M_{\odot}}\right)^{1/3}\left(\frac{M_{\rm p}}{M_{\oplus}}\right)^{-1/3}\left(\frac{R_{\rm p}}{R_{\oplus}}\right) \rm AU,
                    \end{equation}
equation based on the simulations by \cite{Guillochon_2011} for the disruption of hot Jupiters.
                
We found that all systems in our sample have a Roche limit much lower than the semi-axis $a$ value, and, therefore, none of the planets are being disrupted by the action of tidal forces.

\subsection{\label{sec:level46}$q$-correlations}

\begin{figure*}
            	\begin{center}
            		\includegraphics[scale=0.465]{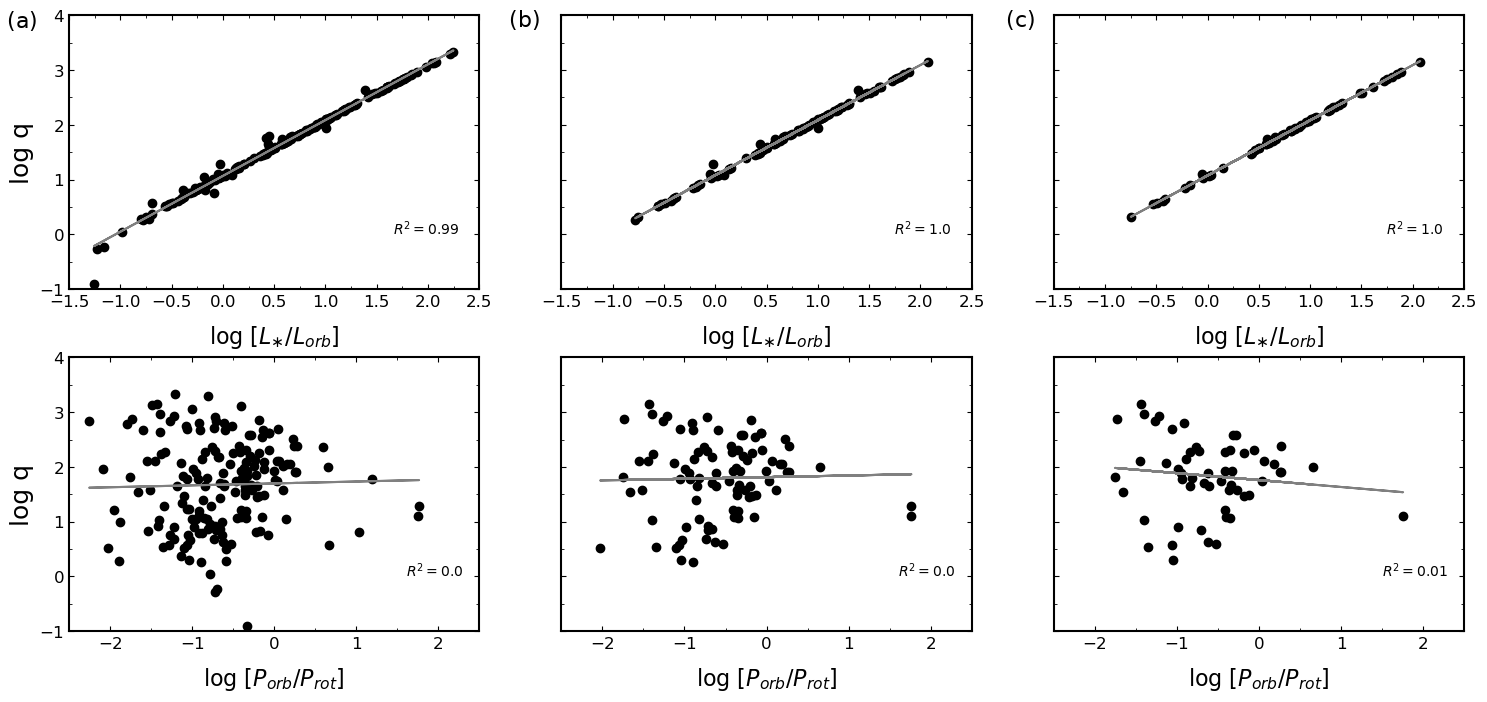}
            	\end{center}
            	\caption{Log-log plot of $q$-index as a function of $L_{*}/L_{\rm orb}$ (\textit{top panel}) and $P_{\rm orb}/P_{\rm rot}$ (\textit{bottom panel}) separated in the three datasets distributed in Figures (a), (b) and (c), respectively. On the top panels, the best-fit is described by relationship $\log q=a+b\log(L_{*}/L_{\rm orb})$ or $q=10^{a}\left(L_{*}/L_{\rm orb}\right)^{b}$ with $a=1.073\pm 0.006$ and $b=1.011\pm 0.007$, whereas in the bottom panel, there is a weak correlation given by $\log q=a+b\log(P_{\rm orb}/P_{\rm rot})$ with $a=1.76\pm 0.06$ and $b=-0.02\pm 0.09$.}
            	\label{fig1a}
        \end{figure*}

The $q$-index is influenced by several parameters such as planetary mass, eccentricity, and $L_{*}/L_{\rm orb}$ ratio (cf. eq. \ref{eq.11}). However, we verified that the parameter that determines the behavior of the $q$-index is the $L_{*}/L_{\rm orb}$ ratio. As we can see in all of the top panels from Fig. \ref{fig1a}, $q$ and $L_{*}/L_{\rm orb}$ are strongly correlated, indicating that the $q$-index depends almost exclusively on the exchange of angular momentum between stars and planets. In contrast, the correlation coefficient $R^{2}$ suggests a negligible relationship between $q$ and $P_{\rm rot}/P_{\rm orb}$. In this way, we can simplify eq. \ref{eq.11} by the expression:
                \begin{equation}
                    \label{eq.12b}
                   q\approx 12\left(\frac{L_{*}}{L_{\rm orb}}\right).
                \end{equation}

As an example, when we consider an exponent that corresponds to \cite{sku}'s law, i.e., $q=3$, the ratio $L_{*}/L_{\rm orb}=1/4$. This scenario occurs when 80\%  of the total angular momentum ($J=L_{*}+L_{\rm orb}$) is in planets. In another example, if we consider Jupiter as an equivalent planet (where $P_{\rm orb}=12$
years, $a=5.2$AU, $M_{p}=317,8M_{\oplus}$ and $e=0.048$), the ratio $L_{\odot}/L_{\rm
orb}=0.06$ and, therefore, $q=1.71$. This example implies that most of the current angular
momentum of the Solar system ($\sim 94\%$) is located in Jupiter, due to its high planetary mass and large orbital distance \citep{lanza2010,sibony2022}.

            \begin{figure*}
                	\begin{center}
                		\includegraphics[scale=0.465]{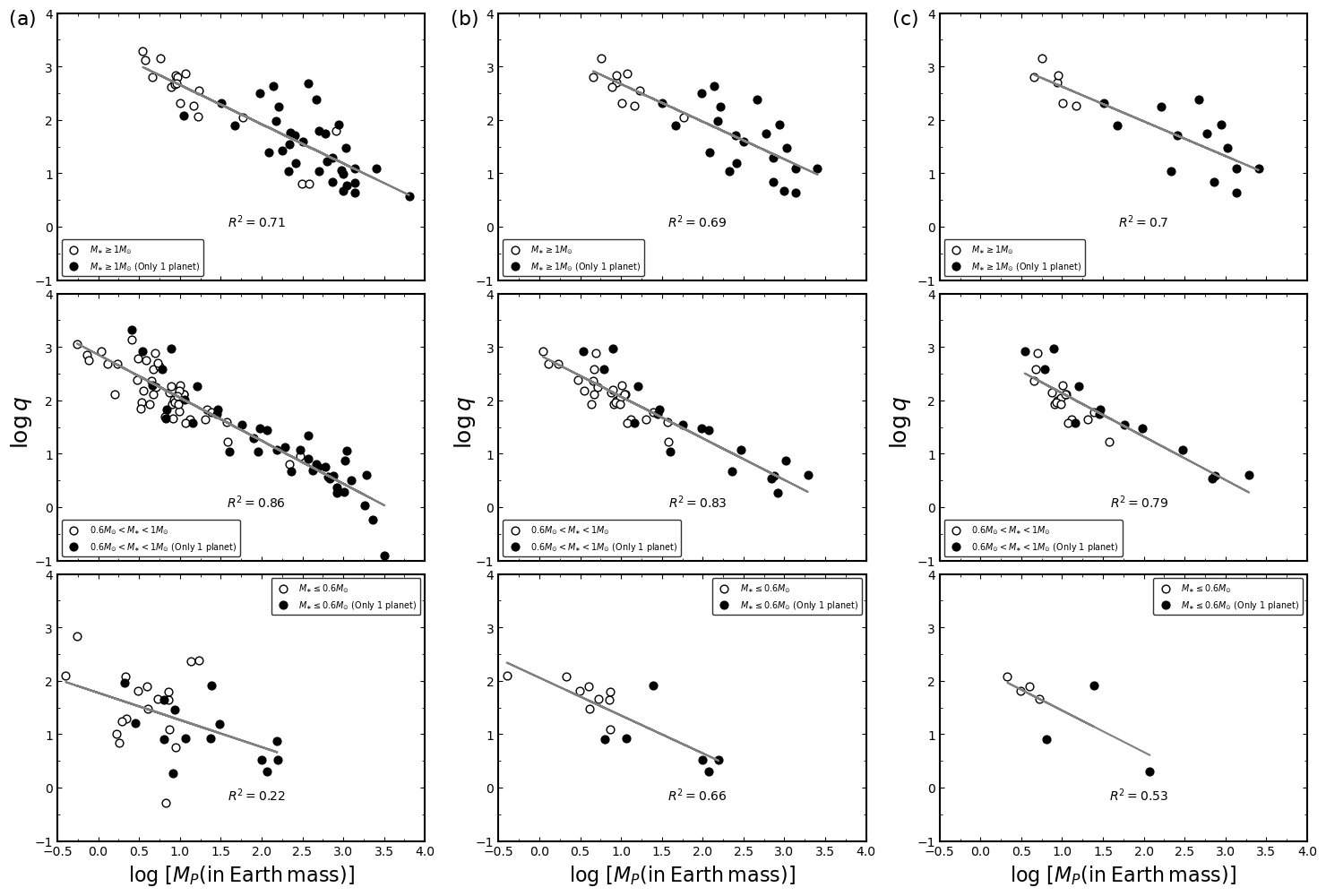}
                	\end{center}
                	\caption{Logarithmic-scale correlation (log-log) plots between $q$ and the planetary mass $M_{p}$ segregated by stellar mass using the equation $\log q=a+b\log M_{p}$ separated in the three datasets distributed in Figures (a), (b) and (c), respectively. The datasets were stratified based on system type, with closed circles representing systems containing a single planet and open circles signifying systems harboring multiple planets. Our results show that for $M\ge 1M_{\odot}$ (with a sample of 53 stars) $\log q=(3.34\pm 0.06)+(-0.70\pm 0.04)\log M_{p}$, while for $0.6<M<1M_{\odot}$ (with a sample of 89 stars)  $\log q=(2.88\pm 0.05)+(-0.80\pm 0.02)\log M_{p}$, and for $M\le 0.6M_{\odot}$ (with a sample of 33 stars) $\log q=(2.01\pm 0.22)+(-0.66\pm 0.14)\log M_{p}$. The values of $R^{2}$ also are highlighted.}
                	\label{fig4}
            \end{figure*}

                    \begin{figure}
                	\begin{center}
                		\includegraphics[scale=0.58]{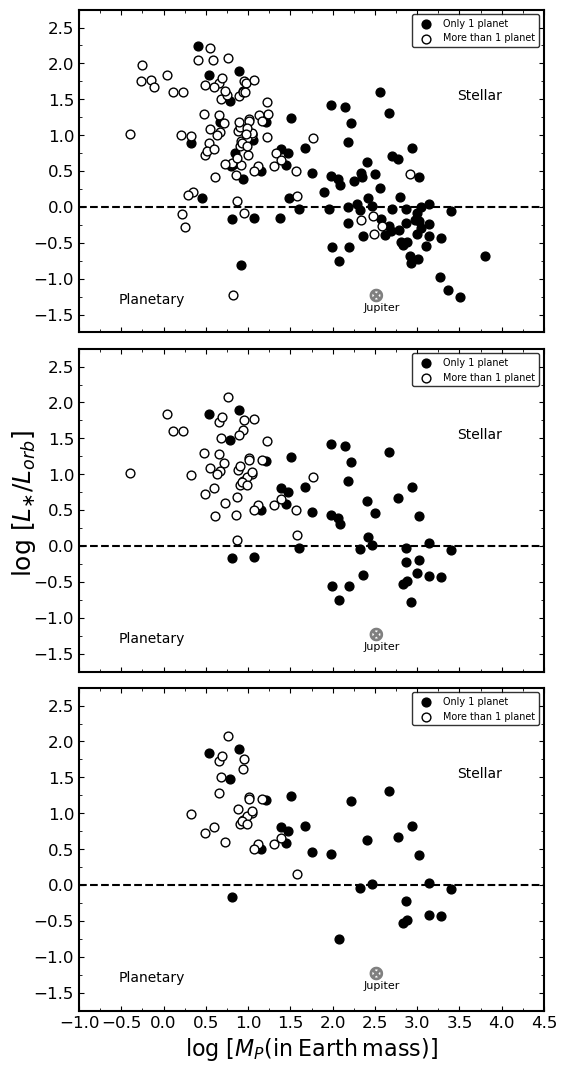}
                	\end{center}
                	\caption{Logarithmic-scale correlation (log-log) plots between $L_{*}$ and $/L_{\rm orb}$ and the planetary mass $M_{p}$ segregated by number of planets and separated in the three datasets. The short dotted line delimits two domains about the consumption of the total angular momentum: the one where the star holds the largest share and another where the planet holds. The symbol $\otimes$ denotes the Jupiter planet.}
                	\label{fig2}
                    \end{figure}

\subsubsection{\label{sec:level47}Planetary mass-$q$ relationship}
Figure \ref{fig4} clearly illustrates three distinct behaviors observed when the sample is segregated by stellar mass repeated in the three datasets. Intermediate-mass stars ($0.6<M<1M_{\odot}$) tend to have a stronger anti-correlation between their $q$-index and planetary mass, as revealed by high correlation coefficient $R^{2}$. 
            
In general terms, the segregation by mass evidences the relevant correlation between the $q$-index and the planetary mass, suggesting a gradual decrease of $q$ when the planetary mass increases. As we have seen in Fig. \ref{fig2}, this effect is mainly due to the ratio of the momenta $L_{*}$ and $/L_{\rm orb}$. We found that systems with only one planet have preferentially smaller $q$ values than those with more than one planet. Systems with more than one planet remove more angular momentum from the host star, causing the rotation period to increase. 
    
                    \begin{figure}
                	\begin{center}
                		\includegraphics[scale=0.58]{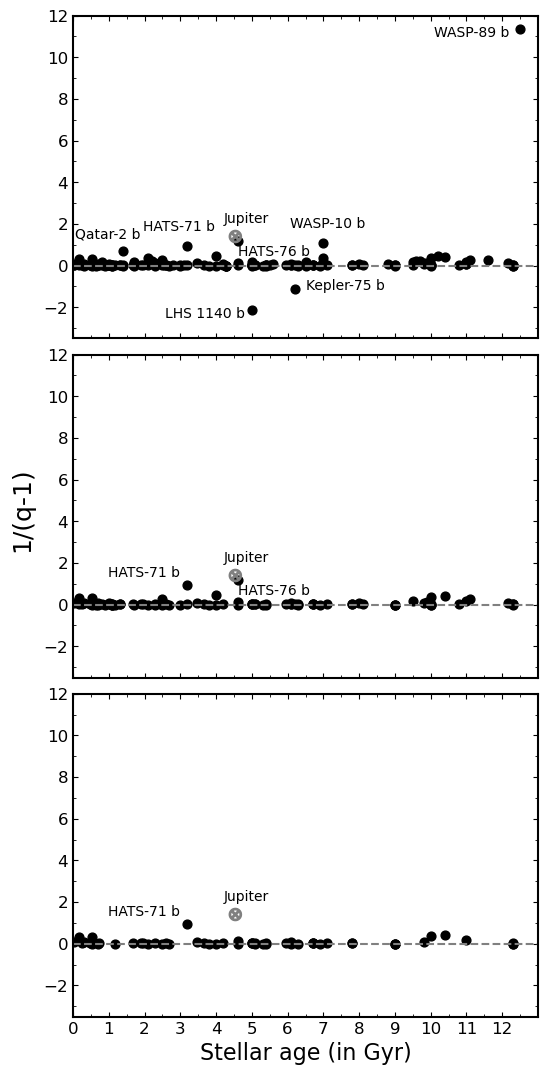}
                	\end{center}
                	\caption{Relation between the stellar age and exponent $1/(q-1)$ for three datasets with well-determined ages.}
                	\label{figAge}
                    \end{figure}
            
                    \begin{figure*}
                	\begin{center}
                		\includegraphics[scale=0.46]{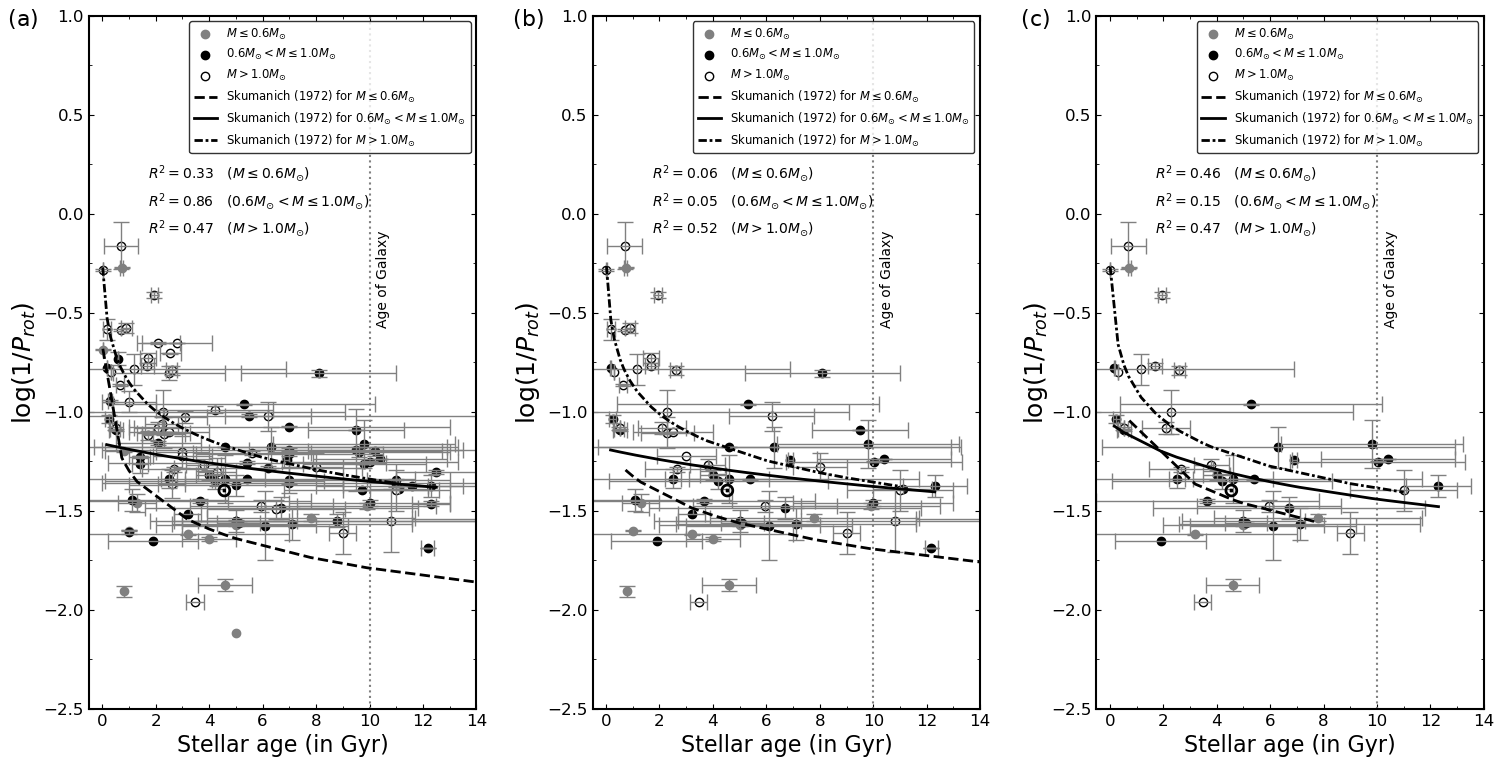}
                	\end{center}
                	\caption{Angular velocity segregated by mass as a function of stellar age separated in the three datasets distributed in Figures (a), (b), and (c), respectively. The symbol $\ odot$ represents the Sun. The curves represent the best fit based on the Skumanich law in agreement with the stellar mass. The vertical line indicates the upper limit of the age of the Galaxy in 10 Gyr.}
                	\label{figProtAge}
                    \end{figure*}

                    \begin{figure*}
                	\begin{center}
                		\includegraphics[scale=0.96]{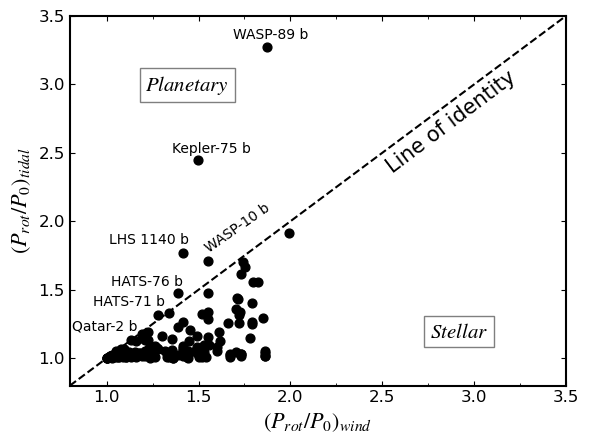}
                	\end{center}
                	\caption{Relationship between ($P_{\rm rot}/P_{0}$)$_{\rm tidal}$ using eq. \ref{eq.12c} (for $q$ described in Table \ref{tab:table_parameters}) and ($P_{\rm rot}/P_{0}$)$_{\rm wind}$ using eq. \ref{eq.12d} (for $q=3$, Skumanich exponent), considering only the \texttt{dataset\#01}. The dashed line denotes the identity between the two ratios. The figure highlights the exoplanets WASP-89b, Kepler-75b, LHS 1140b, WASP-10b, HATS-76b, HATS-71b, and Qatar-2b, which exhibit a greater contribution from the tidal effect than the magnetized stellar wind. }
                	\label{figEqs}
                    \end{figure*}

Three distinct behaviors can be discerned upon examining Fig. \ref{fig2}. Firstly, there is a discernible trend for $L_{*}/L_{\rm orb}$ to decrease as planetary mass increases. Secondly, it is notable that stars hosting multiple planets tend to exhibit higher $L_{*}/L_{\rm orb}$ values compared to systems with a single planet. Additionally, multi-planet systems predominantly retain their angular momentum within the star. These patterns are further supported by the observations in Fig. \ref{fig4}. Given that the tidal index $q$ is predominantly influenced by the $L_{*}/L_{\rm orb}$ ratio, the distinction between systems with one or more planets becomes more pronounced, particularly among stars with masses ranging from $0.6M$ to $1M_{\odot}$.
            
Significant disparities in the existence ranges are evident in comparing the $q$-index values in this study with those reported in \cite{deFreitas2022}. While the previous work constrained the range between values 1 to 3, our findings demonstrate that the span between the minimum and maximum $q$ values can exceed three orders of magnitude. This discrepancy is attributed to the nature of the data in \cite{deFreitas2022}, which focused solely on stars devoid of planets, relying on stellar density and rotation period, as detailed in eq. 17 of their publication. The interplay between these two parameters in the equation restricts the range of $q$ values, as an escalation in stellar density inherently corresponds to a reduction in the rotation period and vice versa.
            
In the current context, the $q$-index's function is linked to the dissipation of angular momentum induced by the presence of planets rather than being influenced by the magnetic wind effects as postulated in the model proposed by \cite{deFreitas2022}. Consequently, the diverse range of $q$ values observed can be attributed to various factors stemming from the interplay of multiple system parameters (cf. eqs. \ref{eq.11} and \ref{eq.12a}).
            
\subsubsection{\label{sec:level48}Age-rotation relationship for the tidal interaction index}

According to eqs. \ref{eq.1} and \ref{eq.11}, the law governing the evolution of rotation period due to tidal interaction effects can be formulated as follows:
           \begin{equation}
                \label{eq.12c}
               \left(\frac{P_{\rm rot}}{P_{\rm 0}}\right)_{\rm tidal}=\left[1+\left(q-1\right)\frac{t}{\tau}\right]^{\frac{1}{q-1}}, \forall\left(q\neq 1\right)
            \end{equation}
where $P_{\rm 0}$ is the stellar rotation period at time $t=0$, $P_{\rm rot}$ is the period at time $t=t_{\rm age}$, and $q$ given by eq. \ref{eq.11}. According to \cite{deFreitas2013}, the $\tau$ value is a function that depends smoothly on the stellar effective temperature. 
        
This equation reveals that if the stars hold most of the total angular momentum, the relation saturates at $P_{\rm rot}=P_{\rm 0}$, meaning that the rotational brake due to planets can be \textit{a priori} neglected and, therefore, the rotational evolution of stars depends mostly on the magnetic braking induced by the stellar wind. On the other hand, if the planets hold the largest share of the total angular momentum, the rotational evolution of the stars depends on both magnetic braking and tidal interaction due to the planets. Figure \ref{figAge} shows that most systems have a tidal exponent $1/(q-1)$ close to zero and do not significantly change the stellar rotation period. Generally speaking, this exponent measures the impact of star-planet interaction on stellar rotation in agreement with eq. \ref{eq.12c}.
            
Still based on Fig. \ref{figAge}, there are two regimes for the values of $q$: one that makes the exponent $1/(q-1)$ positive and another that makes it negative. As we saw in the previous section, $q$-values are primarily a function of the $L_{*}/L_{\rm orb}$ ratio. Therefore, the exponent becomes negative when this ratio is less than $1/12$. When this occurs, the values of $q$ are limited to the range between 0 and 1, as is the case of Kepler-75b and LHS 1140b, as seen in \texttt{dataset\#01}. In systems where the $L_{*}/L_{\rm orb}$ ratio is less than $1/12$, the evolution of the rotation period is asymptotic to a few giga-years, which causes the star to have an extremely high rotation period when compared to those stars that have evolved their period with the positive exponent $1/(q-1)$ for $q$ values greater than 1.

On the other hand, in systems with $L_{*}/L_{\rm orb}$ greater than $1/12$, which result in $q$-values between 1 and 2, the behavior of period is also asymptotic, but smoother than in the previous case. These asymptotic versus smooth profiles are linked to the age of the Galaxy in 10 Gyr, which limits the rotation period's increase. For the case of $q>2$, the $1/(q-1)$ value is also positive but gradually approaches 0 as $q$ increases. For values of $q$ around the canonical Skumanich value, the rotation period grows and reaches a saturation regime.

These different regimes of $1/(q-1)$ can be influenced by some factors originating from the properties of the system or even simplifications imposed on the model. In particular, for the WASP-89b, Kepler-75b, and LHS 1140b systems present in dataset1, the physical implication of this behavior is the $L_{*}/L_{\rm orb}$ ratio. For LHS 1140b, this ratio is 0.06, as seen in Table \ref{tab:table_parameters}. On the other hand, it is also possible that the adopted model (see Section 3) cooperates with this behavior since the multiplicative prefactor depends on the choice of Love number $k_{l,m,n}$.

In general terms, our results for the evolution of the period, considering $1/(q-1)$ values relatively far from 0, reveal that the convective regions of stars can be drastically affected by the tidal effect of planets. According to the Solar System parameters, \cite{Stefani2019} and \cite{scafetta2022} highlighted the efficient synchronization of the solar dynamo by the 11.07-year Venus-Earth-Jupiter tidal model, indicating the potential influence of Jupiter's tidal forces on the Sun's activity. 

Taking eq. \ref{eq.12b}, within the interval of $q$ between 2 and $+\infty$, there is a regime between 2 and 12 that defines the domain of the planets under the stars. Above $q=12$, the angular momentum becomes dominated by the host stars. Figure \ref{fig2} suggests that the three datasets present a reasonable number of systems with $1<q<12$, being more expressive in dataset1, as indicated by the systems below the horizontal dashed line. However, this figure can lead us to conclude that the planets in each system have the same ``strength'' regarding the evolution of rotation. As we will see in the last figure of the paper, this set of systems is reduced to a few, suggesting that it is not just $L_{*}/L_{\rm orb}$ that controls the evolution of the rotation, but there is a small portion of systems in our sample that are affected by the $P_{\rm orb}/P_{\rm rot}$ ratio.

\subsubsection{WASP-89 system}

Based on the \texttt{dataset\#01} exhibited
in the top panel from Figure \ref{figAge}, the system associated with the exoplanet WASP-89b stands out due to its $q$ value, which is approximately $1.09\pm0.02$ (i.e., $1/(q-1)=11.3$). \cite{deFreitas2013} has shown that a $q$ value of 1 signifies a state of thermodynamic equilibrium, indicating saturation of the stellar magnetic field. It's essential to note that this interpretation primarily applies to magnetic braking and may differ when applied to star-planet systems. However, this system disappears in samples \texttt{dataset\#02} and \texttt{dataset\#03} when the conditions highlighted in Section \ref{sec:level2} are applied.  
            
Equation \ref{eq.12b} reveals that $q$ can reach the value of 1 when $L_{*}/L_{\rm orb}$ becomes 1/12. For the WASP-89 system, the calculated $L_{*}/L_{\rm orb}$ value is approximately 1/10, resulting in $q$ being close to unity. However, avoiding analyzing this result in isolation is imperative since the system's eccentricity is 0.19. Considering the non-zero eccentricity, the $P_{\rm orb}/P_{\rm rot}$ ratio is taken into account. With $P_{\rm orb}/P_{\rm rot}=0.17$, the WASP-89 system is subsynchronized. This context suggests that WASP-89b significantly influences stellar rotation due to its classification as a hot Jupiter.

As quoted by \cite{hellier2015}, the enhanced magnetic activity of WASP-89 (K-type star) may be attributed to hosting a massive, short-period planet, a phenomenon often observed in hot Jupiter hosts. Eccentric, short-period orbiting planets are of particular interest because their rotation cannot be fully phase-locked to their orbit, resulting in significant variations in radiative forcing throughout their orbits \citep[cf.][]{poppen2014}. Therefore, they provide valuable insights into the dynamics of giant-planet atmospheres \citep[for further details, see][]{wong2014}.
            
The standard explanation for the presence of such eccentric orbits in short-period hot Jupiters involves a process known as high-eccentricity migration, followed by circularization \citep[][and references therein]{hellier2015}. Despite its short orbit, WASP-89b's enormous mass inhibits circularization in less than 1 Gyr, giving it a circularization period of approximately 2 Gyr. According to the age measured by \cite{bonomo2023}, it is likely that the exoplanet has already achieved circularization. As tidal damping of eccentricity is predicted to happen more quickly than inward orbital decay or damping of obliquity, we would anticipate that the current values of these attributes are a direct result of the high-eccentricity migration \citep{matsu2010}. Finally, we can state that the system with $q\sim 1$ is a vital proxy to explain the influence of hot Jupiters with high eccentricity on stellar rotation.

\subsubsection{\label{sec:level49}Age-rotation relationship for the Skumanich magnetic braking index}

Firstly, we highlight our theoretical findings on the age-$P_{\rm rot}$ relationship, based on a nonextensive formalism described here and by \cite{deFreitas2013}. Cool main-sequence stars experience magnetic braking and spin down as a function of stellar age ($t$) and mass ($M$), according to \cite{sku}. Hotter stars than the Sun ($M>1M_{\odot}$) have very minimal magnetic braking due to their thin convective envelopes and probably weak dynamos. As shown in Fig. \ref{figProtAge}, we classified our sample into three categories depending on stellar mass: lower mass main-sequence dwarfs ($M\le 0.6M_{\odot}$), solar-type stars ($0.6M_{\odot}<M\le 1M_{\odot}$), and higher mass main-sequence dwarfs ($M>1M_{\odot}$). The grid-modeling asteroseismic ages are displayed against the $P_{\rm rot}$ retrieved from our study in Fig. \ref{figProtAge}. These three subsets populate the period-age diagram in a way that is consistent with the theoretical predictions of \cite{deFreitas2013}, demonstrating that the rotation periods of stars in our sample can be interpreted in light of the Skumanich magnetic braking index, neglecting the effects of star-planet interaction. 

According to \cite{deFreitas2013}'s model, the law governing the evolution of rotation period due to magnetic wind effects is given by:
             \begin{equation}
                        \label{eq.12d}
                       \left(\frac{P_{\rm rot}}{P_{0}}\right)_{\rm wind}=\left[1+\left(q-1\right)\frac{t}{\tau}\right]^{\frac{1}{q-1}}, \forall\left(q\neq 1\right)
                    \end{equation}
where $\tau$ is a constant over time, $P_{\rm 0}$ is the stellar rotation period at time $t=0$, $P_{\rm rot}$ is the period at time $t=t_{\rm age}$, and $q=3$ given by eq. \ref{eqSku} \citep[cf. also][]{deFreitas2013}. 

In Fig. \ref{figProtAge}, the curves represent the period-age relationships from Skumanich law, corresponding to roughly from 0.15 to 1.5 $M_{\odot}$ in agreement with eq. \ref{eq.12d}. Stars within this mass range that obey the Skumanich relation should fall between these three curves. The solar-type stars sample displays just this behavior. However, the other samples do not. Many of the stars more massive than the Sun lie at systematically slower rotation periods and older ages than gyrochronology would have predicted when compared to the solar-type stars. Equation \ref{eq.12d} does not consider the effects of star interaction, but only the effects of the magnetized wind that breaks the stars. Based on Fig. \ref{figtSyn}, many of the most massive stars are in the supersynchronization regime.
            
Furthermore, Fig. \ref{figAge} reveals that for a large part of our sample, the exoplanet does not affect the rotation period of the stars. According to this result, the evolution of the rotation period given by eq. \ref{eq.12d} does not need to be corrected by the rotation's evolution due to the planets' presence, as shown in eq. \ref{eq.12c}. Thus, except for the systems highlighted in Fig. \ref{figAge}, the rotation of stars is predominantly governed by the influence of stellar winds. We now aim to elucidate the distinctions between the outcomes derived from eqs. \ref{eq.12c} and \ref{eq.12d}, to distinguish the regimes in which magnetic wind and tidal interaction govern stellar rotation.

Figure \ref{figEqs} illustrates that the influence of star-planet interaction on rotational evolution is generally overshadowed by the impact of magnetic braking from the stellar wind. This figure plays a crucial role in emphasizing the dominant terms in eq. \ref{eq.11}, especially when compared to Fig. \ref{fig2}, which primarily considers the effect of $L_{*}/L_{\rm orb}$ ratio. Figure \ref{figEqs} was derived based on the optimal $\tau$ value that accurately represents the distribution of exponents $1/(q-1)$ illustrated in Fig. \ref{figAge}. This optimal $\tau$ value is intended to indicate the strongest correlation that distinguishes the relationship between the ratios ($P_{\rm rot}/P_{0}$)$_{\rm tidal}$ and ($P_{\rm rot}/P_{0}$)$_{\rm wind}$, facilitating the isolation of the exoplanets specified in the legend of Figure \ref{figAge} when ($P_{\rm rot}/P_{0}$)$_{\rm tidal}>$($P_{\rm rot}/P_{0}$)$_{\rm wind}$. To achieve this, we generated a dataset using eqs. \ref{eq.12c} and \ref{eq.12d} for various $\tau$ values incremented by 0.1 Gyr. Drawing from insights \cite{deFreitas2013} research, we estimate that the most probable value falls within the range of 0.01 to 50 Gyr. The optimal $\tau$ value was determined using an approach grounded in the behavior of the Pearson correlation coefficient \citep{press1992}. Through this methodology, we identified the optimal $\tau$ value as 10 Gyr, coincidentally aligning with the age of the Galaxy, corresponding to the peak value of the aforementioned coefficient.

Analysis from Table \ref{tab:table_parameters} reveals that out of 175 exoplanets in \texttt{dataset\#01}, 49 exoplanets exhibit $L_{*}/L_{\rm orb}$ ratios below 1, with only 7 of these exoplanets showing contrasting behavior in Figure \ref{figEqs}. Among these 7 exoplanets, tidal interactions have a more pronounced effect on the stellar rotation than the magnetic wind. Conversely, the remaining 42 exoplanets experience a decline in rotational dynamics due to a combination of parameters outlined in eq. \ref{eq.11}. Notably, the 7 exoplanets with significant tidal impact have $q$ values ranging from 0 to 3, resulting in higher values of the exponent $1/(q-1)$. In contrast, the other 42 exoplanets have $q$ values exceeding 3, leading to exponents closer to zero. This distinction highlights an effective threshold for assessing the rotational impact of exoplanets. Consequently, our findings indicate that for planetary systems with $L_{*}/L_{\rm orb}$ ratios above 0.2, the rotational evolution of stars hosting planets with $a<1$ AU remains largely unaffected by the tidal interaction. This result implies that in cases where exoplanets retain less than 84\% of the total angular momentum, magnetic braking proves to be more effective than tidal interactions, irrespective of whether the planets' angular momentum surpasses the host star's.

\section{Summary}
The study presented in this paper focuses on investigating the intricate dynamics of star-planet systems, particularly emphasizing the impacts of tidal interaction and magnetic braking in the evolution of stellar rotation. The research delves into how the host star's rotation influences the system's structure and the ensuing interplays by utilizing a sample of well-characterized exoplanets. Notably, the study also explores the influence of orbital changes on stellar rotation and the subsequent angular momentum exchange with planets. By introducing a novel \textit{tidal interaction index}, the research offers a statistical approach to comprehensively consider various parameters affecting angular momentum transport within star-planet systems. 

The findings reveal significant anti-correlations between the tidal index and planetary mass, shedding light on the intricate relationships governing the evolution of stellar rotation in the presence of planets. In general, for our samples, the evolution of the stellar rotation period is not affected by the presence of planets, with exceptions noted in specific systems like WASP-89. In this sense, this work contributes to advancing theoretical models to understand the complex interactions between planetary orbits and stellar rotation, offering valuable insights into the dynamics of exoplanetary systems.

Our primary finding pertains to the physical implications of tidal interaction within the framework of star-planet dynamics on stellar rotation. Our study demonstrates that, in most scrutinized systems, the impact of this interaction, quantified by the parameter $q$, is reduced for values of $L_{*}/L_{\rm orb}>0.2$. From this value onwards, rotational braking is predominantly driven by the magnetized stellar wind, placing tidal interaction in the background, and, in most cases, it can even be neglected when the index $q$ exceeds the value 12. It is noteworthy to highlight that this conclusion is confined to systems characterized by shorter orbital periods than Earth one and that the inclusion of exoplanets with longer orbital periods, as expected with data from the PLATO mission, has the potential to broaden the scope of this investigation and unveil novel insights into tidal influences on stellar rotation.

\section*{Acknowledgments}
DBdeF acknowledges financial support from the Brazilian agency CNPq-PQ2 (Grant No. 305566/2021-0). Continuous grants from the Brazilian agency CNPq support the STELLAR TEAM of the Federal University of Cear\'a's research activities. This research has used the NASA Exoplanet Archive, operated by the California Institute of Technology under contract with the National Aeronautics and Space Administration under the Exoplanet Exploration Program.
    

\bibliographystyle{cas-model2-names}

\bibliography{references}

\providecommand{\noopsort}[1]{}\providecommand{\singleletter}[1]{#1}%
\begin{thebibliography}{65}
\expandafter\ifx\csname natexlab\endcsname\relax\def\natexlab#1{#1}\fi
\providecommand{\url}[1]{\texttt{#1}}
\providecommand{\href}[2]{#2}
\providecommand{\path}[1]{#1}
\providecommand{\DOIprefix}{doi:}
\providecommand{\ArXivprefix}{arXiv:}
\providecommand{\URLprefix}{URL: }
\providecommand{\Pubmedprefix}{pmid:}
\providecommand{\doi}[1]{\href{http://dx.doi.org/#1}{\path{#1}}}
\providecommand{\Pubmed}[1]{\href{pmid:#1}{\path{#1}}}
\providecommand{\bibinfo}[2]{#2}
\ifx\xfnm\relax \def\xfnm[#1]{\unskip,\space#1}\fi
%Type = Article
\bibitem[{{Ahuir} et~al.(2021a){Ahuir}, {Strugarek}, {Brun} and {Mathis}}]{Ahuir}
\bibinfo{author}{{Ahuir}, J.}, \bibinfo{author}{{Strugarek}, A.}, \bibinfo{author}{{Brun}, A.S.}, \bibinfo{author}{{Mathis}, S.}, \bibinfo{year}{2021}a.
\newblock \bibinfo{title}{{Magnetic and tidal migration of close-in planets. Influence of secular evolution on their population}}.
\newblock \bibinfo{journal}{Astronomy \& Astrophysics} \bibinfo{volume}{650}, \bibinfo{pages}{A126}.
\newblock \DOIprefix\doi{10.1051/0004-6361/202040173}, \href{http://arxiv.org/abs/2104.01004}{\tt arXiv:2104.01004}.
%Type = Article
\bibitem[{{Ahuir} et~al.(2021b){Ahuir}, {Strugarek}, {Brun} and {Mathis}}]{ahuir2021}
\bibinfo{author}{{Ahuir}, J.}, \bibinfo{author}{{Strugarek}, A.}, \bibinfo{author}{{Brun}, A.S.}, \bibinfo{author}{{Mathis}, S.}, \bibinfo{year}{2021}b.
\newblock \bibinfo{title}{Magnetic and tidal migration of close-in planets. influence of secular evolution on their population}.
\newblock \bibinfo{journal}{Astronomy \& Astrophysics} \bibinfo{volume}{650}, \bibinfo{pages}{A126}.
\newblock \DOIprefix\doi{10.1051/0004-6361/202040173}, \href{http://arxiv.org/abs/2104.01004}{\tt arXiv:2104.01004}.
%Type = Article
\bibitem[{{Alsubai} et~al.(2017){Alsubai}, {Mislis}, {Tsvetanov}, {Latham}, {Bieryla}, {Buchhave}, {Esquerdo}, {Bramich}, {Pyrzas}, {Vilchez}, {Mancini}, {Southworth}, {Evans}, {Henning} and {Ciceri}}]{alsubai2017}
\bibinfo{author}{{Alsubai}, K.}, \bibinfo{author}{{Mislis}, D.}, \bibinfo{author}{{Tsvetanov}, Z.I.}, \bibinfo{author}{{Latham}, D.W.}, \bibinfo{author}{{Bieryla}, A.}, \bibinfo{author}{{Buchhave}, L.A.}, \bibinfo{author}{{Esquerdo}, G.A.}, \bibinfo{author}{{Bramich}, D.M.}, \bibinfo{author}{{Pyrzas}, S.}, \bibinfo{author}{{Vilchez}, N.P.E.}, \bibinfo{author}{{Mancini}, L.}, \bibinfo{author}{{Southworth}, J.}, \bibinfo{author}{{Evans}, D.F.}, \bibinfo{author}{{Henning}, T.}, \bibinfo{author}{{Ciceri}, S.}, \bibinfo{year}{2017}.
\newblock \bibinfo{title}{{Qatar Exoplanet Survey : Qatar-3b, Qatar-4b, and Qatar-5b}}.
\newblock \bibinfo{journal}{The Astronomical Journal} \bibinfo{volume}{153}, \bibinfo{pages}{200}.
\newblock \DOIprefix\doi{10.3847/1538-3881/aa6340}, \href{http://arxiv.org/abs/1606.06882}{\tt arXiv:1606.06882}.
%Type = Article
\bibitem[{{Alvarado-Montes}(2022)}]{alvarado2022}
\bibinfo{author}{{Alvarado-Montes}, J.A.}, \bibinfo{year}{2022}.
\newblock \bibinfo{title}{{Tidally induced migration of TESS gas giants orbiting M dwarfs}}.
\newblock \bibinfo{journal}{Monthly Notices of the Royal Astronomical Society} \bibinfo{volume}{517}, \bibinfo{pages}{2831--2841}.
\newblock \DOIprefix\doi{10.1093/mnras/stac2741}, \href{http://arxiv.org/abs/2209.11375}{\tt arXiv:2209.11375}.
%Type = Article
\bibitem[{{Alvarado-Montes} et~al.(2021){Alvarado-Montes}, {Sucerquia}, {Garc{\'\i}a-Carmona}, {Zuluaga}, {Spitler} and {Schwab}}]{montes}
\bibinfo{author}{{Alvarado-Montes}, J.A.}, \bibinfo{author}{{Sucerquia}, M.}, \bibinfo{author}{{Garc{\'\i}a-Carmona}, C.}, \bibinfo{author}{{Zuluaga}, J.I.}, \bibinfo{author}{{Spitler}, L.}, \bibinfo{author}{{Schwab}, C.}, \bibinfo{year}{2021}.
\newblock \bibinfo{title}{{The impact of tidal friction evolution on the orbital decay of ultra-short-period planets}}.
\newblock \bibinfo{journal}{Monthly Notices of the Royal Astronomical Society} \bibinfo{volume}{506}, \bibinfo{pages}{2247--2259}.
\newblock \DOIprefix\doi{10.1093/mnras/stab1081}, \href{http://arxiv.org/abs/2104.05967}{\tt arXiv:2104.05967}.
%Type = Article
\bibitem[{{Barker} and {Ogilvie}(2009)}]{barker2009}
\bibinfo{author}{{Barker}, A.J.}, \bibinfo{author}{{Ogilvie}, G.I.}, \bibinfo{year}{2009}.
\newblock \bibinfo{title}{{On the tidal evolution of Hot Jupiters on inclined orbits}}.
\newblock \bibinfo{journal}{Monthly Notices of the Royal Astronomical Society} \bibinfo{volume}{395}, \bibinfo{pages}{2268--2287}.
\newblock \DOIprefix\doi{10.1111/j.1365-2966.2009.14694.x}, \href{http://arxiv.org/abs/0902.4563}{\tt arXiv:0902.4563}.
%Type = Article
\bibitem[{{Barros} et~al.(2022){Barros}, {Demangeon}, {Alibert}, {Leleu}, {Adibekyan}, {Lovis}, {Bossini}, {Sousa}, {Hara}, {Bouchy}, {Lavie}, {Rodrigues}, {Gomes da Silva}, {Lillo-Box}, {Pepe}, {Tabernero}, {Zapatero Osorio}, {Sozzetti}, {Su{\'a}rez Mascare{\~n}o}, {Micela}, {Allende Prieto}, {Cristiani}, {Damasso}, {Di Marcantonio}, {Ehrenreich}, {Faria}, {Figueira}, {Gonz{\'a}lez Hern{\'a}ndez}, {Jenkins}, {Lo Curto}, {Martins}, {Micela}, {Nunes}, {Pall{\'e}}, {Santos}, {Rebolo}, {Seager}, {Twicken}, {Udry}, {Vanderspek} and {Winn}}]{Barros2022}
\bibinfo{author}{{Barros}, S.C.C.}, \bibinfo{author}{{Demangeon}, O.D.S.}, \bibinfo{author}{{Alibert}, Y.}, \bibinfo{author}{{Leleu}, A.}, \bibinfo{author}{{Adibekyan}, V.}, \bibinfo{author}{{Lovis}, C.}, \bibinfo{author}{{Bossini}, D.}, \bibinfo{author}{{Sousa}, S.G.}, \bibinfo{author}{{Hara}, N.}, \bibinfo{author}{{Bouchy}, F.}, \bibinfo{author}{{Lavie}, B.}, \bibinfo{author}{{Rodrigues}, J.}, \bibinfo{author}{{Gomes da Silva}, J.}, \bibinfo{author}{{Lillo-Box}, J.}, \bibinfo{author}{{Pepe}, F.A.}, \bibinfo{author}{{Tabernero}, H.M.}, \bibinfo{author}{{Zapatero Osorio}, M.R.}, \bibinfo{author}{{Sozzetti}, A.}, \bibinfo{author}{{Su{\'a}rez Mascare{\~n}o}, A.}, \bibinfo{author}{{Micela}, G.}, \bibinfo{author}{{Allende Prieto}, C.}, \bibinfo{author}{{Cristiani}, S.}, \bibinfo{author}{{Damasso}, M.}, \bibinfo{author}{{Di Marcantonio}, P.}, \bibinfo{author}{{Ehrenreich}, D.}, \bibinfo{author}{{Faria}, J.}, \bibinfo{author}{{Figueira}, P.}, \bibinfo{author}{{Gonz{\'a}lez Hern{\'a}ndez}, J.I.},
  \bibinfo{author}{{Jenkins}, J.}, \bibinfo{author}{{Lo Curto}, G.}, \bibinfo{author}{{Martins}, C.J.A.P.}, \bibinfo{author}{{Micela}, G.}, \bibinfo{author}{{Nunes}, N.J.}, \bibinfo{author}{{Pall{\'e}}, E.}, \bibinfo{author}{{Santos}, N.C.}, \bibinfo{author}{{Rebolo}, R.}, \bibinfo{author}{{Seager}, S.}, \bibinfo{author}{{Twicken}, J.D.}, \bibinfo{author}{{Udry}, S.}, \bibinfo{author}{{Vanderspek}, R.}, \bibinfo{author}{{Winn}, J.N.}, \bibinfo{year}{2022}.
\newblock \bibinfo{title}{{HD 23472: a multi-planetary system with three super-Earths and two potential super-Mercuries}}.
\newblock \bibinfo{journal}{Astronomy \& Astrophysics} \bibinfo{volume}{665}, \bibinfo{pages}{A154}.
\newblock \DOIprefix\doi{10.1051/0004-6361/202244293}, \href{http://arxiv.org/abs/2209.13345}{\tt arXiv:2209.13345}.
%Type = Article
\bibitem[{{Benbakoura} et~al.(2019){Benbakoura}, {R{\'e}ville}, {Brun}, {Le Poncin-Lafitte} and {Mathis}}]{benba}
\bibinfo{author}{{Benbakoura}, M.}, \bibinfo{author}{{R{\'e}ville}, V.}, \bibinfo{author}{{Brun}, A.S.}, \bibinfo{author}{{Le Poncin-Lafitte}, C.}, \bibinfo{author}{{Mathis}, S.}, \bibinfo{year}{2019}.
\newblock \bibinfo{title}{{Evolution of star-planet systems under magnetic braking and tidal interaction}}.
\newblock \bibinfo{journal}{Astronomy \& Astrophysics} \bibinfo{volume}{621}, \bibinfo{pages}{A124}.
\newblock \DOIprefix\doi{10.1051/0004-6361/201833314}, \href{http://arxiv.org/abs/1811.06354}{\tt arXiv:1811.06354}.
%Type = Article
\bibitem[{{Bonomo} et~al.(2017){Bonomo}, {Desidera}, {Benatti}, {Borsa}, {Crespi}, {Damasso}, {Lanza}, {Sozzetti}, {Lodato}, {Marzari}, {Boccato}, {Claudi}, {Cosentino}, {Covino}, {Gratton}, {Maggio}, {Micela}, {Molinari}, {Pagano}, {Piotto}, {Poretti}, {Smareglia}, {Affer}, {Biazzo}, {Bignamini}, {Esposito}, {Giacobbe}, {H{\'e}brard}, {Malavolta}, {Maldonado}, {Mancini}, {Martinez Fiorenzano}, {Masiero}, {Nascimbeni}, {Pedani}, {Rainer} and {Scandariato}}]{Bonono2017}
\bibinfo{author}{{Bonomo}, A.S.}, \bibinfo{author}{{Desidera}, S.}, \bibinfo{author}{{Benatti}, S.}, \bibinfo{author}{{Borsa}, F.}, \bibinfo{author}{{Crespi}, S.}, \bibinfo{author}{{Damasso}, M.}, \bibinfo{author}{{Lanza}, A.F.}, \bibinfo{author}{{Sozzetti}, A.}, \bibinfo{author}{{Lodato}, G.}, \bibinfo{author}{{Marzari}, F.}, \bibinfo{author}{{Boccato}, C.}, \bibinfo{author}{{Claudi}, R.U.}, \bibinfo{author}{{Cosentino}, R.}, \bibinfo{author}{{Covino}, E.}, \bibinfo{author}{{Gratton}, R.}, \bibinfo{author}{{Maggio}, A.}, \bibinfo{author}{{Micela}, G.}, \bibinfo{author}{{Molinari}, E.}, \bibinfo{author}{{Pagano}, I.}, \bibinfo{author}{{Piotto}, G.}, \bibinfo{author}{{Poretti}, E.}, \bibinfo{author}{{Smareglia}, R.}, \bibinfo{author}{{Affer}, L.}, \bibinfo{author}{{Biazzo}, K.}, \bibinfo{author}{{Bignamini}, A.}, \bibinfo{author}{{Esposito}, M.}, \bibinfo{author}{{Giacobbe}, P.}, \bibinfo{author}{{H{\'e}brard}, G.}, \bibinfo{author}{{Malavolta}, L.}, \bibinfo{author}{{Maldonado}, J.},
  \bibinfo{author}{{Mancini}, L.}, \bibinfo{author}{{Martinez Fiorenzano}, A.}, \bibinfo{author}{{Masiero}, S.}, \bibinfo{author}{{Nascimbeni}, V.}, \bibinfo{author}{{Pedani}, M.}, \bibinfo{author}{{Rainer}, M.}, \bibinfo{author}{{Scandariato}, G.}, \bibinfo{year}{2017}.
\newblock \bibinfo{title}{{The GAPS Programme with HARPS-N at TNG . XIV. Investigating giant planet migration history via improved eccentricity and mass determination for 231 transiting planets}}.
\newblock \bibinfo{journal}{Astronomy \& Astrophysics} \bibinfo{volume}{602}, \bibinfo{pages}{A107}.
\newblock \DOIprefix\doi{10.1051/0004-6361/201629882}, \href{http://arxiv.org/abs/1704.00373}{\tt arXiv:1704.00373}.
%Type = Article
\bibitem[{{Bonomo} et~al.(2023){Bonomo}, {Dumusque}, {Massa}, {Mortier}, {Bongiolatti}, {Malavolta}, {Sozzetti}, {Buchhave}, {Damasso}, {Haywood}, {Morbidelli}, {Latham}, {Molinari}, {Pepe}, {Poretti}, {Udry}, {Affer}, {Boschin}, {Charbonneau}, {Cosentino}, {Cretignier}, {Ghedina}, {Lega}, {L{\'o}pez-Morales}, {Margini}, {Mart{\'\i}nez Fiorenzano}, {Mayor}, {Micela}, {Pedani}, {Pinamonti}, {Rice}, {Sasselov}, {Tronsgaard} and {Vanderburg}}]{bonomo2023}
\bibinfo{author}{{Bonomo}, A.S.}, \bibinfo{author}{{Dumusque}, X.}, \bibinfo{author}{{Massa}, A.}, \bibinfo{author}{{Mortier}, A.}, \bibinfo{author}{{Bongiolatti}, R.}, \bibinfo{author}{{Malavolta}, L.}, \bibinfo{author}{{Sozzetti}, A.}, \bibinfo{author}{{Buchhave}, L.A.}, \bibinfo{author}{{Damasso}, M.}, \bibinfo{author}{{Haywood}, R.D.}, \bibinfo{author}{{Morbidelli}, A.}, \bibinfo{author}{{Latham}, D.W.}, \bibinfo{author}{{Molinari}, E.}, \bibinfo{author}{{Pepe}, F.}, \bibinfo{author}{{Poretti}, E.}, \bibinfo{author}{{Udry}, S.}, \bibinfo{author}{{Affer}, L.}, \bibinfo{author}{{Boschin}, W.}, \bibinfo{author}{{Charbonneau}, D.}, \bibinfo{author}{{Cosentino}, R.}, \bibinfo{author}{{Cretignier}, M.}, \bibinfo{author}{{Ghedina}, A.}, \bibinfo{author}{{Lega}, E.}, \bibinfo{author}{{L{\'o}pez-Morales}, M.}, \bibinfo{author}{{Margini}, M.}, \bibinfo{author}{{Mart{\'\i}nez Fiorenzano}, A.F.}, \bibinfo{author}{{Mayor}, M.}, \bibinfo{author}{{Micela}, G.}, \bibinfo{author}{{Pedani}, M.},
  \bibinfo{author}{{Pinamonti}, M.}, \bibinfo{author}{{Rice}, K.}, \bibinfo{author}{{Sasselov}, D.}, \bibinfo{author}{{Tronsgaard}, R.}, \bibinfo{author}{{Vanderburg}, A.}, \bibinfo{year}{2023}.
\newblock \bibinfo{title}{{Cold Jupiters and improved masses in 38 Kepler and K2 small planet systems from 3661 HARPS-N radial velocities. No excess of cold Jupiters in small planet systems}}.
\newblock \bibinfo{journal}{Astronomy \& Astrophysics} \bibinfo{volume}{677}, \bibinfo{pages}{A33}.
\newblock \DOIprefix\doi{10.1051/0004-6361/202346211}, \href{http://arxiv.org/abs/2304.05773}{\tt arXiv:2304.05773}.
%Type = Article
\bibitem[{{Bonomo} et~al.(2019){Bonomo}, {Zeng}, {Damasso}, {Leinhardt}, {Justesen}, {Lopez}, {Lund}, {Malavolta}, {Silva Aguirre}, {Buchhave}, {Corsaro}, {Denman}, {Lopez-Morales}, {Mills}, {Mortier}, {Rice}, {Sozzetti}, {Vanderburg}, {Affer}, {Arentoft}, {Benbakoura}, {Bouchy}, {Christensen-Dalsgaard}, {Collier Cameron}, {Cosentino}, {Dressing}, {Dumusque}, {Figueira}, {Fiorenzano}, {Garc{\'\i}a}, {Handberg}, {Harutyunyan}, {Johnson}, {Kjeldsen}, {Latham}, {Lovis}, {Lundkvist}, {Mathur}, {Mayor}, {Micela}, {Molinari}, {Motalebi}, {Nascimbeni}, {Nava}, {Pepe}, {Phillips}, {Piotto}, {Poretti}, {Sasselov}, {S{\'e}gransan}, {Udry} and {Watson}}]{Bonomo2019}
\bibinfo{author}{{Bonomo}, A.S.}, \bibinfo{author}{{Zeng}, L.}, \bibinfo{author}{{Damasso}, M.}, \bibinfo{author}{{Leinhardt}, Z.M.}, \bibinfo{author}{{Justesen}, A.B.}, \bibinfo{author}{{Lopez}, E.}, \bibinfo{author}{{Lund}, M.N.}, \bibinfo{author}{{Malavolta}, L.}, \bibinfo{author}{{Silva Aguirre}, V.}, \bibinfo{author}{{Buchhave}, L.A.}, \bibinfo{author}{{Corsaro}, E.}, \bibinfo{author}{{Denman}, T.}, \bibinfo{author}{{Lopez-Morales}, M.}, \bibinfo{author}{{Mills}, S.M.}, \bibinfo{author}{{Mortier}, A.}, \bibinfo{author}{{Rice}, K.}, \bibinfo{author}{{Sozzetti}, A.}, \bibinfo{author}{{Vanderburg}, A.}, \bibinfo{author}{{Affer}, L.}, \bibinfo{author}{{Arentoft}, T.}, \bibinfo{author}{{Benbakoura}, M.}, \bibinfo{author}{{Bouchy}, F.}, \bibinfo{author}{{Christensen-Dalsgaard}, J.}, \bibinfo{author}{{Collier Cameron}, A.}, \bibinfo{author}{{Cosentino}, R.}, \bibinfo{author}{{Dressing}, C.D.}, \bibinfo{author}{{Dumusque}, X.}, \bibinfo{author}{{Figueira}, P.}, \bibinfo{author}{{Fiorenzano}, A.F.M.},
  \bibinfo{author}{{Garc{\'\i}a}, R.A.}, \bibinfo{author}{{Handberg}, R.}, \bibinfo{author}{{Harutyunyan}, A.}, \bibinfo{author}{{Johnson}, J.A.}, \bibinfo{author}{{Kjeldsen}, H.}, \bibinfo{author}{{Latham}, D.W.}, \bibinfo{author}{{Lovis}, C.}, \bibinfo{author}{{Lundkvist}, M.S.}, \bibinfo{author}{{Mathur}, S.}, \bibinfo{author}{{Mayor}, M.}, \bibinfo{author}{{Micela}, G.}, \bibinfo{author}{{Molinari}, E.}, \bibinfo{author}{{Motalebi}, F.}, \bibinfo{author}{{Nascimbeni}, V.}, \bibinfo{author}{{Nava}, C.}, \bibinfo{author}{{Pepe}, F.}, \bibinfo{author}{{Phillips}, D.F.}, \bibinfo{author}{{Piotto}, G.}, \bibinfo{author}{{Poretti}, E.}, \bibinfo{author}{{Sasselov}, D.}, \bibinfo{author}{{S{\'e}gransan}, D.}, \bibinfo{author}{{Udry}, S.}, \bibinfo{author}{{Watson}, C.}, \bibinfo{year}{2019}.
\newblock \bibinfo{title}{{A giant impact as the likely origin of different twins in the Kepler-107 exoplanet system}}.
\newblock \bibinfo{journal}{Nature Astronomy} \bibinfo{volume}{3}, \bibinfo{pages}{416--423}.
\newblock \DOIprefix\doi{10.1038/s41550-018-0684-9}, \href{http://arxiv.org/abs/1902.01316}{\tt arXiv:1902.01316}.
%Type = Article
\bibitem[{{Busetti} et~al.(2018){Busetti}, {Beust} and {Harley}}]{busetti2018}
\bibinfo{author}{{Busetti}, F.}, \bibinfo{author}{{Beust}, H.}, \bibinfo{author}{{Harley}, C.}, \bibinfo{year}{2018}.
\newblock \bibinfo{title}{{Stability of planets in triple star systems}}.
\newblock \bibinfo{journal}{Astronomy \& Astrophysics} \bibinfo{volume}{619}, \bibinfo{pages}{A91}.
\newblock \DOIprefix\doi{10.1051/0004-6361/201833097}, \href{http://arxiv.org/abs/1811.08221}{\tt arXiv:1811.08221}.
%Type = Article
\bibitem[{{Canto Martins} et~al.(2023){Canto Martins}, {Messias}, {Arruda Gon{\c{c}}alves}, {Le{\~a}o}, {Gomes}, {Barraza}, {Fontinele} and {De Medeiros}}]{CantoMartins2023}
\bibinfo{author}{{Canto Martins}, B.L.}, \bibinfo{author}{{Messias}, Y.S.}, \bibinfo{author}{{Arruda Gon{\c{c}}alves}, M.I.}, \bibinfo{author}{{Le{\~a}o}, I.C.}, \bibinfo{author}{{Gomes}, R.L.}, \bibinfo{author}{{Barraza}, L.F.}, \bibinfo{author}{{Fontinele}, D.O.}, \bibinfo{author}{{De Medeiros}, J.R.}, \bibinfo{year}{2023}.
\newblock \bibinfo{title}{{On the behaviour of spin-orbit connection of exoplanets}}.
\newblock \bibinfo{journal}{Nature Astronomy} \bibinfo{volume}{7}, \bibinfo{pages}{900--904}.
\newblock \DOIprefix\doi{10.1038/s41550-023-01976-0}, \href{http://arxiv.org/abs/2305.14455}{\tt arXiv:2305.14455}.
%Type = Article
\bibitem[{{Chebly} et~al.(2022){Chebly}, {Alvarado-G{\'o}mez} and {Poppenhaeger}}]{chebly2022}
\bibinfo{author}{{Chebly}, J.J.}, \bibinfo{author}{{Alvarado-G{\'o}mez}, J.D.}, \bibinfo{author}{{Poppenhaeger}, K.}, \bibinfo{year}{2022}.
\newblock \bibinfo{title}{{Destination exoplanet: Habitability conditions influenced by stellar winds properties}}.
\newblock \bibinfo{journal}{Astronomische Nachrichten} \bibinfo{volume}{343}, \bibinfo{pages}{e10093}.
\newblock \DOIprefix\doi{10.1002/asna.20210093}, \href{http://arxiv.org/abs/2111.09707}{\tt arXiv:2111.09707}.
%Type = Inproceedings
\bibitem[{{Damiani} and {Lanza}(2015a)}]{damiani2015}
\bibinfo{author}{{Damiani}, C.}, \bibinfo{author}{{Lanza}, A.F.}, \bibinfo{year}{2015}a.
\newblock \bibinfo{title}{{Evolution of angular-momentum-losing exoplanetary systems}}, in: \bibinfo{booktitle}{European Planetary Science Congress}, pp. \bibinfo{pages}{EPSC2015--730}.
%Type = Article
\bibitem[{{Damiani} and {Lanza}(2015b)}]{lanza2015}
\bibinfo{author}{{Damiani}, C.}, \bibinfo{author}{{Lanza}, A.F.}, \bibinfo{year}{2015}b.
\newblock \bibinfo{title}{{Evolution of angular-momentum-losing exoplanetary systems. Revisiting Darwin stability}}.
\newblock \bibinfo{journal}{Astronomy \& Astrophysics} \bibinfo{volume}{574}, \bibinfo{pages}{A39}.
\newblock \DOIprefix\doi{10.1051/0004-6361/201424318}, \href{http://arxiv.org/abs/1411.3802}{\tt arXiv:1411.3802}.
%Type = Article
\bibitem[{{Darwin}(1879)}]{darwin1879}
\bibinfo{author}{{Darwin}, G.H.}, \bibinfo{year}{1879}.
\newblock \bibinfo{title}{{VIII}. the determination of the secular effects of tidal friction by a graphical method}.
\newblock \bibinfo{journal}{Proceedings of the Royal Society of London} \bibinfo{volume}{29}, \bibinfo{pages}{168--181}.
\newblock \URLprefix \url{https://doi.org/10.1098/rspl.1879.0028}, \DOIprefix\doi{10.1098/rspl.1879.0028}.
%Type = Article
\bibitem[{Darwin(1880)}]{darwin}
\bibinfo{author}{Darwin, G.H.}, \bibinfo{year}{1880}.
\newblock \bibinfo{title}{{On the Secular Changes in the Elements of the Orbit of a Satellite Revolving about a Tidally Distorted Planet}}.
\newblock \bibinfo{journal}{Philosophical Transactions of the Royal Society of London Series I} \bibinfo{volume}{171}, \bibinfo{pages}{713--891}.
%Type = Article
\bibitem[{{De Freitas}(2021)}]{deFreitas2021}
\bibinfo{author}{{De Freitas}, D.B.}, \bibinfo{year}{2021}.
\newblock \bibinfo{title}{{Stellar age dependence of the nonextensive magnetic braking index: A test for the open cluster {\ensuremath{\alpha}}Per}}.
\newblock \bibinfo{journal}{EPL (Europhysics Letters)} \bibinfo{volume}{135}, \bibinfo{pages}{19001}.
\newblock \DOIprefix\doi{10.1209/0295-5075/ac0dfb}, \href{http://arxiv.org/abs/2104.10248}{\tt arXiv:2104.10248}.
%Type = Article
\bibitem[{{De Freitas} et~al.(2022){De Freitas}, {Cavalcante} and {Santiago}}]{deFreitas2022}
\bibinfo{author}{{De Freitas}, D.B.}, \bibinfo{author}{{Cavalcante}, F.J.}, \bibinfo{author}{{Santiago}, T.M.}, \bibinfo{year}{2022}.
\newblock \bibinfo{title}{{Measuring deviation from Skumanich braking index in active stars observed by Kepler mission}}.
\newblock \bibinfo{journal}{EPL (Europhysics Letters)} \bibinfo{volume}{140}, \bibinfo{pages}{29001}.
\newblock \DOIprefix\doi{10.1209/0295-5075/ac97bc}, \href{http://arxiv.org/abs/2208.11829}{\tt arXiv:2208.11829}.
%Type = Article
\bibitem[{{de Freitas} et~al.(2015){de Freitas}, {Cavalcante}, {Soares} and {Silva}}]{defreitasetal2015}
\bibinfo{author}{{de Freitas}, D.B.}, \bibinfo{author}{{Cavalcante}, F.J.}, \bibinfo{author}{{Soares}, B.B.}, \bibinfo{author}{{Silva}, J.R.P.}, \bibinfo{year}{2015}.
\newblock \bibinfo{title}{{A nonextensive view of the stellar braking indices}}.
\newblock \bibinfo{journal}{EPL (Europhysics Letters)} \bibinfo{volume}{111}, \bibinfo{pages}{39003}.
\newblock \DOIprefix\doi{10.1209/0295-5075/111/39003}, \href{http://arxiv.org/abs/1508.02237}{\tt arXiv:1508.02237}.
%Type = Article
\bibitem[{{De Freitas} and {De Medeiros}(2013)}]{deFreitas2013}
\bibinfo{author}{{De Freitas}, D.B.}, \bibinfo{author}{{De Medeiros}, J.R.}, \bibinfo{year}{2013}.
\newblock \bibinfo{title}{{A non-extensive approach to the stellar rotational evolution - I. F- and G-type stars}}.
\newblock \bibinfo{journal}{Monthly Notices of the Royal Astronomical Society} \bibinfo{volume}{433}, \bibinfo{pages}{1789--1795}.
\newblock \DOIprefix\doi{10.1093/mnras/stt734}, \href{http://arxiv.org/abs/1304.7644}{\tt arXiv:1304.7644}.
%Type = Article
\bibitem[{{de Freitas} et~al.(2014){de Freitas}, {Nepomuceno}, {Soares} and {Silva}}]{defreitas2014}
\bibinfo{author}{{de Freitas}, D.B.}, \bibinfo{author}{{Nepomuceno}, M.M.F.}, \bibinfo{author}{{Soares}, B.B.}, \bibinfo{author}{{Silva}, J.R.P.}, \bibinfo{year}{2014}.
\newblock \bibinfo{title}{{Strong evidences for a nonextensive behavior of the rotation period in open clusters}}.
\newblock \bibinfo{journal}{EPL (Europhysics Letters)} \bibinfo{volume}{108}, \bibinfo{pages}{39001}.
\newblock \DOIprefix\doi{10.1209/0295-5075/108/39001}, \href{http://arxiv.org/abs/1408.0657}{\tt arXiv:1408.0657}.
%Type = Article
\bibitem[{{Do Nascimento} et~al.(2014){Do Nascimento}, {Garc{\'\i}a}, {Mathur}, {Anthony}, {Barnes}, {Meibom}, {da Costa}, {Castro}, {Salabert} and {Ceillier}}]{nascimento2014}
\bibinfo{author}{{Do Nascimento}, J.~D., J.}, \bibinfo{author}{{Garc{\'\i}a}, R.A.}, \bibinfo{author}{{Mathur}, S.}, \bibinfo{author}{{Anthony}, F.}, \bibinfo{author}{{Barnes}, S.A.}, \bibinfo{author}{{Meibom}, S.}, \bibinfo{author}{{da Costa}, J.S.}, \bibinfo{author}{{Castro}, M.}, \bibinfo{author}{{Salabert}, D.}, \bibinfo{author}{{Ceillier}, T.}, \bibinfo{year}{2014}.
\newblock \bibinfo{title}{{Rotation Periods and Ages of Solar Analogs and Solar Twins Revealed by the Kepler Mission}}.
\newblock \bibinfo{journal}{Astrophysics Journal Letters} \bibinfo{volume}{790}, \bibinfo{pages}{L23}.
\newblock \DOIprefix\doi{10.1088/2041-8205/790/2/L23}, \href{http://arxiv.org/abs/1407.2289}{\tt arXiv:1407.2289}.
%Type = Article
\bibitem[{{Dobbs-Dixon} et~al.(2004){Dobbs-Dixon}, {Lin} and {Mardling}}]{dobbs2004}
\bibinfo{author}{{Dobbs-Dixon}, I.}, \bibinfo{author}{{Lin}, D.N.C.}, \bibinfo{author}{{Mardling}, R.A.}, \bibinfo{year}{2004}.
\newblock \bibinfo{title}{{Spin-Orbit Evolution of Short-Period Planets}}.
\newblock \bibinfo{journal}{The Astrophysical Journal} \bibinfo{volume}{610}, \bibinfo{pages}{464--476}.
\newblock \DOIprefix\doi{10.1086/421510}, \href{http://arxiv.org/abs/astro-ph/0408191}{\tt arXiv:astro-ph/0408191}.
%Type = Article
\bibitem[{{Forget} and {Leconte}(2014)}]{forget}
\bibinfo{author}{{Forget}, F.}, \bibinfo{author}{{Leconte}, J.}, \bibinfo{year}{2014}.
\newblock \bibinfo{title}{{Possible climates on terrestrial exoplanets}}.
\newblock \bibinfo{journal}{Philosophical Transactions of the Royal Society of London Series A} \bibinfo{volume}{372}, \bibinfo{pages}{20130084--20130084}.
\newblock \DOIprefix\doi{10.1098/rsta.2013.0084}, \href{http://arxiv.org/abs/1311.3101}{\tt arXiv:1311.3101}.
%Type = Article
\bibitem[{{Gallet} et~al.(2018){Gallet}, {Bolmont}, {Bouvier}, {Mathis} and {Charbonnel}}]{gallet2018}
\bibinfo{author}{{Gallet}, F.}, \bibinfo{author}{{Bolmont}, E.}, \bibinfo{author}{{Bouvier}, J.}, \bibinfo{author}{{Mathis}, S.}, \bibinfo{author}{{Charbonnel}, C.}, \bibinfo{year}{2018}.
\newblock \bibinfo{title}{{Planetary tidal interactions and the rotational evolution of low-mass stars. The Pleiades' anomaly}}.
\newblock \bibinfo{journal}{Astronomy \& Astrophysics} \bibinfo{volume}{619}, \bibinfo{pages}{A80}.
\newblock \DOIprefix\doi{10.1051/0004-6361/201833576}, \href{http://arxiv.org/abs/1808.08728}{\tt arXiv:1808.08728}.
%Type = Article
\bibitem[{{Gallet} et~al.(2017){Gallet}, {Bolmont}, {Mathis}, {Charbonnel} and {Amard}}]{gallet2017}
\bibinfo{author}{{Gallet}, F.}, \bibinfo{author}{{Bolmont}, E.}, \bibinfo{author}{{Mathis}, S.}, \bibinfo{author}{{Charbonnel}, C.}, \bibinfo{author}{{Amard}, L.}, \bibinfo{year}{2017}.
\newblock \bibinfo{title}{{Tidal dissipation in rotating low-mass stars and implications for the orbital evolution of close-in planets. I. From the PMS to the RGB at solar metallicity}}.
\newblock \bibinfo{journal}{Astronomy \& Astrophysics} \bibinfo{volume}{604}, \bibinfo{pages}{A112}.
\newblock \DOIprefix\doi{10.1051/0004-6361/201730661}, \href{http://arxiv.org/abs/1705.10164}{\tt arXiv:1705.10164}.
%Type = Article
\bibitem[{{Guillochon} et~al.(2011){Guillochon}, {Ramirez-Ruiz} and {Lin}}]{Guillochon_2011}
\bibinfo{author}{{Guillochon}, J.}, \bibinfo{author}{{Ramirez-Ruiz}, E.}, \bibinfo{author}{{Lin}, D.}, \bibinfo{year}{2011}.
\newblock \bibinfo{title}{{Consequences of the Ejection and Disruption of Giant Planets}}.
\newblock \bibinfo{journal}{The Astrophysical Journal} \bibinfo{volume}{732}, \bibinfo{pages}{74}.
\newblock \DOIprefix\doi{10.1088/0004-637X/732/2/74}, \href{http://arxiv.org/abs/1012.2382}{\tt arXiv:1012.2382}.
%Type = Article
\bibitem[{Gurumath et~al.(2019)Gurumath, Hiremath and Ramasubramanian}]{Gurumath}
\bibinfo{author}{Gurumath, S.R.}, \bibinfo{author}{Hiremath, K.M.}, \bibinfo{author}{Ramasubramanian, V.}, \bibinfo{year}{2019}.
\newblock \bibinfo{title}{{Angular Momentum of Stars and their Planets}}.
\newblock \bibinfo{journal}{Publications of the Astronomical Society of the Pacific} \bibinfo{volume}{131}, \bibinfo{pages}{014401}.
\newblock \DOIprefix\doi{10.1088/1538-3873/aae6b1}.
%Type = Article
\bibitem[{{Hay} et~al.(2016){Hay}, {Collier-Cameron}, {Doyle}, {H{\'e}brard}, {Skillen}, {Anderson}, {Barros}, {Brown}, {Bouchy}, {Busuttil}, {Delorme}, {Delrez}, {Demangeon}, {D{\'\i}az}, {Gillon}, {G{\'o}mez Maqueo Chew}, {Gonz{\`a}lez}, {Hellier}, {Holmes}, {Jarvis}, {Jehin}, {Joshi}, {Kolb}, {Lendl}, {Maxted}, {McCormac}, {Miller}, {Mortier}, {Pall{\'e}}, {Pollacco}, {Prieto-Arranz}, {Queloz}, {S{\'e}gransan}, {Simpson}, {Smalley}, {Southworth}, {Triaud}, {Turner}, {Udry}, {Vanhuysse}, {West} and {Wilson}}]{hay2016}
\bibinfo{author}{{Hay}, K.L.}, \bibinfo{author}{{Collier-Cameron}, A.}, \bibinfo{author}{{Doyle}, A.P.}, \bibinfo{author}{{H{\'e}brard}, G.}, \bibinfo{author}{{Skillen}, I.}, \bibinfo{author}{{Anderson}, D.R.}, \bibinfo{author}{{Barros}, S.C.C.}, \bibinfo{author}{{Brown}, D.J.A.}, \bibinfo{author}{{Bouchy}, F.}, \bibinfo{author}{{Busuttil}, R.}, \bibinfo{author}{{Delorme}, P.}, \bibinfo{author}{{Delrez}, L.}, \bibinfo{author}{{Demangeon}, O.}, \bibinfo{author}{{D{\'\i}az}, R.F.}, \bibinfo{author}{{Gillon}, M.}, \bibinfo{author}{{G{\'o}mez Maqueo Chew}, Y.}, \bibinfo{author}{{Gonz{\`a}lez}, E.}, \bibinfo{author}{{Hellier}, C.}, \bibinfo{author}{{Holmes}, S.}, \bibinfo{author}{{Jarvis}, J.F.}, \bibinfo{author}{{Jehin}, E.}, \bibinfo{author}{{Joshi}, Y.C.}, \bibinfo{author}{{Kolb}, U.}, \bibinfo{author}{{Lendl}, M.}, \bibinfo{author}{{Maxted}, P.F.L.}, \bibinfo{author}{{McCormac}, J.}, \bibinfo{author}{{Miller}, G.R.M.}, \bibinfo{author}{{Mortier}, A.}, \bibinfo{author}{{Pall{\'e}}, E.},
  \bibinfo{author}{{Pollacco}, D.}, \bibinfo{author}{{Prieto-Arranz}, J.}, \bibinfo{author}{{Queloz}, D.}, \bibinfo{author}{{S{\'e}gransan}, D.}, \bibinfo{author}{{Simpson}, E.K.}, \bibinfo{author}{{Smalley}, B.}, \bibinfo{author}{{Southworth}, J.}, \bibinfo{author}{{Triaud}, A.H.M.J.}, \bibinfo{author}{{Turner}, O.D.}, \bibinfo{author}{{Udry}, S.}, \bibinfo{author}{{Vanhuysse}, M.}, \bibinfo{author}{{West}, R.G.}, \bibinfo{author}{{Wilson}, P.A.}, \bibinfo{year}{2016}.
\newblock \bibinfo{title}{{WASP-92b, WASP-93b and WASP-118b: three new transiting close-in giant planets}}.
\newblock \bibinfo{journal}{Monthly Notices of the Royal Astronomical Society} \bibinfo{volume}{463}, \bibinfo{pages}{3276--3289}.
\newblock \DOIprefix\doi{10.1093/mnras/stw2090}, \href{http://arxiv.org/abs/1607.00774}{\tt arXiv:1607.00774}.
%Type = Article
\bibitem[{{Hellier} et~al.(2015){Hellier}, {Anderson}, {Collier Cameron}, {Delrez}, {Gillon}, {Jehin}, {Lendl}, {Maxted}, {Pepe}, {Pollacco}, {Queloz}, {S{\'e}gransan}, {Smalley}, {Smith}, {Southworth}, {Triaud}, {Turner}, {Udry} and {West}}]{hellier2015}
\bibinfo{author}{{Hellier}, C.}, \bibinfo{author}{{Anderson}, D.R.}, \bibinfo{author}{{Collier Cameron}, A.}, \bibinfo{author}{{Delrez}, L.}, \bibinfo{author}{{Gillon}, M.}, \bibinfo{author}{{Jehin}, E.}, \bibinfo{author}{{Lendl}, M.}, \bibinfo{author}{{Maxted}, P.F.L.}, \bibinfo{author}{{Pepe}, F.}, \bibinfo{author}{{Pollacco}, D.}, \bibinfo{author}{{Queloz}, D.}, \bibinfo{author}{{S{\'e}gransan}, D.}, \bibinfo{author}{{Smalley}, B.}, \bibinfo{author}{{Smith}, A.M.S.}, \bibinfo{author}{{Southworth}, J.}, \bibinfo{author}{{Triaud}, A.H.M.J.}, \bibinfo{author}{{Turner}, O.D.}, \bibinfo{author}{{Udry}, S.}, \bibinfo{author}{{West}, R.G.}, \bibinfo{year}{2015}.
\newblock \bibinfo{title}{{Three WASP-South Transiting Exoplanets: WASP-74b, WASP-83b, and WASP-89b}}.
\newblock \bibinfo{journal}{The Astronomical Journal} \bibinfo{volume}{150}, \bibinfo{pages}{18}.
\newblock \DOIprefix\doi{10.1088/0004-6256/150/1/18}, \href{http://arxiv.org/abs/1410.6358}{\tt arXiv:1410.6358}.
%Type = Article
\bibitem[{{Kawaler}(1988)}]{kawaler1988}
\bibinfo{author}{{Kawaler}, S.D.}, \bibinfo{year}{1988}.
\newblock \bibinfo{title}{{Angular Momentum Loss in Low-Mass Stars}}.
\newblock \bibinfo{journal}{Astrophysical Journal} \bibinfo{volume}{333}, \bibinfo{pages}{236}.
\newblock \DOIprefix\doi{10.1086/166740}.
%Type = Article
\bibitem[{Kraft(1967)}]{Kraft}
\bibinfo{author}{Kraft, R.P.}, \bibinfo{year}{1967}.
\newblock \bibinfo{title}{{Studies of Stellar Rotation. V. The Dependence of Rotation on Age among Solar-Type Stars}}.
\newblock \bibinfo{journal}{The Astrophysical journal} \bibinfo{volume}{150}, \bibinfo{pages}{551}.
\newblock \DOIprefix\doi{10.1086/149359}.
%Type = Article
\bibitem[{{Kramm} et~al.(2011){Kramm}, {Nettelmann}, {Redmer} and {Stevenson}}]{refId0}
\bibinfo{author}{{Kramm}, U.}, \bibinfo{author}{{Nettelmann}, N.}, \bibinfo{author}{{Redmer}, R.}, \bibinfo{author}{{Stevenson}, D.J.}, \bibinfo{year}{2011}.
\newblock \bibinfo{title}{{On the degeneracy of the tidal Love number k$_{2}$ in multi-layer planetary models: application to Saturn and GJ 436b}}.
\newblock \bibinfo{journal}{Astronomy \& Astrophysics} \bibinfo{volume}{528}, \bibinfo{pages}{A18}.
\newblock \DOIprefix\doi{10.1051/0004-6361/201015803}, \href{http://arxiv.org/abs/1101.0997}{\tt arXiv:1101.0997}.
%Type = Article
\bibitem[{{Lanza}(2010)}]{lanza2010}
\bibinfo{author}{{Lanza}, A.F.}, \bibinfo{year}{2010}.
\newblock \bibinfo{title}{{Hot Jupiters and the evolution of stellar angular momentum}}.
\newblock \bibinfo{journal}{Astronomy \& Astrophysics} \bibinfo{volume}{512}, \bibinfo{pages}{A77}.
\newblock \DOIprefix\doi{10.1051/0004-6361/200912789}, \href{http://arxiv.org/abs/0912.4585}{\tt arXiv:0912.4585}.
%Type = Inproceedings
\bibitem[{{Lanza}(2022)}]{lanza2022}
\bibinfo{author}{{Lanza}, A.F.}, \bibinfo{year}{2022}.
\newblock \bibinfo{title}{{The Role of Interactions Between Stars and Their Planets}}, in: \bibinfo{editor}{{Biazzo}, K.}, \bibinfo{editor}{{Bozza}, V.}, \bibinfo{editor}{{Mancini}, L.}, \bibinfo{editor}{{Sozzetti}, A.} (Eds.), \bibinfo{booktitle}{Demographics of Exoplanetary Systems, Lecture Notes of the 3rd Advanced School on Exoplanetary Science}, pp. \bibinfo{pages}{85--140}.
\newblock \DOIprefix\doi{10.1007/978-3-030-88124-5_2}.
%Type = Article
\bibitem[{{Luo} et~al.(2023){Luo}, {He}, {Tong} and {Li}}]{he}
\bibinfo{author}{{Luo}, J.}, \bibinfo{author}{{He}, H.Q.}, \bibinfo{author}{{Tong}, G.S.}, \bibinfo{author}{{Li}, J.}, \bibinfo{year}{2023}.
\newblock \bibinfo{title}{{Quantifying the Key Factors Affecting the Escape of Planetary Atmospheres}}.
\newblock \bibinfo{journal}{The Astrophysical Journal} \bibinfo{volume}{951}, \bibinfo{pages}{136}.
\newblock \DOIprefix\doi{10.3847/1538-4357/acd330}.
%Type = Inproceedings
\bibitem[{{Mardling}(2011a)}]{mardling2011}
\bibinfo{author}{{Mardling}, R.A.}, \bibinfo{year}{2011}a.
\newblock \bibinfo{title}{{Bodily Tides}}, in: \bibinfo{booktitle}{European Physical Journal Web of Conferences}, p. \bibinfo{pages}{03002}.
\newblock \DOIprefix\doi{10.1051/epjconf/20101103002}.
%Type = Inproceedings
\bibitem[{{Mardling}(2011b)}]{mardling2010}
\bibinfo{author}{{Mardling}, R.A.}, \bibinfo{year}{2011}b.
\newblock \bibinfo{title}{{Tidal evolution of star-planet systems}}, in: \bibinfo{editor}{{Sozzetti}, A.}, \bibinfo{editor}{{Lattanzi}, M.G.}, \bibinfo{editor}{{Boss}, A.P.} (Eds.), \bibinfo{booktitle}{The Astrophysics of Planetary Systems: Formation, Structure, and Dynamical Evolution}, pp. \bibinfo{pages}{238--242}.
\newblock \DOIprefix\doi{10.1017/S1743921311020242}.
%Type = Article
\bibitem[{{Matsumura} et~al.(2010){Matsumura}, {Peale} and {Rasio}}]{matsu2010}
\bibinfo{author}{{Matsumura}, S.}, \bibinfo{author}{{Peale}, S.J.}, \bibinfo{author}{{Rasio}, F.A.}, \bibinfo{year}{2010}.
\newblock \bibinfo{title}{{Tidal Evolution of Close-in Planets}}.
\newblock \bibinfo{journal}{Astrophysical Journal} \bibinfo{volume}{725}, \bibinfo{pages}{1995--2016}.
\newblock \DOIprefix\doi{10.1088/0004-637X/725/2/1995}, \href{http://arxiv.org/abs/1007.4785}{\tt arXiv:1007.4785}.
%Type = Article
\bibitem[{{Neron de Surgy} and {Laskar}(1997)}]{neron}
\bibinfo{author}{{Neron de Surgy}, O.}, \bibinfo{author}{{Laskar}, J.}, \bibinfo{year}{1997}.
\newblock \bibinfo{title}{{On the long term evolution of the spin of the Earth.}}
\newblock \bibinfo{journal}{Astronomy \& Astrophysics} \bibinfo{volume}{318}, \bibinfo{pages}{975--989}.
%Type = Article
\bibitem[{{Ogilvie}(2014)}]{ogilvie2014}
\bibinfo{author}{{Ogilvie}, G.I.}, \bibinfo{year}{2014}.
\newblock \bibinfo{title}{{Tidal Dissipation in Stars and Giant Planets}}.
\newblock \bibinfo{journal}{Annual Review of Astronomy and Astrophysics} \bibinfo{volume}{52}, \bibinfo{pages}{171--210}.
\newblock \DOIprefix\doi{10.1146/annurev-astro-081913-035941}, \href{http://arxiv.org/abs/1406.2207}{\tt arXiv:1406.2207}.
%Type = Article
\bibitem[{{Paxton} et~al.(2015){Paxton}, {Marchant}, {Schwab}, {Bauer}, {Bildsten}, {Cantiello}, {Dessart}, {Farmer}, {Hu}, {Langer}, {Townsend}, {Townsley} and {Timmes}}]{Paxton_2015}
\bibinfo{author}{{Paxton}, B.}, \bibinfo{author}{{Marchant}, P.}, \bibinfo{author}{{Schwab}, J.}, \bibinfo{author}{{Bauer}, E.B.}, \bibinfo{author}{{Bildsten}, L.}, \bibinfo{author}{{Cantiello}, M.}, \bibinfo{author}{{Dessart}, L.}, \bibinfo{author}{{Farmer}, R.}, \bibinfo{author}{{Hu}, H.}, \bibinfo{author}{{Langer}, N.}, \bibinfo{author}{{Townsend}, R.H.D.}, \bibinfo{author}{{Townsley}, D.M.}, \bibinfo{author}{{Timmes}, F.X.}, \bibinfo{year}{2015}.
\newblock \bibinfo{title}{{Modules for Experiments in Stellar Astrophysics (MESA): Binaries, Pulsations, and Explosions}}.
\newblock \bibinfo{journal}{The Astrophysical Journal Supplement Series} \bibinfo{volume}{220}, \bibinfo{pages}{15}.
\newblock \DOIprefix\doi{10.1088/0067-0049/220/1/15}, \href{http://arxiv.org/abs/1506.03146}{\tt arXiv:1506.03146}.
%Type = Inproceedings
\bibitem[{{Peale}(1977)}]{1977plsa}
\bibinfo{author}{{Peale}, S.J.}, \bibinfo{year}{1977}.
\newblock \bibinfo{title}{{Rotation Histories of the Natural Satellites}}, in: \bibinfo{booktitle}{IAU Colloq. 28: Planetary Satellites}, p.~\bibinfo{pages}{87}.
%Type = Article
\bibitem[{{Poppenhaeger} and {Wolk}(2014)}]{poppen2014}
\bibinfo{author}{{Poppenhaeger}, K.}, \bibinfo{author}{{Wolk}, S.J.}, \bibinfo{year}{2014}.
\newblock \bibinfo{title}{{Indications for an influence of hot Jupiters on the rotation and activity of their host stars}}.
\newblock \bibinfo{journal}{Astronomy \& Astrophysics} \bibinfo{volume}{565}, \bibinfo{pages}{L1}.
\newblock \DOIprefix\doi{10.1051/0004-6361/201423454}, \href{http://arxiv.org/abs/1404.1073}{\tt arXiv:1404.1073}.
%Type = Book
\bibitem[{{Press} et~al.(1992){Press}, {Teukolsky}, {Vetterling} and {Flannery}}]{press1992}
\bibinfo{author}{{Press}, W.H.}, \bibinfo{author}{{Teukolsky}, S.A.}, \bibinfo{author}{{Vetterling}, W.T.}, \bibinfo{author}{{Flannery}, B.P.}, \bibinfo{year}{1992}.
\newblock \bibinfo{title}{{Numerical recipes in C. The art of scientific computing}}.
%Type = Article
\bibitem[{{Privitera} et~al.(2016){Privitera}, {Meynet}, {Eggenberger}, {Vidotto}, {Villaver} and {Bianda}}]{privitera2016}
\bibinfo{author}{{Privitera}, G.}, \bibinfo{author}{{Meynet}, G.}, \bibinfo{author}{{Eggenberger}, P.}, \bibinfo{author}{{Vidotto}, A.A.}, \bibinfo{author}{{Villaver}, E.}, \bibinfo{author}{{Bianda}, M.}, \bibinfo{year}{2016}.
\newblock \bibinfo{title}{{Star-planet interactions. I. Stellar rotation and planetary orbits}}.
\newblock \bibinfo{journal}{Astronomy \& Astrophysics} \bibinfo{volume}{591}, \bibinfo{pages}{A45}.
\newblock \DOIprefix\doi{10.1051/0004-6361/201528044}, \href{http://arxiv.org/abs/1604.06005}{\tt arXiv:1604.06005}.
%Type = Article
\bibitem[{{Psaridi} et~al.(2023){Psaridi}, {Bouchy}, {Lendl}, {Akinsanmi}, {Stassun}, {Smalley}, {Armstrong}, {Howard}, {Ulmer-Moll}, {Grieves}, {Barkaoui}, {Rodriguez}, {Bryant}, {Su{\'a}rez}, {Guillot}, {Evans}, {Attia}, {Wittenmyer}, {Yee}, {Collins}, {Zhou}, {Galland}, {Parc}, {Udry}, {Figueira}, {Ziegler}, {Mordasini}, {Winn}, {Seager}, {Jenkins}, {Twicken}, {Brahm}, {Jones}, {Abe}, {Addison}, {Brice{\~n}o}, {Briegal}, {Collins}, {Daylan}, {Eigm{\"u}ller}, {Furesz}, {Guerrero}, {Hagelberg}, {Heitzmann}, {Hounsell}, {Huang}, {Krenn}, {Law}, {Mann}, {McCormac}, {M{\'e}karnia}, {Mounzer}, {Nielsen}, {Osborn}, {Reinarz}, {Sefako}, {Steiner}, {Str{\o}m}, {Triaud}, {Vanderspek}, {Vanzi}, {Vines}, {Watson}, {Wright} and {Zapata}}]{psaridi2023}
\bibinfo{author}{{Psaridi}, A.}, \bibinfo{author}{{Bouchy}, F.}, \bibinfo{author}{{Lendl}, M.}, \bibinfo{author}{{Akinsanmi}, B.}, \bibinfo{author}{{Stassun}, K.G.}, \bibinfo{author}{{Smalley}, B.}, \bibinfo{author}{{Armstrong}, D.J.}, \bibinfo{author}{{Howard}, S.}, \bibinfo{author}{{Ulmer-Moll}, S.}, \bibinfo{author}{{Grieves}, N.}, \bibinfo{author}{{Barkaoui}, K.}, \bibinfo{author}{{Rodriguez}, J.E.}, \bibinfo{author}{{Bryant}, E.M.}, \bibinfo{author}{{Su{\'a}rez}, O.}, \bibinfo{author}{{Guillot}, T.}, \bibinfo{author}{{Evans}, P.}, \bibinfo{author}{{Attia}, O.}, \bibinfo{author}{{Wittenmyer}, R.A.}, \bibinfo{author}{{Yee}, S.W.}, \bibinfo{author}{{Collins}, K.A.}, \bibinfo{author}{{Zhou}, G.}, \bibinfo{author}{{Galland}, F.}, \bibinfo{author}{{Parc}, L.}, \bibinfo{author}{{Udry}, S.}, \bibinfo{author}{{Figueira}, P.}, \bibinfo{author}{{Ziegler}, C.}, \bibinfo{author}{{Mordasini}, C.}, \bibinfo{author}{{Winn}, J.N.}, \bibinfo{author}{{Seager}, S.}, \bibinfo{author}{{Jenkins}, J.M.},
  \bibinfo{author}{{Twicken}, J.D.}, \bibinfo{author}{{Brahm}, R.}, \bibinfo{author}{{Jones}, M.I.}, \bibinfo{author}{{Abe}, L.}, \bibinfo{author}{{Addison}, B.}, \bibinfo{author}{{Brice{\~n}o}, C.}, \bibinfo{author}{{Briegal}, J.T.}, \bibinfo{author}{{Collins}, K.I.}, \bibinfo{author}{{Daylan}, T.}, \bibinfo{author}{{Eigm{\"u}ller}, P.}, \bibinfo{author}{{Furesz}, G.}, \bibinfo{author}{{Guerrero}, N.M.}, \bibinfo{author}{{Hagelberg}, J.}, \bibinfo{author}{{Heitzmann}, A.}, \bibinfo{author}{{Hounsell}, R.}, \bibinfo{author}{{Huang}, C.X.}, \bibinfo{author}{{Krenn}, A.}, \bibinfo{author}{{Law}, N.M.}, \bibinfo{author}{{Mann}, A.W.}, \bibinfo{author}{{McCormac}, J.}, \bibinfo{author}{{M{\'e}karnia}, D.}, \bibinfo{author}{{Mounzer}, D.}, \bibinfo{author}{{Nielsen}, L.D.}, \bibinfo{author}{{Osborn}, A.}, \bibinfo{author}{{Reinarz}, Y.}, \bibinfo{author}{{Sefako}, R.R.}, \bibinfo{author}{{Steiner}, M.}, \bibinfo{author}{{Str{\o}m}, P.A.}, \bibinfo{author}{{Triaud}, A.H.M.J.}, \bibinfo{author}{{Vanderspek}, R.},
  \bibinfo{author}{{Vanzi}, L.}, \bibinfo{author}{{Vines}, J.I.}, \bibinfo{author}{{Watson}, C.A.}, \bibinfo{author}{{Wright}, D.J.}, \bibinfo{author}{{Zapata}, A.}, \bibinfo{year}{2023}.
\newblock \bibinfo{title}{{Three Saturn-mass planets transiting F-type stars revealed with TESS and HARPS. TOI-615b, TOI-622b, and TOI-2641b}}.
\newblock \bibinfo{journal}{Astronomy \& Astrophysics} \bibinfo{volume}{675}, \bibinfo{pages}{A39}.
\newblock \DOIprefix\doi{10.1051/0004-6361/202346406}, \href{http://arxiv.org/abs/2303.15080}{\tt arXiv:2303.15080}.
%Type = Article
\bibitem[{{Scafetta} and {Bianchini}(2022)}]{scafetta2022}
\bibinfo{author}{{Scafetta}, N.}, \bibinfo{author}{{Bianchini}, A.}, \bibinfo{year}{2022}.
\newblock \bibinfo{title}{{The Planetary Theory of Solar Activity Variability: A Review}}.
\newblock \bibinfo{journal}{Frontiers in Astronomy and Space Sciences} \bibinfo{volume}{9}, \bibinfo{pages}{937930}.
\newblock \DOIprefix\doi{10.3389/fspas.2022.937930}, \href{http://arxiv.org/abs/2208.09293}{\tt arXiv:2208.09293}.
%Type = Article
\bibitem[{Schatzman(1962)}]{Schatzman}
\bibinfo{author}{Schatzman, E.}, \bibinfo{year}{1962}.
\newblock \bibinfo{title}{{A theory of the role of magnetic activity during star formation}}.
\newblock \bibinfo{journal}{Annales d'Astrophysique} \bibinfo{volume}{25}, \bibinfo{pages}{18}.
%Type = Incollection
\bibitem[{{Shkolnik} and {Llama}(2018)}]{Shkolnik}
\bibinfo{author}{{Shkolnik}, E.L.}, \bibinfo{author}{{Llama}, J.}, \bibinfo{year}{2018}.
\newblock \bibinfo{title}{{Signatures of Star-Planet Interactions}}, in: \bibinfo{editor}{{Deeg}, H.J.}, \bibinfo{editor}{{Belmonte}, J.A.} (Eds.), \bibinfo{booktitle}{Handbook of Exoplanets}, p.~\bibinfo{pages}{20}.
\newblock \DOIprefix\doi{10.1007/978-3-319-55333-7_20}.
%Type = Article
\bibitem[{{Sibony} et~al.(2022){Sibony}, {Helled} and {Feldmann}}]{sibony2022}
\bibinfo{author}{{Sibony}, Y.}, \bibinfo{author}{{Helled}, R.}, \bibinfo{author}{{Feldmann}, R.}, \bibinfo{year}{2022}.
\newblock \bibinfo{title}{{The rotation of planet-hosting stars}}.
\newblock \bibinfo{journal}{Monthly Notices of the Royal Astronomical Society} \bibinfo{volume}{513}, \bibinfo{pages}{2057--2075}.
\newblock \DOIprefix\doi{10.1093/mnras/stac951}, \href{http://arxiv.org/abs/2204.01421}{\tt arXiv:2204.01421}.
%Type = Article
\bibitem[{Silva et~al.(2013)Silva, Nepomuceno, Soares and de~Freitas}]{Silvaetal13}
\bibinfo{author}{Silva, J.R.P.}, \bibinfo{author}{Nepomuceno, M.M.F.}, \bibinfo{author}{Soares, B.B.}, \bibinfo{author}{de~Freitas, D.B.}, \bibinfo{year}{2013}.
\newblock \bibinfo{title}{{Time-dependent Nonextensivity Arising from the Rotational Evolution of Solar-type Stars}}.
\newblock \bibinfo{journal}{The Astrophysical journal} \bibinfo{volume}{777}, \bibinfo{pages}{20}.
\newblock \DOIprefix\doi{10.1088/0004-637X/777/1/20}, \href{http://arxiv.org/abs/1308.6047}{\tt arXiv:1308.6047}.
%Type = Article
\bibitem[{Skumanich(1972)}]{sku}
\bibinfo{author}{Skumanich, A.}, \bibinfo{year}{1972}.
\newblock \bibinfo{title}{{Time Scales for Ca II Emission Decay, Rotational Braking, and Lithium Depletion}}.
\newblock \bibinfo{journal}{The Astrophysical Journal} \bibinfo{volume}{171}, \bibinfo{pages}{565}.
\newblock \DOIprefix\doi{10.1086/151310}.
%Type = Article
\bibitem[{{Spada} et~al.(2011){Spada}, {Lanzafame}, {Lanza}, {Messina} and {Collier Cameron}}]{spada2011}
\bibinfo{author}{{Spada}, F.}, \bibinfo{author}{{Lanzafame}, A.C.}, \bibinfo{author}{{Lanza}, A.F.}, \bibinfo{author}{{Messina}, S.}, \bibinfo{author}{{Collier Cameron}, A.}, \bibinfo{year}{2011}.
\newblock \bibinfo{title}{{Modelling the rotational evolution of solar-like stars: the rotational coupling time-scale}}.
\newblock \bibinfo{journal}{Monthly Notices of the Royal Astronomical Society} \bibinfo{volume}{416}, \bibinfo{pages}{447--456}.
\newblock \DOIprefix\doi{10.1111/j.1365-2966.2011.19052.x}, \href{http://arxiv.org/abs/1105.3125}{\tt arXiv:1105.3125}.
%Type = Article
\bibitem[{Stefani et~al.(2019)Stefani, Giesecke and Weier}]{Stefani2019}
\bibinfo{author}{Stefani, F.}, \bibinfo{author}{Giesecke, A.}, \bibinfo{author}{Weier, T.}, \bibinfo{year}{2019}.
\newblock \bibinfo{title}{{A Model of a Tidally Synchronized Solar Dynamo}}.
\newblock \bibinfo{journal}{Solar Physics} \bibinfo{volume}{294}, \bibinfo{pages}{60}.
\newblock \DOIprefix\doi{10.1007/s11207-019-1447-1}, \href{http://arxiv.org/abs/1803.08692}{\tt arXiv:1803.08692}.
%Type = Article
\bibitem[{{Strugarek} et~al.(2014){Strugarek}, {Brun}, {Matt} and {R{\'e}ville}}]{Strugarek}
\bibinfo{author}{{Strugarek}, A.}, \bibinfo{author}{{Brun}, A.S.}, \bibinfo{author}{{Matt}, S.P.}, \bibinfo{author}{{R{\'e}ville}, V.}, \bibinfo{year}{2014}.
\newblock \bibinfo{title}{{On the Diversity of Magnetic Interactions in Close-in Star-Planet Systems}}.
\newblock \bibinfo{journal}{The Astrophysical Journal} \bibinfo{volume}{795}, \bibinfo{pages}{86}.
\newblock \DOIprefix\doi{10.1088/0004-637X/795/1/86}, \href{http://arxiv.org/abs/1409.5268}{\tt arXiv:1409.5268}.
%Type = Article
\bibitem[{{Strugarek} et~al.(2015){Strugarek}, {Brun}, {Matt} and {R{\'e}ville}}]{Strug2015}
\bibinfo{author}{{Strugarek}, A.}, \bibinfo{author}{{Brun}, A.S.}, \bibinfo{author}{{Matt}, S.P.}, \bibinfo{author}{{R{\'e}ville}, V.}, \bibinfo{year}{2015}.
\newblock \bibinfo{title}{{Magnetic Games between a Planet and Its Host Star: The Key Role of Topology}}.
\newblock \bibinfo{journal}{Astrophysical Journal} \bibinfo{volume}{815}, \bibinfo{pages}{111}.
\newblock \DOIprefix\doi{10.1088/0004-637X/815/2/111}, \href{http://arxiv.org/abs/1511.02837}{\tt arXiv:1511.02837}.
%Type = Article
\bibitem[{Tsallis(1988)}]{tsallis1988}
\bibinfo{author}{Tsallis, C.}, \bibinfo{year}{1988}.
\newblock \bibinfo{title}{{Possible generalization of Boltzmann-Gibbs statistics}}.
\newblock \bibinfo{journal}{Journal of Statistical Physics} \bibinfo{volume}{52}, \bibinfo{pages}{479--487}.
\newblock \DOIprefix\doi{10.1007/BF01016429}.
%Type = Inproceedings
\bibitem[{{Vidotto}(2020)}]{vidotto2020}
\bibinfo{author}{{Vidotto}, A.A.}, \bibinfo{year}{2020}.
\newblock \bibinfo{title}{{Different types of star-planet interactions}}, in: \bibinfo{editor}{{Kosovichev}, A.}, \bibinfo{editor}{{Strassmeier}, S.}, \bibinfo{editor}{{Jardine}, M.} (Eds.), \bibinfo{booktitle}{Solar and Stellar Magnetic Fields: Origins and Manifestations}, pp. \bibinfo{pages}{259--267}.
\newblock \DOIprefix\doi{10.1017/S1743921319009979}, \href{http://arxiv.org/abs/1911.10915}{\tt arXiv:1911.10915}.
%Type = Inproceedings
\bibitem[{Vidotto(2020)}]{vidotto1}
\bibinfo{author}{Vidotto, A.A.}, \bibinfo{year}{2020}.
\newblock \bibinfo{title}{{Different types of star-planet interactions}}, in: \bibinfo{editor}{Kosovichev, A.}, \bibinfo{editor}{Strassmeier, S.}, \bibinfo{editor}{Jardine, M.} (Eds.), \bibinfo{booktitle}{Solar and Stellar Magnetic Fields: Origins and Manifestations}, pp. \bibinfo{pages}{259--267}.
\newblock \DOIprefix\doi{10.1017/S1743921319009979}, \href{http://arxiv.org/abs/1911.10915}{\tt arXiv:1911.10915}.
%Type = Article
\bibitem[{{Vidotto} et~al.(2014){Vidotto}, {Gregory}, {Jardine}, {Donati}, {Petit}, {Morin}, {Folsom}, {Bouvier}, {Cameron}, {Hussain}, {Marsden}, {Waite}, {Fares}, {Jeffers} and {do Nascimento}}]{vidotto2014}
\bibinfo{author}{{Vidotto}, A.A.}, \bibinfo{author}{{Gregory}, S.G.}, \bibinfo{author}{{Jardine}, M.}, \bibinfo{author}{{Donati}, J.F.}, \bibinfo{author}{{Petit}, P.}, \bibinfo{author}{{Morin}, J.}, \bibinfo{author}{{Folsom}, C.P.}, \bibinfo{author}{{Bouvier}, J.}, \bibinfo{author}{{Cameron}, A.C.}, \bibinfo{author}{{Hussain}, G.}, \bibinfo{author}{{Marsden}, S.}, \bibinfo{author}{{Waite}, I.A.}, \bibinfo{author}{{Fares}, R.}, \bibinfo{author}{{Jeffers}, S.}, \bibinfo{author}{{do Nascimento}, J.D.}, \bibinfo{year}{2014}.
\newblock \bibinfo{title}{{Stellar magnetism: empirical trends with age and rotation}}.
\newblock \bibinfo{journal}{Monthly Notices of the Royal Astronomical Society} \bibinfo{volume}{441}, \bibinfo{pages}{2361--2374}.
\newblock \DOIprefix\doi{10.1093/mnras/stu728}, \href{http://arxiv.org/abs/1404.2733}{\tt arXiv:1404.2733}.
%Type = Article
\bibitem[{{Winn} and {Holman}(2005)}]{Winn_2005}
\bibinfo{author}{{Winn}, J.N.}, \bibinfo{author}{{Holman}, M.J.}, \bibinfo{year}{2005}.
\newblock \bibinfo{title}{{Obliquity Tides on Hot Jupiters}}.
\newblock \bibinfo{journal}{The Astrophysical Journal Letters} \bibinfo{volume}{628}, \bibinfo{pages}{L159--L162}.
\newblock \DOIprefix\doi{10.1086/432834}, \href{http://arxiv.org/abs/astro-ph/0506468}{\tt arXiv:astro-ph/0506468}.
%Type = Article
\bibitem[{{Wong} et~al.(2014){Wong}, {Knutson}, {Cowan}, {Lewis}, {Agol}, {Burrows}, {Deming}, {Fortney}, {Fulton}, {Langton}, {Laughlin} and {Showman}}]{wong2014}
\bibinfo{author}{{Wong}, I.}, \bibinfo{author}{{Knutson}, H.A.}, \bibinfo{author}{{Cowan}, N.B.}, \bibinfo{author}{{Lewis}, N.K.}, \bibinfo{author}{{Agol}, E.}, \bibinfo{author}{{Burrows}, A.}, \bibinfo{author}{{Deming}, D.}, \bibinfo{author}{{Fortney}, J.J.}, \bibinfo{author}{{Fulton}, B.J.}, \bibinfo{author}{{Langton}, J.}, \bibinfo{author}{{Laughlin}, G.}, \bibinfo{author}{{Showman}, A.P.}, \bibinfo{year}{2014}.
\newblock \bibinfo{title}{{Constraints on the Atmospheric Circulation and Variability of the Eccentric Hot Jupiter XO-3b}}.
\newblock \bibinfo{journal}{Astrophysical Journal} \bibinfo{volume}{794}, \bibinfo{pages}{134}.
\newblock \DOIprefix\doi{10.1088/0004-637X/794/2/134}, \href{http://arxiv.org/abs/1407.1313}{\tt arXiv:1407.1313}.

\end{thebibliography}



\end{document}